**Inferring selective constraint from population genomic data suggests recent regulatory turnover in the human brain**

Daniel R. Schrider[1,*] and Andrew D. Kern[1,2]

[1]Department of Genetics, Rutgers University, Piscataway, NJ

[2]Human Genetics Institute of New Jersey, Piscataway, NJ

[*]Corresponding author: Department of Genetics, Rutgers University, 604 Allison Rd., Piscataway, NJ 08854. E-mail: dan.schrider@rutgers.edu

## ABSTRACT

The comparative genomics revolution of the past decade has enabled the discovery of functional elements in the human genome via sequence comparison. While that is so, an important class of elements, those specific to humans, is entirely missed by searching for sequence conservation across species. Here we present an analysis based on variation data among human genomes that utilizes a supervised machine learning approach for the identification of human specific purifying selection in the genome. Using only allele frequency information from the complete low coverage 1000 Genomes Project dataset in conjunction with a support vector machine trained from known functional and non-functional portions of the genome, we are able to accurately identify portions of the genome constrained by purifying selection. Our method identifies previously known human-specific gains or losses of function and uncovers many novel candidates. Candidate targets for gain and loss of function along the human lineage include numerous putative regulatory regions of genes essential for normal development of the central nervous system, including a significant enrichment of gain of function events near neurotransmitter receptor genes. These results are consistent with regulatory turnover being a key mechanism in the evolution of human-specific characteristics of brain development. Finally, we show that the majority of the genome is unconstrained by natural selection currently, in agreement with what has been estimated from phylogenetic methods but in sharp contrast to estimates based on transcriptomics or other high-throughput functional methods.

## INTRODUCTION

While computational and experimental approaches have identified the majority of protein-coding genes in humans, these coding sequences only account for ~1% of the genome. Determining the extent to which the remaining ~99% of the genome may be functional remains a major challenge for biology. To this end, recent experimental advances have facilitated the identification of regulatory regions (Johnson et al. 2007), non-coding RNAs (Guttman et al. 2010), histone modifications (Barski et al. 2007), and accessible chromatin (Boyle et al. 2008). Collectively, these experiments suggest that a substantial number of functional genomic elements reside in non-coding regions.

While these experimental approaches represent a promising avenue towards identifying non-coding functional elements in the genome, many of the putatively functional non-coding regions they identify may be inconsequential to the organism. For example, the ENCODE project (Dunham et al. 2012) integrated data from a variety of genome-wide experiments assessing expression, transcription factor binding, and other biochemical activities and concluded that 80.4% of the human genome is functional. However, if we define function as biochemical activity with fitness consequences for the organism, then evolutionary analyses tell a very different story (Graur et al. 2013). Under this definition, which we adopt here, functional regions of the genome will experience purifying (or negative) selection, which removes deleterious mutations from populations. Comparative genomic studies have identified regions of the human genome where substitutions occur less often than expected in the absence of selection, and have concluded that on the order of 5% of the human genome is functional (Chinwalla et al. 2002; Siepel et al. 2005; Lunter et al. 2006; Birney et al. 2007; Pollard et al. 2010)—far less than estimated by ENCODE. This disparity demonstrates that knowledge of purifying selection is essential for identifying functional regions of the genome.

One limitation of purely comparative genomic approaches to detect purifying selection is that selective constraint may not be detected if it is present in only a small portion of the phylogenetic tree being examined. A particularly interesting class of elements is therefore missed by these techniques: elements that have acquired selective constraint only recently in a single species (e.g., human-specific gains-of-function). Conversely, genomic regions experiencing a recent loss of selective constraint in only a single lineage may be misidentified as conserved throughout the phylogeny. Identifying these species-specific gain and loss of function events is critical to illuminating the genetic bases for species-specific biology. Yet while comparative genomic data may not be able to detect these events, population genetic data can be used to infer the current action of purifying selection within a single species. Within a population, purifying selection will confine deleterious mutations to relatively rare allele frequencies or eliminate them altogether. This process will also reduce variation at linked sites via background selection (Charlesworth et al. 1993). Together negative and background selection decrease the number of polymorphisms and the average derived allele frequencies of polymorphisms within and surrounding functional elements (fig. 1). Indeed, the marked reduction in diversity seen within and around coding regions in the human genome is consistent with the effects of background selection (McVicker et al. 2009; Hernandez et al. 2011; Lohmueller et al. 2011).

Here we describe a method exploiting the impact of negative selection on genetic diversity within populations to identify functional regions of the human genome. While recent studies have been able to leverage population genetic data to identify differences in the amount of purifying selection acting on different classes of sites (Pierron et al. 2012; Ward and Kellis 2012; Somel et al. 2013), we attempt to classify individual genomic regions as constrained or unconstrained by selection. In principle, this could be accomplished by comparing observed patterns of diversity to theoretical expectations. However, these expectations depend on the demographic history of the populations examined as well as the distribution of selection coefficients encountered by new mutations. Given that there is considerable uncertainty surrounding these selective and demographic parameters (Marth et al. 2003; Stajich and Hahn 2005; Eyre-Walker and Keightley 2007; Boyko et al. 2008), and given the extensive heterogeneity in recombination rates (McVean et al. 2004), as well as variation in mutation rate and data quality across the genome (Green and Ewing 2013), here we adopt a supervised machine learning approach to classification—where genomic windows of known class (i.e. functional or not) are used to algorithmically learn a set of criteria to predict the classes of genomic windows whose class membership is unknown.

In particular, we use a support vector machine (SVM) approach to classify sliding windows of the human genome as either experiencing purifying/background selection or as unconstrained based on the density and allele frequencies of single nucleotide polymorphisms (SNPs) in the 1000 Genomes dataset (Altshuler et al. 2012). SVMs are trained by finding the hyperplane that optimally separates two classes of data points from a training set (where the true class of each datum is known) (Vapnik and Lerner 1963), with each data point represented by a vector of multiple measured attributes or "features." The SVM can then be used to classify data points whose classes are not known *a priori* according to the side of the hyperplane on which their feature vectors are located. This classification is often performed after implicitly mapping feature vectors to a higher-dimensional space where the two classes are easier to separate (the "kernel trick"; Aizerman et al. 1964; Boser et al. 1992), allowing for non-linear discrimination.

Modern support vector machines can also learn hyperplanes that do not perfectly separate the entire training set (Cortes and Vapnik 1995)—a necessity when some of the training data themselves may have been misclassified. SVMs have proven highly effective in a variety of biological applications (Byvatov and Schneider 2003), yet have only begun to be applied to evolutionary questions (e.g. Pavlidis et al. 2010; Lin et al. 2011; Ronen et al. 2013; Schrider et al. 2015).

Because we use genomic variation data (shaped by demographic history) to train our classifier, it will be robust to non-equilibrium demographic events provided they typically have a similar effect on patterns of variation in constrained and unconstrained regions. Thus, this supervised machine learning approach allows us to sidestep the problem of learning a parameter-rich model of demography and selection. This is a particular strength of our method in that we can use the most comprehensive dataset on genomic variation, the 1000 Genomes collection, without having to fit a model consisting of dozens if not hundreds of parameters. Importantly, using real population genetic data to train our classifier will expose it to heterogeneity in mutation rate, recombination rate, and read depth.

Our resulting SVM is very effective on both simulated data and human population genetic data. Examining regions classified with high confidence, we find that the majority of the genome is unconstrained. Finally, by contrasting our classifications with phylogenetic conservation (fig. 1), we identify regions that appear to have experienced human-specific changes in selective constraint. Such regions are disproportionately found near genes involved in development of the central nervous system (CNS), and may point to important regulatory changes affecting the human brain. These results underscore the utility of population genetic data for revealing function within the human genome.

## METHODS

### Single nucleotide polymorphism data

We downloaded single nucleotide polymorphism (SNP) genotypes from Phase 1 of The 1000 Genomes Project (Altshuler et al. 2012); we ignored SNPs discovered in the exome and/or trio data but not the low-coverage whole-genome data in order to minimize variation in read depth across the genome, which affects the probability of discovering a polymorphism (Ajay et al. 2011). This data set contains 1,092 low coverage genomes; however, 28 pairs of individuals in this set are close relatives to one another. We removed one individual from each of these 28 pairs leaving a set of 1,064 unrelated individuals. These individuals and their populations of origin are listed in supplementary table S1.

### Genomes, gene annotations and other genomic features

For the purposes of counting SNPs and monomorphic sites in an unbiased manner, creating training sets, and performing various downstream analyses we downloaded a variety of data from version hg19 of the UCSC Genome Browser database (Kent et al. 2002; Meyer et al. 2013). These data included version GRCh37 of the human genome (Lander et al. 2001; Collins et al. 2004), with bases masked by RepeatMasker (http://www.repeatmasker.org) appearing in lower case, the UCSC gene annotation (Hsu et al. 2006), human-chimpanzee and human-macaque pairwise whole-genome alignments generated by BLASTZ (Schwartz et al. 2003), "mappability" scores for 50 bp reads (Derrien et al. 2012), regulatory regions from ORegAnno (Montgomery et al. 2006; Griffith et al. 2008), transcription factor binding sites from ENCODE (Dunham et al. 2012), lincRNAs (Trapnell et al. 2010; Cabili et al. 2011), small noncoding RNAs from miRBase (Griffiths-Jones et al. 2006; Lestrade and Weber 2006), gene-disease associations from the Genetic Association Database (Becker et al. 2004), disease-associated SNPs from genome-wide association studies compiled by Hindorff et al. (2009), and phastCons elements (Siepel et al. 2005). We also used phastCons elements called from an alignment of 29 mammalian genomes but ignoring the human state (Lindblad-Toh et al. 2011). Most of these data were downloaded using The UCSC Table Browser (Karolchik et al. 2004). We also downloaded the GENCODE v7 annotation including non-coding RNAs (Harrow et al. 2012) from www.gencodegenes.org, and Gene Ontology (GO) data from www.geneontology.org, and used the set of regulatory elements inferred to be gained or lost on the human lineage by Cotney et al. (Cotney et al. 2013).

**Inferring ancestral states and removing uninformative sites**

Because we sought to use the derived (or "unfolded") site frequency spectrum, we attempted to determine the ancestral state of each site containing a SNP. This was done by parsimony using whole genome alignments of human and chimpanzee (Mikkelsen et al. 2005) and human and rhesus macaque (Gibbs et al. 2007). For each SNP, we compared the chimpanzee and macaque genomes. If both genomes exhibited the same nucleotide as one another and as one of the two human alleles, we inferred that this nucleotide was the ancestral state. Otherwise, we considered the ancestral state to be ambiguous and ignored the SNP. If only one of the chimpanzee or macaque genomes had a base call at the site, we inferred that this base was the ancestral state if it agreed with either human allele and considered the ancestral state to be ambiguous otherwise. We also considered the ancestral state to be ambiguous if neither chimpanzee nor macaque had a base call at the site. All SNPs whose ancestral state could not be inferred unambiguously according to these rules were considered as uninformative. While our ancestral state inferences may contain errors, our machine learning strategy should be robust if such mis-orientation errors also appear in our training set.

We aimed to use not only SNP allele frequencies, but also the fraction of monomorphic sites in a given region in order to classify it as constrained or unconstrained. Thus, eliminating biases affecting the fraction of sites within a genomic region inferred to be polymorphic was essential for our analysis. Because we eliminated SNPs with ambiguous ancestral states, we therefore eliminated monomorphic sites with ambiguous ancestral states to prevent the failure of ancestral state reconstruction from biasing the density of polymorphisms. This was done by attempting to infer the ancestral state at each site in the genome using rules similar to those used for SNPs as described above, but with no requirement that the sole human allele equal the chimpanzee/macaque allele(s). I.e., we considered sites where chimpanzee and macaque alleles were both found but differed from one another, or where neither were found as having ambiguous ancestral states and considered these sites as uninformative.

In order to prevent biases related to accuracy of mapping short read sequences from affecting our analysis, we examined "mappability" scores calculated by Derrien et al. (2012). The mappability score for a given site is $1/n$, where $n$ is the number of distinct positions in the genome from which a read mapped to this site could be derived (allowing two mismatches). For example, a site lying in a sequence motif occurring three times in the genome would have a score of 1/3, while a site in unique sequence would have a score of 1. We examined all adjacent 1 kb windows across the human genome and found a significant positive correlation with average mappability score and the number of SNPs called from the 1000 Genomes data ($\rho$=0.068; $P<2.2\times10^{-16}$). Windows in the lowest mappability score bin contained 7.9 SNPs on average, while windows with a mappability score of one averaged 13.6 SNPs (supplementary fig. S1).

The lack of SNP calls within regions of low mappability shows that poor mapping quality prevents high confidence SNP detection—this underscores the importance of accounting for mappability when examining the density of SNPs or other polymorphisms. We therefore considered only sites with mappability scores of 1 to be informative. Similarly, sites masked by RepeatMasker were considered uninformative. All uninformative sites were ignored when calculating the site frequency spectrum for a given window as described in the following section, and therefore had no impact on SVM training or classification.

**Estimating a modified site frequency spectrum in genomic windows**

Our goal in this study was to accurately classify genomic windows of a given size as constrained or unconstrained by purifying selection. The practical utility of this approach depends on the size of the windows: small windows may be difficult to classify accurately as they have fewer informative sites, while larger windows provide lower resolution. To find an appropriate balance between accuracy and resolution, we attempted to train classifiers using 5 kb, 10 kb, and 20 kb windows; windows of these sizes contain 65, 130, and 260 SNPs and 2,176, 4,352, and 8,703 informative sites on average in the 1000 Genomes data, respectively.

We represented each window with the same modified version of the site frequency spectrum (SFS) used for simulated data set (as described in supplementary text S1): $\boldsymbol{\xi}$=[$\xi_0$ $\xi_1$ $\xi_2$ … $\xi_{n\text{-}1}$] where $\xi_i$ is the fraction of informative sites in the window having a SNP whose derived allele is present in *i* chromosomes, and *n* is the number of chromosomes in the sample (i.e., twice the number of diploid individuals). As with the simulated data, sites containing a fixed derived allele were included in $\xi_0$, as our goal was to use only polymorphism data to perform classification. However we did experiment with including derived fixations during training (as described below), finding that the gains in accuracy were quite modest (typically on the order of 1% or less; supplementary table S2).

We estimated the modified SFS for each window only from informative sites as defined above. As a consequence, for some windows the SFS was estimated from only a small number of sites. To prevent elevated uncertainty around these SFS estimates from confounding our classifier, we arbitrarily removed windows comprised of ≤25% informative sites. We refer to the remaining windows as informative windows.

Because SVMs allow for a large number of features, we are able to use the complete SFS rather than a small number of summary statistics to perform classification—this is an important advantage of our method insofar as condensing the entire SFS into a summary statistic such as Tajima's *D* (Tajima 1989) might remove valuable information. However, the full SFS in the 1000 Genomes data is quite sparse, containing 2,128 frequency bins but only ~130 SNPs per 10 kb window on average. We therefore experimented with grouping the SFS into different numbers of bins: 10, 25, 50, 100, 250, 500, 1,000, and 2,128 (no binning), in addition to the different genomic window sizes listed above. We found that classification was most effective with 1,000 bins, and that 10 kb windows yielded a good balance between resolution and accuracy (supplementary table S2).

**Training a support vector machine classifier**

For the purposes of extracting a training set from the human genome, we subdivided the genome into adjacent windows. We then labeled windows as constrained if they were composed of >25% sites conserved across vertebrates according to phastCons (Siepel et al. 2005), or unconstrained if they contained zero base pairs within vertebrate phastCons elements, GENCODE v7 exons including non-coding RNAs (Harrow et al. 2012), UCSC exons (Hsu et al. 2006), ENCODE transcription factor binding sites (Dunham et al. 2012), or ORegAnno regulatory elements (Montgomery et al. 2006; Griffith et al. 2008). Though the >25% phastCons cutoff for functional training data is arbitrary, only ~5% of the human genome is conserved across species; windows that are 25% conserved according to phastCons are thus very likely to encode important functions. Because the amount of observed divergence on the human branch will correlate with the amount of observed polymorphism within humans due to ascertainment bias (Kern 2009), when building our training set we used phastCons conserved elements obtained from examining only non-human mammals (Lindblad-Toh et al. 2011). The 25% conserved sequence cutoff was adjusted for 5 kb and 20 kb windows sizes to achieve appropriate sized training sets (supplementary table S2). To construct an unbiased training set, we included the same number of conserved and unconserved windows. Because for each training set examined below there were more unconserved than conserved windows, windows meeting the unconserved criteria were randomly selected until a set matching the conserved set in size was obtained (i.e., a balanced training set). For 10 kb windows, this training set contained 1,482 windows in total—741 windows met the criterion for inclusion in the functional set, and 741 of the 11,439 that met the nonfunctional criteria were randomly selected for inclusion in the nonfunctional set.

For each combination of bin size and window size, we conducted a grid search of the $C$ and $\gamma$ hyperparameters and assessed the accuracy of the resulting SVMs in the same manner as for our simulated data sets. The results of these grid searches are shown in supplementary table S2. Prior to training the SVM, we used LIBSVM's svm-scale to rescale the training data (with default parameters), saving the scalars for re-use prior to prediction. We then used LIBSVM's svm-train to learn an SVM from the entire training data set using the optimal number of bins (1,000) for 10 kb windows. The -b 1 option was used to allow estimation class membership probabilities during prediction. We used LIBSVM's plotroc.py python script to generate the ROC curve (supplementary fig. S2) for this SVM using 10-fold cross validation. We also used plotroc.py to generate the ROC curve on a balanced independent test set and calculate the area under the curve. For this test set windows with between 20 and 25 percent phastCons elements were labeled as functional while only windows with no phastCons conservation were labeled as nonfunctional.

**Predictions and element calls**

After training the SVM, we formatted all overlapping 10 kb windows (100 bp step size) for classification, and rescaled these windows using the same scalars used for the training set. We then used svm-pred to perform classification, using the -b 1 option to perform class probability estimates for each window. Next, we then combined all overlapping windows assigned to a given class with probability >0.95; LIBSVM calculates these probability estimates using Algorithm 2 from Wu *et al.* (2004). We refer to these regions as popCons elements when made up of windows classified as constrained, and as popUncons elements when made up of windows classified as unconstrained. We imposed this 95% probability cutoff in order to focus on windows classified with high-confidence. Finally, we removed elements having ≥20% of informative sites masked by the 1000 Genomes Project for having elevated or reduced read-depth or low mapping quality in order to limit the effect of these sources of error on our predictions. This was done using the strictMask files which impose stringent filters devised for population genetic analysis (available at http://www.1000genomes.org/). Note that because we performed classification on overlapping windows, it was possible for popCons elements and popUncons elements to overlap.

**Searching for evidence of human-specific gain and loss of function**

In order to find genomic regions experiencing gain or loss of selective pressure in humans only, we contrasted phylogenetic evidence for selective constraint from phastCons with population genetic evidence from popCons and popUncons elements. To find human-specific losses of function we examined popUncons elements made up of at least 15% vertebrate phastCons elements and cross-referenced this list with UCSC genes (Hsu et al. 2006) to search for compelling candidates. For human-specific gains of selective constraint, we examined popCons elements composed of <1% vertebrate phastCons elements, cross-referencing this list with UCSC genes and ORegAnno elements to find candidate regions. For this analysis, we only included elements with informative windows (on which classification was performed) within at most 100 kb of the element in each direction. Thus, the element must be flanked by regions that contain enough informative sites to be classified but do not exhibit a strong enough signal of selective constraint to be classified as popCons elements. This step is necessary to ensure that the target of purifying selection resides within the gain of function candidate element itself rather than some flanking functional element lacking enough informative sites to be classified. Candidate gain of function regions singled out in the text were also examined manually via the UCSC Genome Browser (Kent et al. 2002) to ensure that no flanking but unclassified element appeared to be the true target of selection. Patterns of phylogenetic conservation among primates, mammals, and vertebrates were examined using the phastCons (Siepel et al. 2005) and GERP (Davydov et al. 2010) tracks in the UCSC Genome Browser.

**Testing for enrichment of element calls with various genomic features**

To ask whether popCons elements overlapped more often than expected by chance with exons and other features listed in supplementary table S3, we first counted the number of base pairs lying within both a popCons element and within one of the features being tested for enrichment. Next, we permuted the popCons coordinates such that no two elements in the permuted data set overlapped (just as in the true set). For our popCons permutations, we ensured that every permuted element consisted entirely of windows that were classified one way or another by our SVM (i.e. "informative windows"); this step ensures that any systematic differences between informative and uninformative regions (e.g. repeat content or read mappability) will not produce spurious enrichment/depletion results. We were unable to meet this constraint when permuting popUncons elements, as our permutation algorithm of randomly placing the largest remaining element in an unoccupied portion of the genome and repeating would run out of available room to randomly place elements before terminating. Fortunately, this limitation likely makes our depletion results conservative, as our informative windows are enriched for many of the functional annotation categories listed in supplementary tables S3 and S4. For both popCons and popUncons permutations, we also ensured that no permuted elements had fewer than 80% of base pairs passing the 1000 Genomes Project's coverage and quality cutoffs in the same manner as described above for our filtering of popCons and popUncons elements.

We constructed 1000 such permuted data sets, and then compared each of these permuted sets with each of the data sets listed in supplementary table S3. For each comparison we counted the total number of base pairs lying within both sets. The *P*-value for each enrichment test was simply the number of permuted data sets exhibiting equal or greater overlap with the genomic feature being examined than the real popCons data set. For popUncons elements, we performed a similar test but counted permuted data sets exhibiting lesser or equal overlap to obtain a *P*-value for depletion.

We performed similar tests for gain and loss of function (GOF and LOF, respectively) candidate regions and sets of genomic features listed in supplementary table S4. These sets were obtained by applying the phastCons cutoffs we used to define GOF and LOF regions to our permuted sets. Specifically, each permuted GOF set was constructed by removing all elements from the corresponding permuted popCons set except those with <1% phastCons bases. Similarly, each permuted LOF set was constructed by removing all elements from the corresponding permuted popUncons set but those with >15% phastCons bases. In each case, the permuted set yielded more regions than the true candidate set, so we randomly sampled permuted sets of the correct size. Before testing our GOF candidates for enrichment of the genomic features in supplementary table S4, we removed from these sets of features all elements comprised of ≥1% phastCons bases. Similarly, we removed all genomic features comprised of ≤15% phastCons bases before testing for LOF candidates for enrichment.

We also used version 2.0.2 of GREAT (McLean et al. 2010) to ask whether GOF and LOF elements were preferentially located near genes of particular functional categories, relative to the set of all popCons or popUncons elements, respectively. We then repeated these tests on our permuted data, asking how often terms significantly enriched in our true data were enriched in the permuted data sets.

**Synonymous and nonsynonymous variation within popCons and popUnCons elements**

For orthogonal evidence that popCons and popUncons elements were correctly classified as conserved or unconserved, respectively, we examined coding SNPs within genes found in these regions. We counted the number of nonsynonymous and synonymous SNPs in each gene using the GENCODE annotation. Singleton SNPs were omitted from this analysis to limit the influence of sequencing/genotyping error.

**Recombination rates in popCons elements, popUncons elements, and training data**

We downloaded sex-averaged recombination rates calculated by Kong et al. (2010) from the UCSC Genome Browser Database (Meyer et al. 2013). These data show the average recombination rates within 10 kilobase windows. These rates are adjusted so that a rate value of 1 is the genome-wide average. We calculated the average rate for each element as the sum of the rates of each 10 kb window overlapping the element, with each the rate of each window weighted by the fraction of the element overlapped by the window.

**Human-specific substitutions from a four-way ape whole-genome alignment**

To locate human-specific substitutions and indels we first obtained an alignment consisting of human (hg19), chimpanzee (panTro2), gorilla (gorGor1), and orangutan (ponAbe2). To do this we obtained the multiz46way alignment from the UCSC genome browser (http://genome.ucsc.edu) and then extracted only these four sequences. Using this four-way alignment we then located human-specific changes using parsimony criteria requiring invariance in the other three great apes. To obtain counts of substitutions or indels per window of a given size throughout the genome we used the featureBits tool from the Kent source tree available from the UCSC Genome Browser group.

**Data availability**

Our popCons, popUncons, GOF, and LOF predictions are available in BED format on GitHub (https://github.com/kern-lab/popCons). We have also made these data accessible as a UCSC Genome Browser track hub (http://kerndev.rutgers.edu/~dan/popCons/hub.txt).

## RESULTS AND DISCUSSION

### Detecting negative selection in simulated data

We assessed the effectiveness of our SVM-based approach to detect selective constraint by performing forward simulations of functional 10 kb windows containing constrained elements of various sizes, and experiencing varying strengths of negative selection, as well as nonfunctional 10 kb windows evolving entirely under drift. Each simulation utilized one of three different mutation and recombination rates (supplementary text S1). These simulations were performed under the demographic model learned from Tennessen *et al.* (2012) as described in the supplementary text S1. This scenario models the divergence of Europeans and Africans and their subsequent population size dynamics. This demographic model is not meant to perfectly match the demographic history of our dataset, which contains samples from a variety of subpopulations across the globe. Rather, it was chosen simply because it models some events common to many human subpopulations (e.g. migration out of Africa, and recent exponential population size expansion). For the purposes of training and testing our SVMs, we represented the output from each simulation by a feature vector consisting of the window's site frequency spectrum (supplementary text S1).

After training, we assessed the accuracy of each SVM using an independent test set; this estimate typically closely matched that obtained from cross-validation during the grid search (1.6% lower on average; supplementary table S5), showing that our grid search does not lead to substantial overfitting. This important result implies that our cross-validation accuracies estimated from real data (see below) are probably reliable indicators of our method's effectiveness, even though the demographic and selective history of the 1000 Genomes population sample differs from that of our simulated populations. Moreover, we found that after imposing a >95% posterior probability cutoff (as we did when calling putative constrained and unconstrained elements from the 1000 Genomes data as discussed below), classification accuracy typically well-exceeded 95% (supplementary table S5).

Next, we assessed the effectiveness of a single SVM classifier on test sets with varying selection coefficients, selected element lengths, mutation rates, and recombination rates. For this analysis we used the classifier learned from regions evolving under drift from those with 75% of sites under selection with a selection coefficient of $2Ns$ of 100, and with variable mutation and recombination rates; we chose to test the classifier learned from these data because this SVM's cross-validation accuracy closely mirrored that of the SVM we learned from real genomic data (see below). Perhaps unsurprisingly, we found that accuracy with which we could discriminate between simulated functional and nonfunctional windows varied according to the fraction of the 10 kb window experiencing selective constraint. When a 2.5 kb subset of the region was under negative selection, accuracy was quite low, ranging from 50-60% and varying only slightly according to the strength of selection (supplementary fig. S3A). When the entire window was constrained accuracy was much higher (supplementary fig. S3A), typically ~90% or greater (supplementary table S6), with the one exception being cases where the mutation rate was low— in these cases unselected regions were often misclassified as selected. However, for these and other parameter combinations, the number of unselected regions misclassified as constrained decreases dramatically after imposing a 95% confidence cutoff (supplementary fig. S3B). On the other hand, we find that regions with a smaller number of selected sites may often be classified as unconstrained even after imposing this cutoff (supplementary fig. S3C). Thus, it may be difficult, using our approach, to confidently assert that a genomic window contains no functional sequence—a window experiencing selection at relatively few selected sites will be difficult to distinguish from an unconstrained window.

Because our SVM classifies every genomic window as either evolving under selective constraint or under drift, we reasoned that regions experiencing positive selection might be classified as constrained, as positive selection reduces diversity at linked sites (Maynard Smith and Haigh 1974). We thus simulated regions experiencing adaptive mutations (supplementary text S1), and asked how often each SVM described above classified such regions as experiencing selective constraint. The fraction of positively selected regions classified as constrained exhibited considerable variation across SVMs, governed in part by the extent to which diversity within the negatively selected regions used to train the SVM mirrored that within positively selected simulations: the absolute difference in average $\pi$ in the positively and negatively selected simulations was negatively correlated with the fraction of positively selected regions classified as constrained, (Spearman's $\rho$=-0.87; $P<2.2\times10^{-16}$). Thus, it appears that, depending on the strength and amount of negative selection acting on putatively functional windows used to train our SVM and the strength of recent selective sweeps occurring in the human genome, positively selected regions may often be classified as constrained.

In summary, extensive forward population genetic simulations show that our SVM approach is able to detect negative selection even in the face of the confounding effects of non-equilibrium demography. While we have greater ability to classify as functional those windows that are comprised more completely of selected sites and sites under stronger selection, we have very high specificity when detecting functional windows after imposing a strict 95% posterior probability cutoff, though we may classify windows with smaller numbers of functional base pairs as unconstrained. With these encouraging results in hand, we turn attention to empirical human data.

**Accurate classification of functional and nonfunctional windows**

We trained an SVM to classify 10 kb genomic windows as either constrained or unconstrained according to the same modified SFS used to classify simulated data (Methods) using LIBSVM (Chang and Lin 2011). For this we used data from 1,064 unrelated whole genome sequences included in Phase 1 of the 1000 Genomes Project (http://www.1000genomes.org; Altshuler et al. 2012; Methods). This data set contains one SNP every 76.9 bp on average—we hypothesized that this high density of polymorphism would allow for the detection of regions under purifying selection at high enough resolution to be of practical utility. We then trained our SVM as described in the Methods. Because cross validation accuracies achieved on the X were relatively low, perhaps due to limited training data (supplementary table S2), we only performed classification on the autosomal portion of the genome.

The optimal hyperparameter combination ($C$=2; $\gamma$=0.125) from the autosomal grid search resulted in a cross-validation accuracy of 87.79% (supplementary table S2; area under ROC curve=0.94; supplementary fig. S2). The full results of this grid search are shown in supplementary fig. S4. That many of the other parameter values neighboring the optimal combination were nearly as accurate suggests that we did not significantly overfit our training data. Moreover, we achieve high accuracy on an independent test set not used in the selection of hyperparameter values or training (area under curve=0.88). Further, simulation results (see above) demonstrate that cross-validation accuracy for our SVM is reflective of true accuracies under a broad range of models, suggesting that we are not dramatically overestimating our accuracy due to overfitting. Moreover, these high accuracies show that although levels of genetic diversity are impacted by forces other than natural selection such as drift and variation in mutation and recombination rates, supporting the notion that population genetic data can be used to distinguish constrained from unconstrained DNA (Schrider and Kern 2014).

We then used the optimal hyperparameters to train an SVM from the entire training set; this SVM was in turn used to classify every 10 kb window (with 100 bp step size) in the genome comprised of at least 25% informative sites as either constrained or unconstrained. Of 22,358,126 such genomic windows covering a total of 86.5% of the genome, the majority (16,836,483, or 75.3% of windows) were classified as unconstrained, in general agreement with comparative genomic studies (Shabalina et al. 2001; Chinwalla et al. 2002; Siepel et al. 2005; Lunter et al. 2006; Birney et al. 2007; Pollard et al. 2010). LIBSVM can be used to estimate posterior probabilities for classifications according to the distances between the classified feature vector and the discriminating hyperplane during cross-validation. In order to focus on windows classified with high-confidence, we imposed a 95% probability cutoff for windows assigned as constrained or unconstrained, a cutoff that we show to be quite conservative in our simulation study (see above). Overlapping windows classified as constrained with high-confidence were merged together into regions we refer to as popCons elements, and overlapping high-confidence unconstrained windows were merged into popUncons elements.

Because we trained our SVM to discriminate between regions with a fairly large fraction of conserved sites according to phastCons (>25%) and regions with zero conservation according to phastCons, regions with lower levels of conservation may not be properly classified. Indeed, this appears to often be the case in simulated data as discussed above. We therefore sought to directly assess our method's accuracy on regions with fewer functional sites by constructing several test sets with different amounts of selective constraint. We found windows with between 0% and 5% conserved sites according to phastCons are classified as popUncons elements by our classifier 33.2% of the time, while 16.4% of windows with 5-10% conservation are classified as popUncons elements, versus 8.9% of windows with 10-15% conservation and 5.3% of windows with 15-20% conservation (Table 1). These results imply that many of our popUncons elements may have a relatively small number of selected sites. While we do not have power to classify 10 kb windows as completely unconstrained by negative selection, the results from Table 1 imply that our popCons elements probably contain a substantially greater density of selected sites than popUncons on average.

Crucially, we sought to minimize the impact of variation in read depth and mapping quality on our predictions. We therefore only retained elements for which >80% of all informative sites met the strict read depth and mapping quality constraints imposed by the 1000 Genomes Consortium (Altshuler et al. 2012) for population genetic analyses using these data (Methods); these criteria enforce both strict minimum and maximum read depth as well as minimum mapping quality thresholds. This step may not be sufficient to completely eliminate the impact of variation in read depth on our predictions (Green and Ewing 2013). Such variation may thus contribute to the error rates that we have measured on our empirical test data sets.

We examined the amount and spectrum of genetic variation found in popCons and popUncons elements. Consistent with purifying and background selection acting on popCons elements popCons elements exhibit a much greater skew in the SFS toward lower frequency variants than do popUncons elements (fig. 2A) as well as much lower nucleotide diversity in ($\pi$=4.02×$10^{-4}$ in popCons elements and $\pi$=1.12×$10^{-3}$ in popUncons elements; fig. 2B). Thus, our classifier is segmenting the genome based on the amount and spectrum of genetic diversity, as expected.

**PopCons elements are enriched for features indicative of functionality**

To test if our predictions recover previously known functional elements, we asked whether popCons elements were enriched for various genomic features that may experience selective constraint, including coding sequences, phylogenetically conserved regions of the genome (phastCons elements), regulatory elements gained or lost on the human lineage, transcription factor binding sites and other oRegAnno regulatory elements, small noncoding RNAs, lincRNAs, disease-associated genes, and candidate SNPs from GWAS studies (Methods). The results of these enrichment tests are shown in supplementary table S3. After Bonferroni correction, PopCons elements were significantly enriched for, and popUncons elements depleted of, all of these features except of lincRNAs, GWAS SNPs, and regulatory elements lost in humans. These results show that our classifier correctly identifies constrained and unconstrained genomic regions as expected from current annotations, providing further evidence that our approach is not severely confounded by nonselective factors that impact genetic diversity.

Moreover, these results confirm that our predictions have practical utility despite their relatively coarse resolution in comparison to phylogenetic methods such as GERP (Davydov et al. 2010) and phastCons (Siepel et al. 2005).

As stated above, many more genomic windows were classified as unconstrained than constrained. When using only high-confidence windows, more than half of the genome lies within popUncons elements (50,378 elements; 53.8% of the autosomes); far more than in popCons elements (17,551 elements; 11.1% of the autosomes). popUncons elements are also much larger than popCons elements on average (28,695.2 bp versus 16,999.4 bp; $P<2.2\times10^{-16}$; Mann-Whitney *U*-test; fig. 2C). At face value this result seems to strongly reject the possibility that 80% of the human genome is functional (Dunham et al. 2012). However, our classifier does not have enough resolution to predict precisely which base pairs are functional and which are not—popUncons elements may be experiencing purifying selection weak enough to go undetected, and popCons elements probably contain many base pairs not directly under purifying selection but instead linked to sites undergoing negative selection (or recent positive selection; see simulation results). Nonetheless, our results suggest that only a small fraction of the genome is experiencing strong purifying selection, again in general agreement with comparative genomic analyses (Shabalina et al. 2001; Chinwalla et al. 2002; Siepel et al. 2005; Lunter et al. 2006; Birney et al. 2007; Pollard et al. 2010; Gulko et al. 2015).

**Identifying human-specific loss of function**

Comparative genomic studies have identified many genes lost in humans but present in other primates (Wang et al. 2006); these loss events are typically caused by a missense or other inactivating mutation and leave behind a pseudogene remnant (Schrider et al. 2009). It has been hypothesized that these loss of function (LOF) events often confer fitness advantages (Olson 1999), and there are several examples of putative adaptive losses occurring since the human-chimpanzee split (e.g. Hayakawa et al. 2006; Wang et al. 2006; Xue et al. 2006).

Using evidence of phylogenetic conservation in conjunction with our population genetic based predictions of conservation should allow for discovery of LOF events in the genome. That is, LOF events should have strong signatures of phylogenetic conservation but also reside within popUncons elements. Indeed, our classifier was able to recover several previously identified cases of putatively adaptive pseudogenization events. For example, *MYH16*, which encodes a protein that is found in the temporalis and masseter muscles and increases bite strength, has been inactivated in the human lineage (Stedman et al. 2004). It has been hypothesized that the loss of this protein has allowed for cranial expansion in humans (Stedman et al. 2004). This gene exhibits strong phylogenetic evidence for conservation within primates according to phastCons, but is largely contained within a popUncons element, consistent with human-specific loss of selective constraint. Additional human-specific losses of *CASP12* (Fischer et al. 2002) and *CMAH* (Chou et al. 1998; Irie et al. 1998), both of which appear to have been fixed by positive selection (Hayakawa et al. 2006; Wang et al. 2006; Xue et al. 2006), occur in regions conserved across species according to phastCons but are contained entirely in popUncons elements.

Perhaps the most striking pattern to emerge from studies of human-specific pseudogenization events is the large number of nonfunctional olfactory receptors (ORs) in the human genome (Rouquier et al. 1998). ORs appear to have experienced diminished selective constraint in primates (Rouquier et al. 1998; Young et al. 2002; Zhang and Firestein 2002), perhaps due to reduced dependence on olfaction after the gain of trichromatic vision (Gilad et al. 2004). This reduction appears to be particularly pronounced in humans (Gilad et al. 2003), with roughly two-thirds of human ORs being pseudogenes (Glusman et al. 2001). Many of these inactivation events are still segregating in human populations (Menashe et al. 2003), suggesting that the loss of these genes is ongoing.

We asked whether there was greater than expected overlap between popUncons elements and OR genes and found substantial and significant enrichment (1.23-fold enrichment; $P<0.001$, one-tailed permutation test; Methods). In fact, 272 of 395 autosomal ORs not annotated as pseudogenes by GENCODE were contained entirely within a popUncons element (versus 144.25 expected; $P<0.001$; one-tailed permutation test), while only 17 OR genes reside even partially within popCons elements (versus 46.43 expected; $P<0.001$; one-tailed permutation test). Given that background selection may cause a gene to exhibit reduced diversity even if it is not itself the target of purifying selection, our results imply that vast majority of OR genes in the human genome are currently experiencing little if any selective constraint. This is consistent with the elevated fraction of nonsynonsymous SNPs predicted to disrupt protein function in OR genes recently observed by Pierron et al. (2012).

We searched for previously unknown cases of human-specific LOF by examining popUncons elements with strong phylogenetic evidence for conservation. We identified a total of 496 popUncons elements of which at least 15% was conserved across vertebrates according to phastCons; we refer to this set of elements and candidate LOF regions. This heuristic cutoff of 15% conservation is three times the genome-wide average and four times the average within popUncons elements (supplementary fig. S4A), implying that these regions were subject to considerable selective constraint for the majority of vertebrate evolution. As discussed above, many of our popUncons elements may contain a small fraction of sites under selective constraint. This hinders our ability to detect complete loss of function with high confidence. However, given that we have defined LOF candidates as having >15% conservation across vertebrates (and they exhibit 18.56% conservation on average; supplementary fig. S4B), and that our classifier labels less than 5% of regions with this level of conservation as popUncons elements (Table 1), many of our 496 LOF candidates may have lost selective constraint at some of these previously conserved sites. This finding suggests that the loss of selective constraint on the human branch may have been a common occurrence, as suggested by Olson (1999).

Because we defined LOF candidates as regions where phylogenetic and population genetic signatures of purifying selection disagree (phastCons and popCons, respectively), they may be enriched for false positives, especially if functional turnover is a rare event. It is necessary to seek orthogonal evidence that these candidates may represent true losses of functional constraint. For this reason, we asked whether these candidates were enriched for any ontology categories. Such information can also aid in the separation of biologically meaningful candidates from spurious ones (i.e. candidates associated with an enriched functional category may more often represent true positives). This same line of reasoning also holds for gain of function candidates (discussed below).

First, we used GREAT (McLean et al. 2010) to determine whether these candidate LOF regions were enriched for particular functional categories compared to the set of all popUncons elements (though the results described below hold qualitatively when using the entire human genome as a background). Because GREAT examines genes and their flanking regions, it is able to identify the enrichment of elements within cis-regulatory regions of genes with a particular annotation (e.g. McLean et al. 2011) as well as the genes themselves. Using GREAT, we found that a variety of annotation terms were significantly enriched after correcting for multiple testing using $q$-values (false discovery rates). However, the most striking result was the enrichment of candidate LOF regions near genes expressed in the nervous system during various developmental stages in mice, including the developing forebrain, telencephalon, diencephalon, medulla oblongata, and optic stalk (all enriched structures shown in supplementary table S7). We repeated this analysis on our permuted data sets (Methods) and found that most of these terms very rarely, if ever, exhibited significant enrichment (at $q<0.05$) in the permuted data (supplementary table S7). The enrichment of these categories is driven largely by a set of transcription factors annotated with the Zinc finger, C2H2-type/integrase, DNA-binding InterPro domain, which is also enriched for the presence of nearby LOF candidates (2.27-fold enrichment; false discovery rate $q=2.54\times10^{-4}$). This result suggests that changes in the transcriptional regulation of genes may have been a common feature on the lineage leading to humans (King and Wilson 1975), with regulators of brain development playing an especially important role. We also found that LOF candidates were significantly depleted of various genomic features, including exons, disease-associated mutations, noncoding RNAs, and transcription factor binding sites (supplementary table S4; Methods). Together these results provide additional evidence that at least a portion of sites within many of our LOF candidates have recently lost selective constraint.

Several interesting candidate loci emerged from the GREAT analysis. For example, we found a LOF candidate located <150 bp downstream of the homeobox gene *EMX2* (fig. 3A). This gene is expressed in the cerebral cortex during embryonic development in mice (Simeone et al. 1992), where it is required for the proper assignment of area identity to neocortical cells, as is *PAX6* (Bishop et al. 2000), another homeobox gene which itself has two upstream LOF candidates. *EMX2* also plays a role in the development of the sensory and motor regions (Hamasaki et al. 2004). The gene is one of two human homologs of the Drosophila gene *empty* *spiracles* (or *ems*) which is required for development of the head as well as the posterior spiracles (Walldorf and Gehring 1992). We also find a LOF candidate region overlapping the 3' exon of *SIM1* (fig. 3B), the homolog of *sim* (*single-minded*), which is essential for proper neurogenesis in Drosophila (Thomas et al. 1988). *SIM1* is associated with obesity in humans (Holder et al. 2000) and in mice (Michaud et al. 2001) where it is required for the development of the paraventricular nucleus, which is responsible for appetite regulation among other functions (Michaud et al. 1998). Another candidate LOF region lies 7.5 kb downstream of *NR4A2* (also known as *NURR1*), a transcription factor expressed in the brain (Law et al. 1992) where it is involved in the production of dopamine neurons in mice (Saucedo-Cardenas et al. 1998). Mutations in this gene have been implicated in schizophrenia (Chen et al. 2001), Parkinson's disease (Le et al. 2002), and bipolar disorder (Buervenich et al. 2000). Intriguingly, *NR4A2* has experienced a human-specific change in the expression pattern it exhibits over the course of the lifespan in the lateral cerebellar cortex (Liu et al. 2012), which may be involved in language and other cognitive functions (Rilling 2006).

Additional transcription factors expressed in the mouse brain and involved in nervous system development and that are flanked or overlapped by candidate LOF regions include: two zinc finger homeobox genes involved in neuronal differentiation, *ZFHX3* (Miura et al. 1995), whose first coding exon overlaps a LOF region, and *ZFHX4* (Hemmi et al. 2006); myelin transcription factor 1 (*MYT1*), which is important for oligodenderocyte differentiation (Nielsen et al. 2004); *LMX1B*, which plays a role in hindbrain roof plate development (Mishima et al. 2009); *NEUROG3*, a gene that is important for neuronal determination (Sommer et al. 1996); and *PAX2*, which can result in brain defects in mice when deleted (Favor et al. 1996), and whose first 3 exons are contained within a LOF candidate. The presence of LOF candidate regions near these transcription factors suggests recent functional turnover at their regulatory regions. *NR4A2*'s human-specific expression pattern in the brain is consistent with this hypothesis.

One notable LOF candidate region not associated with an enriched category is found within the protocadherin β (PCDHB) cluster on chromosome 5, containing most of *PCDHB14* and the *PCDHB18* pseudogene. In addition to this LOF region, the PCDHB cluster contains four additional popUncons elements, three of which contain a fair amount of conserved sequence according to phastCons, though less than our 15% cutoff for LOF candidates: one element containing *PCDHB4* is made up of 10.3% conserved sequence (across vertebrates); a second element encompassing *PCDHB6* and the *PCDHB17* pseudogene is 8.3% conserved; and a third element covering most of *PCDHB15* is 7% conserved. In total, 6 of the 19 PCDHB genes are mostly contained within these five popUnCons elements which encompass over one-third of the nearly 200 kb gene cluster.

Protocadherin genes, including the PCDHB cluster, encode cell-cell adhesion molecules that are believed to play a role in the formation of synaptic connections (Frank and Kemler 2002). The large number of and functional diversity among these genes may contribute to the complexity of the network of synapses in the human brain (Shapiro and Colman 1999). Despite strong phylogenetic evidence of purifying selection—each of the 19 PCDHB genes is largely comprised of vertebrate phastCons elements—there is a fairly high rate of gene turnover in this cluster among mammals (Vanhalst et al. 2001). Indeed, 3 of the 19 genes in this cluster in humans are known to be pseudogenes. The prevalence of popUncons elements and pseudogenes among the PCDHB genes implies that their selective constraint is considerably reduced in humans. Such a change in selective pressure may have allowed for changes to the neural network in the human brain.

**Candidate human-specific gain of function events**

As our extensive simulation and cross-validation experiments show, we should have excellent specificity for detecting human specific gains of function (GOFs). We use a complementary approach to that described above to find GOFs—by searching the genome for those regions that show no signs of phylogenetic conservation but are contained within popCons elements. Unfortunately there are relatively few well-studied examples of previously nonfunctional sequences acquiring function recently in humans. We examined three known human-specific *de novo* genes identified by Knowles and McLysaght (2009), *CLLU1*, *C22orf45*, and *DNAH10OS*, to see if our approach could identify these candidates. Two of these genes *C22orf45* and *DNAH10Os*, were largely contained within popCons elements. However, these genes are found on the opposite strand of more ancient and conserved genes, thus negative selection on these older genes may be responsible for the popCons classification. Interestingly, the other gene, *CLLU1,* was found within a popUncons element and exhibits a ratio of nonsynonymous to synonymous SNPs in the 92nd percentile among all genes (Methods), suggesting that it may not be experiencing strong selective constraint.

While there are not enough known examples of *de novo* human functional elements for us to systematically assess our strategy, we can identify candidate GOF regions in a similar vein as our search for LOF regions. To this end we searched for popCons elements with little phylogenetic evidence of conservation and found 700 popCons elements composed of <1% of base pairs within phastCons elements. On average, 0.54% of nucleotides within these regions are conserved across vertebrates, versus 9.54% of nucleotides lying within the full set of popCons elements (supplementary fig. S5C-D). These candidate GOF regions are enriched for promoters/enhancers identified by Cotney et al. (2013) as present in humans but absent from mice, as well as small noncoding RNAs, with the latter remaining significant after Bonferonni correction (supplementary table S4; Methods).

As before for loss of function, we ran GREAT to identify functional categories of genes either overlapping or neighboring candidate GOF regions more often than expected by chance (using the set of all popCons elements as the background, though again we recover similar terms when using the whole genome as the background). Here we found a striking pattern: we observed significant enrichment of genes annotated with the Gene Ontology (GO) molecular function term "extracellular ligand-gated ion channel activity" (false discovery rate $q$=0.045). Indeed all enriched molecular function terms were related to GABA or other neurotransmitters. This enrichment was driven primarily by GOFs near genes annotated with the GO molecular function "GABA-A receptor activity" ($q$=0.022). GABA (γ-Aminobutyric acid) is the nervous system's primary inhibitory neurotransmitter (Petroff 2002), and GABA receptor expression patterns are known to play a key role in brain development (Lujan et al. 2005). As for LOFs, we found that these two terms were enriched at $q$<0.05 in only a small fraction of our permuted data sets (0.3% and 2.2% of permuted sets, respectively). Human-specific changes in function affecting either GABA sequences themselves or their flanking regions could thus have profound effects on the CNS. We therefore examined these GOF candidates more closely for evidence that they may have affected the human CNS after the split with chimpanzees.

We found five GOF regions within a cluster of three GABA receptor subunit genes (*GABRB3*, *GABRA5*, *GABRG3*) on chromosome 15. Three of these GOF candidates are located downstream of *GABRB3*, which Liu et al. (Liu et al. 2012) identified as having evolved a human-specific temporal expression pattern in the prefrontal cortex (PFC) after the human-chimpanzee divergence. *GABRB3* alleles have also been associated with autism (Buxbaum et al. 2002; Kim et al. 2007), savant skills (Nurmi et al. 2003), and epilepsy (Tanaka et al. 2008). The other two GOF candidates are located within introns of *GABRG3*. These GOFs contain several transcription factor binding sites identified by ENCODE ChIP-seq, including one ~400 bp peak observed in brain cancer cell lines among other tissues and containing 7 human-specific substitutions in an alignment of great apes (Methods). This is a relatively high density of changes occurring on the human branch: fewer than 2.5% of adjacent 500 bp windows in a whole-genome great ape alignment exhibit 7 or more human-specific substitutions or indels. We also observed three GOF regions within a cluster of four GABA receptors on chromosome 5. One of these appears within an intron of *GABRB2*, while the other two flank either side of *GABRG2*, which evolved a novel temporal expression pattern in the human PFC according to Liu et al. (Liu et al. 2012). Dysfunction of *GABRG2* appears to play a role in epilepsy (Hirose 2006; Tanaka et al. 2008) and alcohol dependence (Radel et al. 2005). Another GOF candidate is located upstream of *GABRA2* on chromosome 4, which like *GABRG2* and *GABRB3* experienced a human-specific change in PFC temporal expression pattern (Liu et al. 2012). *GABRA2* is also up-regulated following neuronal stimulation via exposure to potassium chloride (Liu et al. 2012), and has been associated with alcohol dependence (Edenberg et al. 2004; Dick et al. 2006). It is also worth noting that we found a LOF candidate within an intron of *GABBR2*, also singled out by Liu et al. as having evolved a human-specific expression pattern in the PFC (Liu et al. 2012).

The proximity of GOF and LOF candidates around GABA receptor genes implies that these candidate regions may be the site of regulatory turnover responsible for human-specific expression patterns of these genes in the prefrontal cortex. Moreover, the association of these genes with neurological phenotypes such as autism suggests that they play a crucial role in central nervous system development. Thus our findings, combined with Liu et al.'s observation that GABA receptors have experienced an unusually high rate of such changes in expression (Liu et al. 2012), strongly suggests that human-specific changes in selective pressure in these candidate regions may underlie important developmental differences between the brains of humans and chimpanzees.

The signal of GOFs near neurotransmitter receptors is not limited to GABA receptors—we also find several GOFs near subunits of receptors of glutamate, the primary excitatory neurotransmitter in the CNS. Glutamate is a GABA precursor (Petroff 2002), and glutamate signaling is vital for CNS development (Lujan et al. 2005). For example, we observe a GOF candidate upstream of *GRIK1* which encodes a glutamate receptor subunit. This gene has been associated with autism (Haldeman-Englert et al. 2010), Down syndrome (Ghosh et al. 2009), and juvenile absence epilepsy (Sander et al. 1997), and its expression levels are altered in patients with schizophrenia and bipolar disorder (Woo et al. 2007). We also find a GOF region within an intron and another downstream of *GRIK2*, a glutamate receptor subunit (fig. 4A). *GRIK2* has been linked to mental retardation (Motazacker et al. 2007), autism (Jamain et al. 2002), and schizophrenia (Bah et al. 2004), suggesting an important developmental role in the CNS. In addition, we find five GOF candidates in the vicinity of *GRID2* (two upstream and three intronic; fig. 4B), another glutamate receptor subunit which interacts directly with *GRIK2* (Kohda et al. 2003). Deletions in *GRID2* can result in cerebellar ataxia and related motor deficits (Utine et al. 2013) and delays in cognition and speech (Hills et al. 2013). Both *GRID2* and *GRIK2* were identified by Liu et al. as evolving a human-specific temporal expression profile in the lateral cerebellar cortex (Liu et al. 2012).

Although zinc finger genes were not enriched for GOF candiates according to GREAT, two GOFs located upstream of the brain-expressed *ZNF131* (Trappe et al. 2002) are notable because they harbor regulatory elements that may modulate its expression (fig. 5). The GOF candidate closest to the gene, ~9 kb upstream, encompasses a 1,051 bp ORegAnno element. Examining the great ape alignment we find 11 human-specific substitutions or indels within the ORegAnno element—this number is within the upper 2.5% tail of the empirical distribution of all adjacent 1 kb windows in the genome. These substitutions may have created regulatory features unique to humans. A second GOF element is located another 19 kb further upstream containing another ORegAnno element along with two noncoding RNAs with no annotated function. In addition, *ZNF131* is predicted by UNIPROT to function in the brain.

Overall, our results suggest the possibility that a substantial number of regions flanking or overlapping genes functioning in the CNS may have gained selective constraint specifically in humans. We see this pattern from not only the compelling individual cases presented above, but also from genome-wide enrichments of our predicted GOF elements. This pattern could result from the gain or modification of regulatory regions bringing about novel expression patterns. Such changes could in part be responsible for the dramatic differences in structure and function between the human brain and that of other primates. The fact that many of these genes have recently changed expression patterns in the human brain, combined with the significant enrichment of our GOF candidates for human-specific regulatory elements, shows the power of our approach of contrasting phylogenetic and population genetic data to find human-specific change of function.

## Concluding remarks

Understanding which portions of the human genome are functional is a central goal in modern biology. Here we have developed a supervised machine learning framework to detect purifying selection from population genetic data alone. Because our approach does not examine phylogenetic evidence for sequence conservation, it can be used to detect recent lineage-specific changes in selective pressure. We found through extensive simulations and cross-validation on the 1000 Genomes dataset that our method is highly accurate and can be used to identify candidate regions experiencing either gain or loss of function occurring after the human-chimpanzee divergence, successfully recovering known examples of the latter. Moreover, because our supervised machine learning approach does not depend on heavily parameterized models of human demographic history and selection, we are able to leverage all available human sequence data in our search.

While it has many advantages, our method does come with some caveats. Because we utilize the fraction of segregating sites in a region as well as their allele frequencies, variation in the spontaneous mutation rate across the genome could impact predictions. However, because we used supervised learning our classifier should be robust to such variation if it is well-represented in our training set or if its effect is modest compared to the impact of purifying selection. Our high accuracy rates show that this is the case.

On the other hand, our method does appear to be confounded by balancing selection, which is expected to increase variability within the population. For example, the *HLA* loci, the *ABO* locus, and the hemoglobin *HBB* gene, which are all highly polymorphic and believed to be experiencing balancing selection (Allison 1954; Hedrick and Thomson 1983; Saitou and Yamamoto 1997; Stajich and Hahn 2005), are all classified as unconstrained by our method. This limitation of our method is probably a minor one, as balancing selection in the human genome appears to be the exception rather than the rule (Bubb et al. 2006; Leffler et al. 2013).

Our method may also be confounded by selective sweeps, which we suspect will be classified as constrained because sweeps reduce the number of segregating sites and skew the SFS away from intermediate-frequency variants (though an excess of high-frequency variants is also observed at flanking sites; Fay and Wu 2000). This issue may not greatly affect accuracy as regions experiencing selective sweeps must contain functional DNA, and as with balancing selection, such sweeps seem to have little impact on human polymorphism genome-wide in any case (Hernandez et al. 2011; Lohmueller et al. 2011). However, strong selective sweeps can reduce diversity in large regions, potentially greatly inflating the inferred size of the functional region. Given sufficient numbers of examples of targets of positive or balancing selection, one could in principle train an SVM to identify these types of loci as well. Finally, our approach may not be able to differentiate recent changes in selective pressure affecting multiple lineages (e.g. occurring prior to the human-chimpanzee split) from truly lineage-specific changes. Dense polymorphism data from multiple species would allow us to discriminate between these two cases.

Despite these limitations, our approach appears to be quite useful for identifying candidate human-specific gains and losses of function. Indeed, while we cannot directly show that these candidate regions have experienced recent changes in selective constraint, the clustering of such candidates in loci affecting CNS development and exhibiting novel expression patterns in the human brain suggest that many of these candidates represent true gains or losses of function responsible for key human-specific traits, and that such functional turnover is common on evolutionary timescales. Furthermore, our method is complementary to previous strategies for identifying lineage-specific changes in selective pressure. For example, searches for sequences highly conserved in other species but evolving rapidly in humans reveals regions likely responsible for important human-specific adaptations (Pollard et al. 2006a; Pollard et al. 2006b; Kostka et al. 2012); however, the acquisition of new functional elements need not occur in previously conserved regions or be accompanied by a burst of substitution. Our approach does not depend on either of these two assumptions. Unfortunately, our resolution is currently limited by the relatively low density of polymorphism in humans. Nonetheless, the results presented here demonstrate the promise of leveraging population genetic data to detect selective constraint, an approach whose power will improve as more human genomes are sequenced.

## ACKNOWLEDGEMENTS

We thank K. Pollard for the phastCons elements from Linbald-Toh et al. (2011), and M. W. Hahn and D. J. Begun for comments on the manuscript. D. R. S. was supported by the National Institutes of Health under Ruth L. Kirschstein National Research Service Award F32 GM105231. A. D. K. was supported in part by Rutgers University and National Science Foundation Award MCB-1161367.

## REFERENCES

Aizerman A, Braverman EM and Rozoner L. 1964. Theoretical foundations of the potential function method in pattern recognition learning. Automation and remote control 25: 821-837.

Ajay SS, Parker SC, Abaan HO, Fajardo KVF and Margulies EH. 2011. Accurate and comprehensive sequencing of personal genomes. Genome research 21: 1498-1505.

Allison AC. 1954. Protection afforded by sickle-cell trait against subtertian malarial infection. British medical journal 1: 290.

Altshuler DM, Durbin RM, Abecasis GR, et al. 2012. An integrated map of genetic variation from 1,092 human genomes. Nature 491: 56-65.

Bah J, Quach H, Ebstein R, et al. 2004. Maternal transmission disequilibrium of the glutamate receptor GRIK2 in schizophrenia. Neuroreport 15: 1987-1991.
Barski A, Cuddapah S, Cui K, Roh T-Y, Schones DE, Wang Z, Wei G, Chepelev I and Zhao K. 2007. High-resolution profiling of histone methylations in the human genome. Cell 129: 823-837.
Becker KG, Barnes KC, Bright TJ and Wang SA. 2004. The genetic association database. Nature genetics 36: 431-432.
Birney E, Stamatoyannopoulos JA, Dutta A, et al. 2007. Identification and analysis of functional elements in 1% of the human genome by the ENCODE pilot project. Nature 447: 799-816.
Bishop KM, Goudreau G and O'Leary DD. 2000. Regulation of area identity in the mammalian neocortex by Emx2 and Pax6. Science 288: 344-349.
Boser BE, Guyon IM and Vapnik VN. 1992. A training algorithm for optimal margin classifiers. Proceedings of the fifth annual workshop on Computational learning theory: 144-152.
Boyko AR, Williamson SH, Indap AR, et al. 2008. Assessing the evolutionary impact of amino acid mutations in the human genome. PLoS genetics 4: e1000083.
Boyle AP, Davis S, Shulha HP, Meltzer P, Margulies EH, Weng Z, Furey TS and Crawford GE. 2008. High-resolution mapping and characterization of open chromatin across the genome. Cell 132: 311-322.
Bubb KL, Bovee D, Buckley D, et al. 2006. Scan of human genome reveals no new loci under ancient balancing selection. Genetics 173: 2165-2177.
Buervenich S, Carmine A, Arvidsson M, et al. 2000. NURR1 Mutations in cases of schizophrenia and manic-depressive disorder. American journal of medical genetics 96: 808-813.
Buxbaum J, Silverman J, Smith C, et al. 2002. Association between a GABRB3 polymorphism and autism. Molecular psychiatry 7: 311-316.
Byvatov E and Schneider G. 2003. Support vector machine applications in bioinformatics. Applied bioinformatics 2: 67.
Cabili MN, Trapnell C, Goff L, Koziol M, Tazon-Vega B, Regev A and Rinn JL. 2011. Integrative annotation of human large intergenic noncoding RNAs reveals global properties and specific subclasses. Genes & development 25: 1915-1927.
Chang C-C and Lin C-J. 2011. LIBSVM: a library for support vector machines. ACM Transactions on Intelligent Systems and Technology (TIST) 2: 27.
Charlesworth B, Morgan M and Charlesworth D. 1993. The effect of deleterious mutations on neutral molecular variation. Genetics 134: 1289-1303.
Chen YH, Tsai MT, Shaw CK and Chen CH. 2001. Mutation analysis of the human NR4A2 gene, an essential gene for midbrain dopaminergic neurogenesis, in schizophrenic patients. American journal of medical genetics 105: 753-757.
Chinwalla AT, Cook LL, Delehaunty KD, et al. 2002. Initial sequencing and comparative analysis of the mouse genome. Nature 420: 520-562.
Chou H-H, Takematsu H, Diaz S, et al. 1998. A mutation in human CMP-sialic acid hydroxylase occurred after the Homo-Pan divergence. Proceedings of the National Academy of Sciences 95: 11751-11756.
Collins F, Lander E, Rogers J, Waterston R and Conso I. 2004. Finishing the euchromatic sequence of the human genome. Nature 431: 931-945.

Cortes C and Vapnik V. 1995. Support-vector networks. Machine learning 20: 273-297.
Cotney J, Leng J, Yin J, Reilly SK, DeMare LE, Emera D, Ayoub AE, Rakic P and Noonan JP. 2013. The Evolution of Lineage-Specific Regulatory Activities in the Human Embryonic Limb. Cell 154: 185-196.
Davydov EV, Goode DL, Sirota M, Cooper GM, Sidow A and Batzoglou S. 2010. Identifying a high fraction of the human genome to be under selective constraint using GERP++. PLoS computational biology 6: e1001025.
Derrien T, Estellé J, Sola SM, Knowles DG, Raineri E, Guigó R and Ribeca P. 2012. Fast computation and applications of genome mappability. PLoS One 7: e30377.
Dick DM, Agrawal A, Schuckit MA, et al. 2006. Marital status, alcohol dependence, and GABRA2: evidence for gene-environment correlation and interaction. Journal of Studies on Alcohol and Drugs 67: 185.
Dunham I, Kundaje A, Aldred SF, et al. 2012. An integrated encyclopedia of DNA elements in the human genome. Nature 489: 57-74.
Edenberg HJ, Dick DM, Xuei X, et al. 2004. Variations in GABRA2, Encoding the α2 Subunit of the GABA(A) Receptor, Are Associated with Alcohol Dependence and with Brain Oscillations. The American Journal of Human Genetics 74: 705-714.
Eyre-Walker A and Keightley PD. 2007. The distribution of fitness effects of new mutations. Nature Reviews Genetics 8: 610-618.
Favor J, Sandulache R, Neuhäuser-Klaus A, et al. 1996. The mouse Pax21Neu mutation is identical to a human PAX2 mutation in a family with renal-coloboma syndrome and results in developmental defects of the brain, ear, eye, and kidney. Proceedings of the National Academy of Sciences 93: 13870-13875.
Fay JC and Wu C-I. 2000. Hitchhiking under positive Darwinian selection. Genetics 155: 1405-1413.
Fischer H, Koenig U, Eckhart L and Tschachler E. 2002. Human caspase 12 has acquired deleterious mutations. Biochemical and biophysical research communications 293: 722-726.
Frank M and Kemler R. 2002. Protocadherins. Current opinion in cell biology 14: 557-562.
Ghosh D, Sinha S, Chatterjee A and Nandagopal K. 2009. A study of GluK1 kainate receptor polymorphisms in Down syndrome reveals allelic non-disjunction at 1173 (C/T). Disease Markers 27: 45-54.
Gibbs RA, Rogers J, Katze MG, et al. 2007. Evolutionary and biomedical insights from the rhesus macaque genome. science 316: 222-234.
Gilad Y, Man O, Pääbo S and Lancet D. 2003. Human specific loss of olfactory receptor genes. Proceedings of the National Academy of Sciences 100: 3324-3327.
Gilad Y, Wiebe V, Przeworski M, Lancet D and Pääbo S. 2004. Loss of olfactory receptor genes coincides with the acquisition of full trichromatic vision in primates. PLoS biology 2: e5.
Glusman G, Yanai I, Rubin I and Lancet D. 2001. The complete human olfactory subgenome. Genome research 11: 685-702.
Graur D, Zheng Y, Price N, Azevedo RB, Zufall RA and Elhaik E. 2013. On the immortality of television sets:"function" in the human genome according to the evolution-free gospel of ENCODE. Genome biology and evolution 5: 578-590.
Green P and Ewing B. 2013. Comment on "Evidence of Abundant Purifying Selection in Humans for Recently Acquired Regulatory Functions". Science 340: 682-682.

Griffith OL, Montgomery SB, Bernier B, et al. 2008. ORegAnno: an open-access community-driven resource for regulatory annotation. Nucleic acids research 36: D107-D113.
Griffiths-Jones S, Grocock RJ, Van Dongen S, Bateman A and Enright AJ. 2006. miRBase: microRNA sequences, targets and gene nomenclature. Nucleic acids research 34: D140-D144.
Gulko B, Hubisz MJ, Gronau I and Siepel A. 2015. A method for calculating probabilities of fitness consequences for point mutations across the human genome. Nat Genet 47: 276-283.
Guttman M, Garber M, Levin JZ, et al. 2010. Ab initio reconstruction of cell type-specific transcriptomes in mouse reveals the conserved multi-exonic structure of lincRNAs. Nature biotechnology 28: 503-510.
Haldeman-Englert CR, Chapman KA, Kruger H, Geiger EA, McDonald-McGinn DM, Rappaport E, Zackai EH, Spinner NB and Shaikh TH. 2010. A de novo 8.8-Mb deletion of 21q21. 1–q21. 3 in an autistic male with a complex rearrangement involving chromosomes 6, 10, and 21. American Journal of Medical Genetics Part A 152: 196-202.
Hamasaki T, Leingärtner A, Ringstedt T and O'Leary DD. 2004. EMX2 regulates sizes and positioning of the primary sensory and motor areas in neocortex by direct specification of cortical progenitors. Neuron 43: 359-372.
Harrow J, Frankish A, Gonzalez JM, et al. 2012. GENCODE: The reference human genome annotation for The ENCODE Project. Genome research 22: 1760-1774.
Hayakawa T, Aki I, Varki A, Satta Y and Takahata N. 2006. Fixation of the human-specific CMP-N-acetylneuraminic acid hydroxylase pseudogene and implications of haplotype diversity for human evolution. Genetics 172: 1139-1146.
Hedrick PW and Thomson G. 1983. Evidence for balancing selection at HLA. Genetics 104: 449-456.
Hemmi K, Ma D, Miura Y, Kawaguchi M, Sasahara M, Hashimoto-Tamaoki T, Tamaoki T, Sakata N and Tsuchiya K. 2006. A homeodomain-zinc finger protein, ZFHX4, is expressed in neuronal differentiation manner and suppressed in muscle differentiation manner. Biological and Pharmaceutical Bulletin 29: 1830-1835.
Hernandez RD, Kelley JL, Elyashiv E, Melton SC, Auton A, McVean G, Sella G and Przeworski M. 2011. Classic selective sweeps were rare in recent human evolution. Science 331: 920-924.
Hills LB, Masri A, Konno K, et al. 2013. Deletions in GRID2 lead to a recessive syndrome of cerebellar ataxia and tonic upgaze in humans. Neurology: 10.1212/WNL. 1210b1013e3182a1841a1213.
Hindorff LA, Sethupathy P, Junkins HA, Ramos EM, Mehta JP, Collins FS and Manolio TA. 2009. Potential etiologic and functional implications of genome-wide association loci for human diseases and traits. Proceedings of the National Academy of Sciences 106: 9362-9367.
Hirose S. 2006. A new paradigm of channelopathy in epilepsy syndromes: intracellular trafficking abnormality of channel molecules. Epilepsy research 70: 206-217.
Holder JL, Butte NF and Zinn AR. 2000. Profound obesity associated with a balanced translocation that disrupts the SIM1 gene. Human molecular genetics 9: 101-108.
Hsu F, Kent WJ, Clawson H, Kuhn RM, Diekhans M and Haussler D. 2006. The UCSC known genes. Bioinformatics 22: 1036-1046.

Irie A, Koyama S, Kozutsumi Y, Kawasaki T and Suzuki A. 1998. The Molecular Basis for the Absence ofN-Glycolylneuraminic Acid in Humans. Journal of Biological Chemistry 273: 15866-15871.

Jamain S, Betancur C, Quach H, Philippe A, Fellous M, Giros B, Gillberg C, Leboyer M and Bourgeron T. 2002. Linkage and association of the glutamate receptor 6 gene with autism. Molecular psychiatry 7: 302.

Johnson DS, Mortazavi A, Myers RM and Wold B. 2007. Genome-wide mapping of in vivo protein-DNA interactions. Science 316: 1497-1502.

Karolchik D, Hinrichs AS, Furey TS, Roskin KM, Sugnet CW, Haussler D and Kent WJ. 2004. The UCSC Table Browser data retrieval tool. Nucleic acids research 32: D493-D496.

Kent WJ, Sugnet CW, Furey TS, Roskin KM, Pringle TH, Zahler AM and Haussler D. 2002. The human genome browser at UCSC. Genome research 12: 996-1006.

Kern AD. 2009. Correcting the site frequency spectrum for divergence-based ascertainment. PLoS One 4: e5152.

Kim SA, Kim JH, Park M, Cho IH and Yoo HJ. 2007. Association of GABRB3 polymorphisms with autism spectrum disorders in Korean trios. Neuropsychobiology 54: 160-165.

King M-C and  Wilson AC. 1975. Evolultion at Two Levels in Humand and Chimpanzees. Science 188: 107-116.

Knowles DG and  McLysaght A. 2009. Recent de novo origin of human protein-coding genes. Genome research 19: 1752-1759.

Kohda K, Kamiya Y, Matsuda S, Kato K, Umemori H and Yuzaki M. 2003. Heteromer formation of δ2 glutamate receptors with AMPA or kainate receptors. Molecular brain research 110: 27-37.

Kong A, Frigge ML, Masson G, et al. 2012. Rate of de novo mutations and the importance of father/'s age to disease risk. Nature 488: 471-475.

Kong A, Thorleifsson G, Gudbjartsson DF, et al. 2010. Fine-scale recombination rate differences between sexes, populations and individuals. Nature 467: 1099-1103.

Kostka D, Hubisz MJ, Siepel A and Pollard KS. 2012. The role of GC-biased gene conversion in shaping the fastest evolving regions of the human genome. Molecular biology and evolution 29: 1047-1057.

Lander ES, Linton LM, Birren B, et al. 2001. Initial sequencing and analysis of the human genome. Nature 409: 860-921.

Law SW, Conneely O, DeMayo F and O'malley B. 1992. Identification of a new brain-specific transcription factor, NURR1. Molecular Endocrinology 6: 2129-2135.

Le W-d, Xu P, Jankovic J, Jiang H, Appel SH, Smith RG and Vassilatis DK. 2002. Mutations in NR4A2 associated with familial Parkinson disease. Nature genetics 33: 85-89.

Leffler EM, Gao Z, Pfeifer S, et al. 2013. Multiple instances of ancient balancing selection shared between humans and chimpanzees. Science 339: 1578-1582.

Lestrade L and  Weber MJ. 2006. snoRNA-LBME-db, a comprehensive database of human H/ACA and C/D box snoRNAs. Nucleic acids research 34: D158-D162.

Lin K, Li H, Schlötterer C and Futschik A. 2011. Distinguishing positive selection from neutral evolution: boosting the performance of summary statistics. Genetics 187: 229-244.

Lindblad-Toh K, Garber M, Zuk O, et al. 2011. A high-resolution map of human evolutionary constraint using 29 mammals. Nature 478: 476-482.

Liu X, Somel M, Tang L, et al. 2012. Extension of cortical synaptic development distinguishes humans from chimpanzees and macaques. Genome research 22: 611-622.
Lohmueller KE, Albrechtsen A, Li Y, et al. 2011. Natural selection affects multiple aspects of genetic variation at putatively neutral sites across the human genome. PLoS genetics 7: e1002326.
Lujan R, Shigemoto R and Lopez-Bendito G. 2005. Glutamate and GABA receptor signalling in the developing brain. Neuroscience 130: 567-580.
Lunter G, Ponting CP and Hein J. 2006. Genome-wide identification of human functional DNA using a neutral indel model. PLoS computational biology 2: e5.
Marth G, Schuler G, Yeh R, et al. 2003. Sequence variations in the public human genome data reflect a bottlenecked population history. Proceedings of the National Academy of Sciences 100: 376-381.
Maynard Smith J and Haigh J. 1974. The hitch-hiking effect of a favourable gene. Genetical Research 23: 23-35.
McLean CY, Bristor D, Hiller M, Clarke SL, Schaar BT, Lowe CB, Wenger AM and Bejerano G. 2010. GREAT improves functional interpretation of cis-regulatory regions. Nature biotechnology 28: 495-501.
McLean CY, Reno PL, Pollen AA, et al. 2011. Human-specific loss of regulatory DNA and the evolution of human-specific traits. Nature 471: 216-219.
McVean GA, Myers SR, Hunt S, Deloukas P, Bentley DR and Donnelly P. 2004. The fine-scale structure of recombination rate variation in the human genome. Science 304: 581-584.
McVicker G, Gordon D, Davis C and Green P. 2009. Widespread genomic signatures of natural selection in hominid evolution. PLoS genetics 5: e1000471.
Menashe I, Man O, Lancet D and Gilad Y. 2003. Different noses for different people. Nature genetics 34: 143-144.
Messer PW. 2013. SLiM: simulating evolution with selection and linkage. Genetics 194: 1037-1039.
Meyer LR, Zweig AS, Hinrichs AS, et al. 2013. The UCSC Genome Browser database: extensions and updates 2013. Nucleic acids research 41: D64-D69.
Michaud JL, Boucher F, Melnyk A, Gauthier F, Goshu E, Lévy E, Mitchell GA, Himms-Hagen J and Fan C-M. 2001. Sim1 haploinsufficiency causes hyperphagia, obesity and reduction of the paraventricular nucleus of the hypothalamus. Human molecular genetics 10: 1465-1473.
Michaud JL, Rosenquist T, May NR and Fan C-M. 1998. Development of neuroendocrine lineages requires the bHLH–PAS transcription factor SIM1. Genes & development 12: 3264-3275.
Mikkelsen TS, Hillier LW, Eichler EE, et al. 2005. Initial sequence of the chimpanzee genome and comparison with the human genome. Nature 437: 69-87.
Mishima Y, Lindgren AG, Chizhikov VV, Johnson RL and Millen KJ. 2009. Overlapping function of Lmx1a and Lmx1b in anterior hindbrain roof plate formation and cerebellar growth. The Journal of Neuroscience 29: 11377-11384.
Miura Y, Tam T, Ido A, Morinaga T, Miki T, Hashimoto T and Tamaoki T. 1995. Cloning and characterization of an ATBF1 isoform that expresses in a neuronal differentiation-dependent manner. Journal of Biological Chemistry 270: 26840-26848.

Montgomery S, Griffith OL, Sleumer MC, Bergman CM, Bilenky M, Pleasance E, Prychyna Y, Zhang X and Jones SJ. 2006. ORegAnno: an open access database and curation system for literature-derived promoters, transcription factor binding sites and regulatory variation. Bioinformatics 22: 637-640.

Motazacker MM, Rost BR, Hucho T, et al. 2007. A Defect in the Ionotropic Glutamate Receptor 6 Gene (*GRIK2*) Is Associated with Autosomal Recessive Mental Retardation. The American Journal of Human Genetics 81: 792-798.

Nielsen JA, Berndt JA, Hudson LD and Armstrong RC. 2004. Myelin transcription factor 1 (Myt1) modulates the proliferation and differentiation of oligodendrocyte lineage cells. Molecular and Cellular Neuroscience 25: 111-123.

Nurmi EL, Dowd M, Tadevosyan-Leyfer O, Haines JL, Folstein SE and Sutcliffe JS. 2003. Exploratory subsetting of autism families based on savant skills improves evidence of genetic linkage to 15q11-q13. Journal of the American Academy of Child & Adolescent Psychiatry 42: 856-863.

Olson MV. 1999. When less is more: gene loss as an engine of evolutionary change. American journal of human genetics 64: 18.

Pavlidis P, Jensen JD and Stephan W. 2010. Searching for footprints of positive selection in whole-genome SNP data from nonequilibrium populations. Genetics 185: 907-922.

Petroff OA. 2002. Book Review: GABA and glutamate in the human brain. The Neuroscientist 8: 562-573.

Pierron D, Cortés NG, Letellier T and Grossman LI. 2012. Current relaxation of selection on the human genome: Tolerance of deleterious mutations on olfactory receptors. Molecular Phylogenetics and Evolution.

Pollard KS, Hubisz MJ, Rosenbloom KR and Siepel A. 2010. Detection of nonneutral substitution rates on mammalian phylogenies. Genome research 20: 110-121.

Pollard KS, Salama SR, King B, et al. 2006a. Forces shaping the fastest evolving regions in the human genome. PLoS Genetics 2: e168.

Pollard KS, Salama SR, Lambert N, et al. 2006b. An RNA gene expressed during cortical development evolved rapidly in humans. Nature 443: 167-172.

Radel M, Vallejo RL, Iwata N, Aragon R, Long JC, Virkkunen M and Goldman D. 2005. Haplotype-based localization of an alcohol dependence gene to the 5q34 {gamma}-aminobutyric acid type A gene cluster. Archives of general psychiatry 62: 47.

Rilling JK. 2006. Human and nonhuman primate brains: Are they allometrically scaled versions of the same design? Evolutionary Anthropology: Issues, News, and Reviews 15: 65-77.

Ronen R, Udpa N, Halperin E and Bafna V. 2013. Learning natural selection from the site frequency spectrum. Genetics: doi: 10.1534/genetics.1113.152587.

Rouquier S, Taviaux S, Trask BJ, Brand-Arpon V, van den Engh G, Demaille J and Giorgi D. 1998. Distribution of olfactory receptor genes in the human genome. Nature genetics 18: 243-250.

Saitou N and  Yamamoto F-I. 1997. Evolution of primate ABO blood group genes and their homologous genes. Molecular biology and evolution 14: 399-411.

Sander T, Hildmann T, Kretz R, et al. 1997. Allelic association of juvenile absence epilepsy with a GluR5 kainate receptor gene (GRIK1) polymorphism. American journal of medical genetics 74: 416-421.

Saucedo-Cardenas O, Quintana-Hau JD, Le W-D, Smidt MP, Cox JJ, De Mayo F, Burbach JPH and Conneely OM. 1998. Nurr1 is essential for the induction of the dopaminergic phenotype and the survival of ventral mesencephalic late dopaminergic precursor neurons. Proceedings of the National Academy of Sciences 95: 4013-4018.

Schrider DR, Costello JC and Hahn MW. 2009. All human-specific gene losses are present in the genome as pseudogenes. Journal of Computational Biology 16: 1419-1427.

Schrider DR and Kern AD. 2014. Discovering functional DNA elements using population genomic information: A proof of concept using human mtDNA. Genome Biol Evol 6: 1542-1548.

Schrider DR, Mendes FK, Hahn MW and Kern AD. 2015. Soft shoulders ahead: spurious signatures of soft and partial selective sweeps result from linked hard sweeps. Genetics: genetics. 115.174912.

Schwartz S, Kent WJ, Smit A, Zhang Z, Baertsch R, Hardison RC, Haussler D and Miller W. 2003. Human–mouse alignments with BLASTZ. Genome research 13: 103-107.

Shabalina SA, Ogurtsov AY, Kondrashov VA and Kondrashov AS. 2001. Selective constraint in intergenic regions of human and mouse genomes. Trends in Genetics 17: 373-376.

Shapiro L and Colman DR. 1999. The diversity of cadherins and implications for a synaptic adhesive code in the CNS. Neuron 23: 427-430.

Siepel A, Bejerano G, Pedersen JS, et al. 2005. Evolutionarily conserved elements in vertebrate, insect, worm, and yeast genomes. Genome research 15: 1034-1050.

Simeone A, Gulisano M, Acampora D, Stornaiuolo A, Rambaldi M and Boncinelli E. 1992. Two vertebrate homeobox genes related to the Drosophila empty spiracles gene are expressed in the embryonic cerebral cortex. The EMBO journal 11: 2541.

Somel M, Sayres MW, Jordan G, Huerta-Sanchez E, Fumagalli M, Ferrer-Admetlla A and Nielsen R. 2013. A scan for human-specific relaxation of negative selection reveals unexpected polymorphism in proteasome genes. Molecular biology and evolution 30: 1808-1815.

Sommer L, Ma Q and Anderson DJ. 1996. *neurogenins*, a Novel Family of *atonal*-Related bHLH Transcription Factors, Are Putative Mammalian Neuronal Determination Genes That Reveal Progenitor Cell Heterogeneity in the Developing CNS and PNS. Molecular and Cellular Neuroscience 8: 221-241.

Stajich JE and Hahn MW. 2005. Disentangling the effects of demography and selection in human history. Molecular biology and evolution 22: 63-73.

Stedman HH, Kozyak BW, Nelson A, et al. 2004. Myosin gene mutation correlates with anatomical changes in the human lineage. Nature 428: 415-418.

Tajima F. 1989. Statistical method for testing the neutral mutation hypothesis by DNA polymorphism. Genetics 123: 585-595.

Tanaka M, Olsen RW, Medina MT, et al. 2008. Hyperglycosylation and Reduced GABA Currents of Mutated< i> GABRB3</i> Polypeptide in Remitting Childhood Absence Epilepsy. The American Journal of Human Genetics 82: 1249-1261.

Tennessen JA, Bigham AW, O'Connor TD, et al. 2012. Evolution and functional impact of rare coding variation from deep sequencing of human exomes. science 337: 64-69.

Thomas JB, Crews ST and Goodman CS. 1988. Molecular genetics of the< i> single-minded</i> locus: A gene involved in the development of the Drosophila nervous system. Cell 52: 133-141.

Trapnell C, Williams BA, Pertea G, Mortazavi A, Kwan G, van Baren MJ, Salzberg SL, Wold BJ and Pachter L. 2010. Transcript assembly and quantification by RNA-Seq reveals unannotated transcripts and isoform switching during cell differentiation. Nature biotechnology 28: 511-515.

Trappe R, Buddenberg P, Uedelhoven J, Gläser B, Buck A, Engel W and Burfeind P. 2002. The murine BTB/POZ zinc finger gene *ZNF131*: predominant expression in the developing central nervous system, in adult brain, testis, and thymus. Biochemical and biophysical research communications 296: 319-327.

Utine GE, Haliloğlu G, Salancı B, Çetinkaya A, Kiper PÖ, Alanay Y, Aktaş D, Boduroğlu K and Alikaşifoğlu M. 2013. A Homozygous Deletion in GRID2 Causes a Human Phenotype With Cerebellar Ataxia and Atrophy. Journal of child neurology 28: 926-932.

Vanhalst K, Kools P, Vanden E and van Roy F. 2001. The human and murine protocadherin-beta one-exon gene families show high evolutionary conservation, despite the difference in gene number. FEBS Lett 495: 120-125.

Vapnik V and Lerner A. 1963. Pattern recognition using generalized portrait method. Automation and Remote Control 24: 774-780.

Walldorf U and Gehring W. 1992. Empty spiracles, a gap gene containing a homeobox involved in Drosophila head development. The EMBO journal 11: 2247.

Wang X, Grus WE and Zhang J. 2006. Gene losses during human origins. PLoS biology 4: e52.

Ward LD and Kellis M. 2012. Evidence of abundant purifying selection in humans for recently acquired regulatory functions. Science 337: 1675-1678.

Woo T-UW, Shrestha K, Amstrong C, Minns MM, Walsh JP and Benes FM. 2007. Differential alterations of kainate receptor subunits in inhibitory interneurons in the anterior cingulate cortex in schizophrenia and bipolar disorder. Schizophrenia research 96: 46-61.

Wu T-F, Lin C-J and Weng RC. 2004. Probability estimates for multi-class classification by pairwise coupling. The Journal of Machine Learning Research 5: 975-1005.

Xue Y, Daly A, Yngvadottir B, et al. 2006. Spread of an inactive form of caspase-12 in humans is due to recent positive selection. The American Journal of Human Genetics 78: 659-670.

Young JM, Friedman C, Williams EM, Ross JA, Tonnes-Priddy L and Trask BJ. 2002. Different evolutionary processes shaped the mouse and human olfactory receptor gene families. Human Molecular Genetics 11: 535-546.

Zhang X and Firestein S. 2002. The olfactory receptor gene superfamily of the mouse. Nature neuroscience 5: 124-133.

**Table 1: SVM accuracies when discriminating between simulated constrained and unconstrained genomic regions in independent test sets.**

| Fraction of selected sites | Overall accuracy | Accuracy of popCons calls (95% confidence) | Fraction of unconstrained windows classified as popCons elements | Accuracy of popUnconsCalls (95% confidence) | Fraction of constrained windows classified as popUncons elements |
|---|---|---|---|---|---|
| 0-5% (*n*=2000) | 54.45% | 29/46=63.04% | 17/1000=1.70% | 495/827=59.85% | 332/1000=33.20% |
| 5-10% (*n*=2000) | 62.70% | 54/67=80.60% | 13/1000=1.30% | 475/639=74.33% | 164/1000=16.40% |
| 10-15% (*n*=2000) | 69.45% | 114/125=91.20% | 11/1000=1.10% | 472/561=84.13% | 89/1000=8.90% |
| 15-20% (*n*=2000) | 76.25% | 184/201=91.50% | 17/1000=1.70% | 477/530=90.00% | 53/1000=5.30% |
| 20-25% (*n*=1652) | 81.17% | 219/231=94.81% | 12/826=1.45% | 385/413=93.22% | 28/826=3.39% |

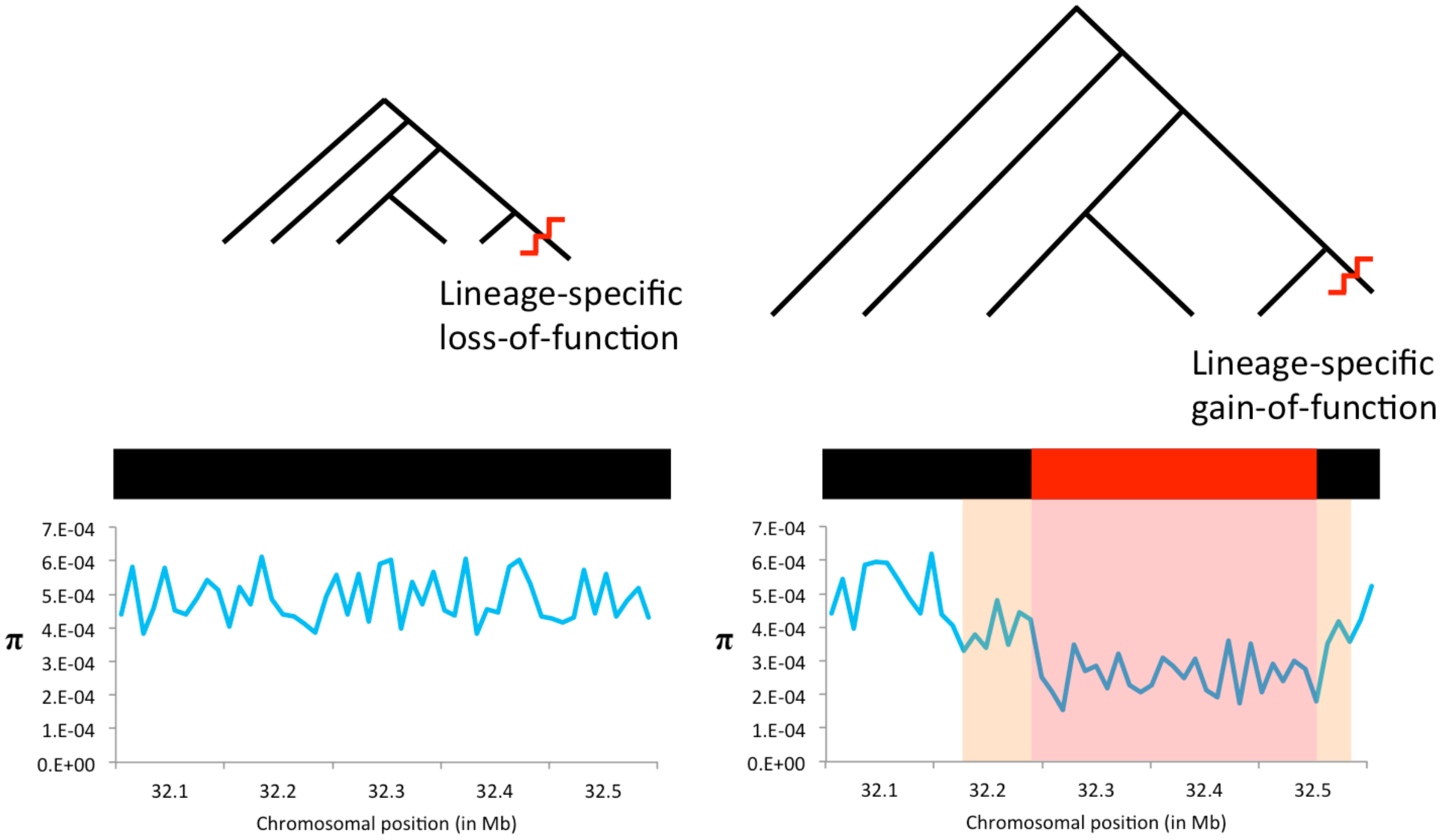

**fig. 1. Using phylogenetic and population genetic data to find lineage-specific changes in selective constraint.** In a genomic region (black bar) experiencing a lineage-specific loss of function (left), the presence of purifying selection in the majority of the phylogeny reduces divergence (short branch lengths). However, because the genomic region no longer performs a function with fitness consequences in one species, population genetic data from this species shows no reduction in diversity (as measured by nucleotide diversity, $\pi$) in this region. In the case of a lineage-specific gain of function, the majority of the phylogeny has experienced no purifying selection, and therefore divergence is higher (long branch lengths). In the species experiencing the gain of function, purifying selection reduces genetic variation in the functional region (red portion of the black bar), and background selection lowers diversity at flanking sites.

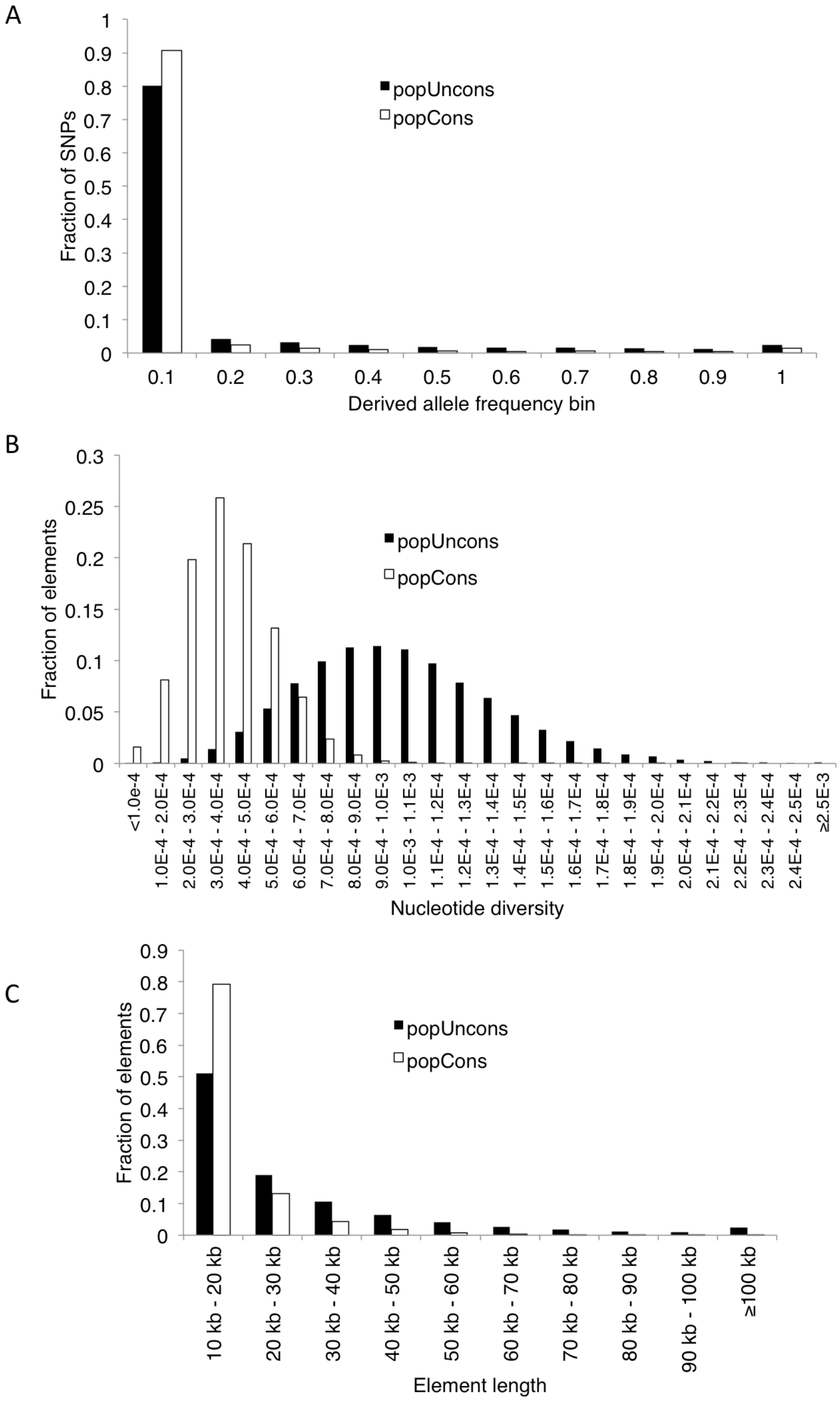

**fig. 2. Reduced genetic variation in popCons versus popUncons elements.** (A) Site frequency spectra (SFS) of popCons (white) and popUncons elements (black). The bars show the fraction of SNPs in a given element type found within each derived allele frequency bin. (B) Histogram of values of $\pi$ within popCons (white) and popUncons (black) elements. (C) Histogram of lengths of popCons (white) and popUncons (black) elements.

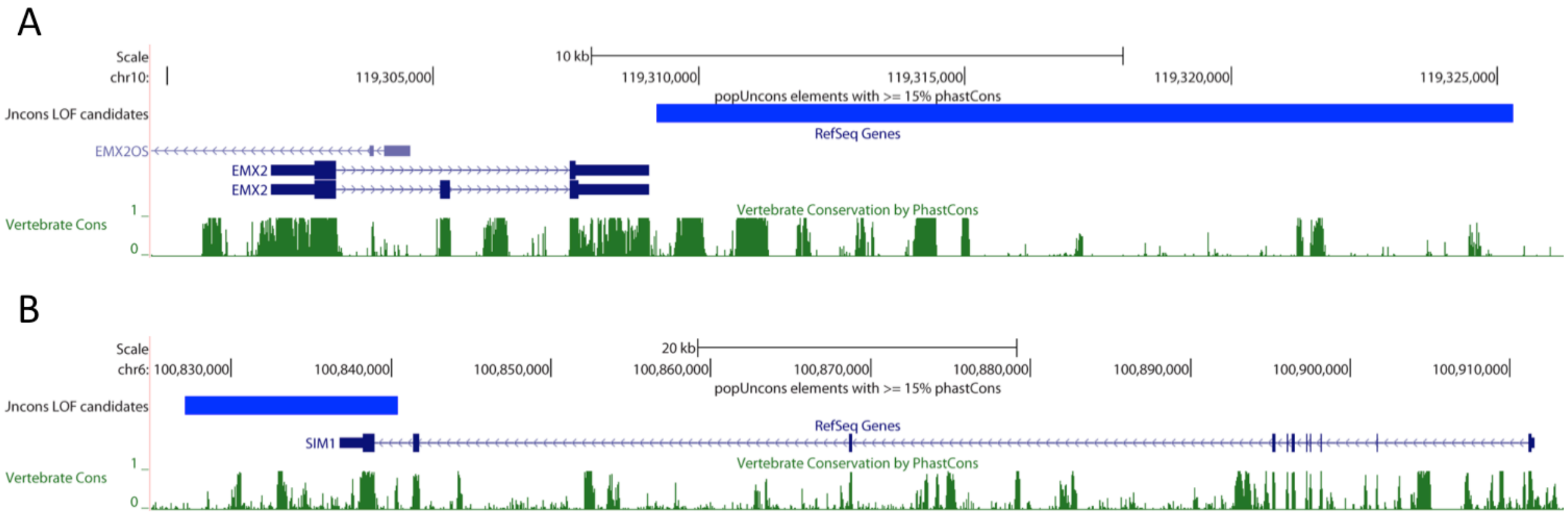

**fig. 3. Candidate loss of function regions.** (A) A diagram of *EMX2* and the downstream flanking region generated by the UCSC Genome Browser shows a popUncons LOF candidate region (large blue bar) with a strong phylogenetic signal of conservation (high phastCons posterior probabilities, green). (B) A diagram of *SIM1* and its downstream flanking region.

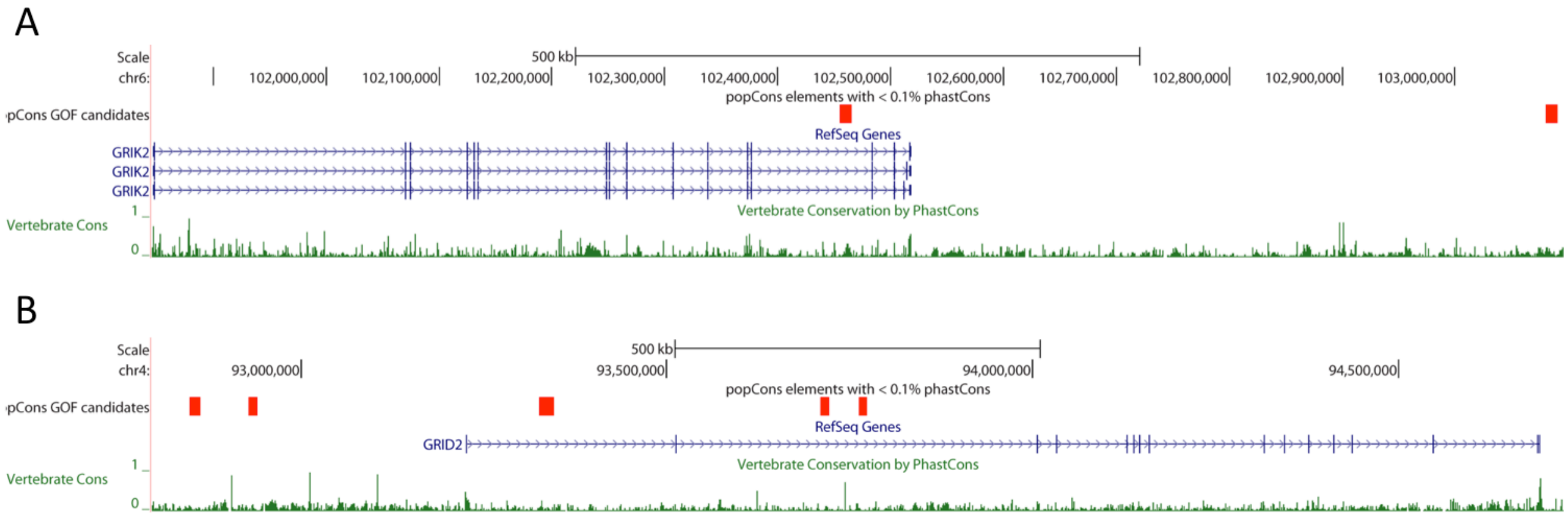

**fig. 4. Gain of function candidates near glutamate receptor genes.** (A) A diagram of *GRIK2* and its downstream flanking region generated by the UCSC Genome Browser. PopCons GOF candidate regions, shown in red, show little evidence for selective constraint across vertebrates (low phastCons posterior probabilities, green). (B) A diagram of GRID2 and its upstream region.

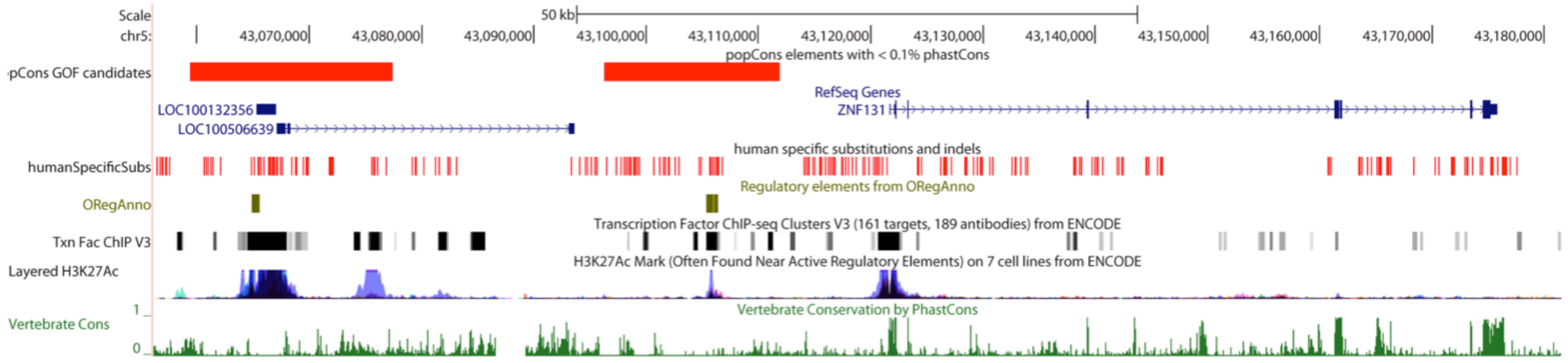

**fig. 5. Gain of function candidates upstream of *ZNF131*.** A diagram of *ZNF131* and its upstream flanking region generated by the UCSC Genome Browser. PopCons GOF candidate regions are shown in red. Each of these GOF regions contains an ORegAnno regulatory element, with the element closer to ZNF131 having a high density of human-specific substitutions (red tick marks). ChIP-seq peaks indicative of transcription factor binding sites are also shown (black and grey bars), as are H3K27Ac peaks (blue graph), both from ENCODE.

## SUPPLEMENTARY METHODS

### Assessing classification accuracy using forward simulations

In order to assess the accuracy of our classification approach we used SLiM (Messer 2013) to perform forward simulations of 10 kilobase regions (the same window size we used for real data; see below) evolving strictly under drift, containing a region experiencing purifying selection, or experiencing positive selection across the entire region. For our simulations including purifying selection, we set the selection coefficient ($2Ns$, where $N$ is the initial total population size and $1\text{-}s$ and $1\text{-}0.5s$ are the fitnesses of homozygotes and heterozygotes, respectively) to either 50, 100, or 500, and set the length of the constrained region $L$ to either 2.5 kb, 5 kb, 7.5 kb, or the full 10 kb window. The start of this constrained region was always located on the left end of the simulated chromosome. Each of these simulations followed the demographic scenario reported by Tennessen *et al.* (2012), which models the divergence of Europeans from the ancestral African population, and subsequent population size changes for these two populations. For these simulations we set the mutation rate to $1.2\times10^{-8}$ mutations per base pair (Kong et al. 2012) in some instances, or to reduced ($6\times10^{-9}$) or elevated ($2.4\times10^{-8}$) rates in others. Similarly, we set the recombination rate to either $1\times10^{-8}$ crossovers per base pair (or 1 cM/Mb), or to reduced or elevated values of $1\times10^{-6}$ or $1\times10^{-7}$, respectively.

For each combination of $2Ns$, $L$, mutation rate, and recombination rate, we performed 500 independent replicate simulations for training and testing our classifier. Each of these simulations was initialized from a burn-in simulation proceeding for 4,000,000 generations under equilibrium demography, using the same combination of $2Ns$, $L$, mutation rate, and recombination rate. From each of the 500 replicate simulations we sampled 1000 chromosomes without replacement form each of the two subpopulations simulated in the Tennessen *et al.*

model (2012), yielding a total sample size of 2,000 individuals. We ran all burn-in and replicate simulations using a version of SLiM that we modified to simulate additive rather than multiplicative fitness effects. We then summarized the output of each simulated window by the feature vector $\boldsymbol{\xi}=[\xi_0\ \xi_1\ \xi_2\ \ldots\ \xi_{n\text{-}1}]$ where $n$ is the number of chromosomes in the sample, $\xi_i$ is the fraction of sites in the window with a derived allele segregating at frequency $i$, except $\xi_0$, which is the fraction of sites in the window that are monomorphic (including derived fixations). The feature vector $\boldsymbol{\xi}$ is thus the site frequency spectrum (SFS) modified to divide the value in each bin by the number of sites rather than the number of polymorphisms, and to include the fraction of sites that are monomorphic.

Next, for each combination of $2Ns$, $L$, mutation, and recombination rates, we sought to train an SVM to classify simulated regions as constrained or unconstrained based on their feature vectors $\boldsymbol{\xi}$. First, we constructed a training set using the output from 300 simulations with no selection and 300 windows including a region experiencing negative selection. We performed additional simulations to construct a balanced test set (200 selected and 200 unselected windows) with the same parameter combination for later use. For each combination of $2Ns$ and $L$, we constructed training and test sets were also with variable mutation rates, variable recombination rates, or both, by drawing equal numbers of simulated examples from each of the three rate values listed above. Next, we collapsed these feature vectors to contain 1,000 bins as we found that this amount of binning improved cross-validation accuracy on our training set made from real genomic data (see below). We then formatted feature vectors for these training and test sets for use by LIBSVM (Chang and Lin 2011), and rescaled them using the svm-scale command with default scaling parameters; the training and test sets were concatenated together for this step (along with positively selected simulations; see below) to ensure the same scaling parameters were used for both sets.

We then used LIBSVM's svm-train command with a radial basis kernel function to learn the hyperplane optimally separating the conserved and unconserved training data according to the SVM's *C* parameter (Cortes and Vapnik 1995). The hyperplane chosen, and therefore its accuracy when classifying data not included in the training set, depend on this *C* parameter and the radial basis function's $\gamma$ parameter. We therefore performed a grid search of these two parameter values, examining all powers of two between $2^{-11}$ and $2^{9}$ for each parameter. For each combination of *C* and $\gamma$, we performed 10-fold cross validation in order to assess the SVM's accuracy. We then used the optimal combination of hyperparameters to train an SVM from the entire training set, and assessed the accuracy of this SVM using the test set.

In some cases, the optimal combination of the *C* and gamma $\gamma$ yielded poor accuracy on the test set only when using LIBSVM's option to compute posterior classification probabilities: in such cases the classification probability for the feature vector being classified was nearly always exactly 0.5 for both classes. We only observed this behavior when one or both of the *C* or $\gamma$ hyperparameters was very small (i.e. $2^{-6}$ or less). We therefore slightly modified our grid search procedure to obtain all hyperparameter combinations with a cross-validation accuracy value within 1% of that of the optimal combination, and then selected from these the combination with the smallest sum of $|\log_2 C|+|\log_2 \lambda|$, thereby punishing hyperparameters whose base-2 exponent differed greatly from zero. After this modification, all grid searches produced SVMs that had similar performance on the test set as on the training set and emitted a more continuous range of probability estimates. Crucially, the poor posterior probability estimation described above was not exhibited by the final SVM learned from the 1000 Genomes data (as described below), which produced a more uniform range of probability estimates and whose performance we also assessed using independent test sets.

We also simulated a set of 400 population samples experiencing recurrent positive selection with each mutation at each site in the chromosome being positively selected with $2Ns=100$ (where homozygote and heterozygote fitness values are $1+s$ and $1+0.5s$, respectively). We then asked what fraction of these positively selected samples were classified as negatively selected or unselected by each SVM. For these simulations, the mutation rate was set to $1.2\times10^{-8}$ and the recombination rate to $1\times10^{-8}$.

## SUPPLEMENTARY REFERENCES

Chang C-C and Lin C-J. 2011. LIBSVM: a library for support vector machines. ACM Transactions on Intelligent Systems and Technology (TIST) 2: 27.

Cortes C and Vapnik V. 1995. Support-vector networks. Machine learning 20: 273-297.

Kong A, Frigge ML, Masson G, et al. 2012. Rate of de novo mutations and the importance of father/'s age to disease risk. Nature 488: 471-475.

Messer PW. 2013. SLiM: simulating evolution with selection and linkage. Genetics 194: 1037-1039.

Tennessen JA, Bigham AW, O'Connor TD, et al. 2012. Evolution and functional impact of rare coding variation from deep sequencing of human exomes. science 337: 64-69.

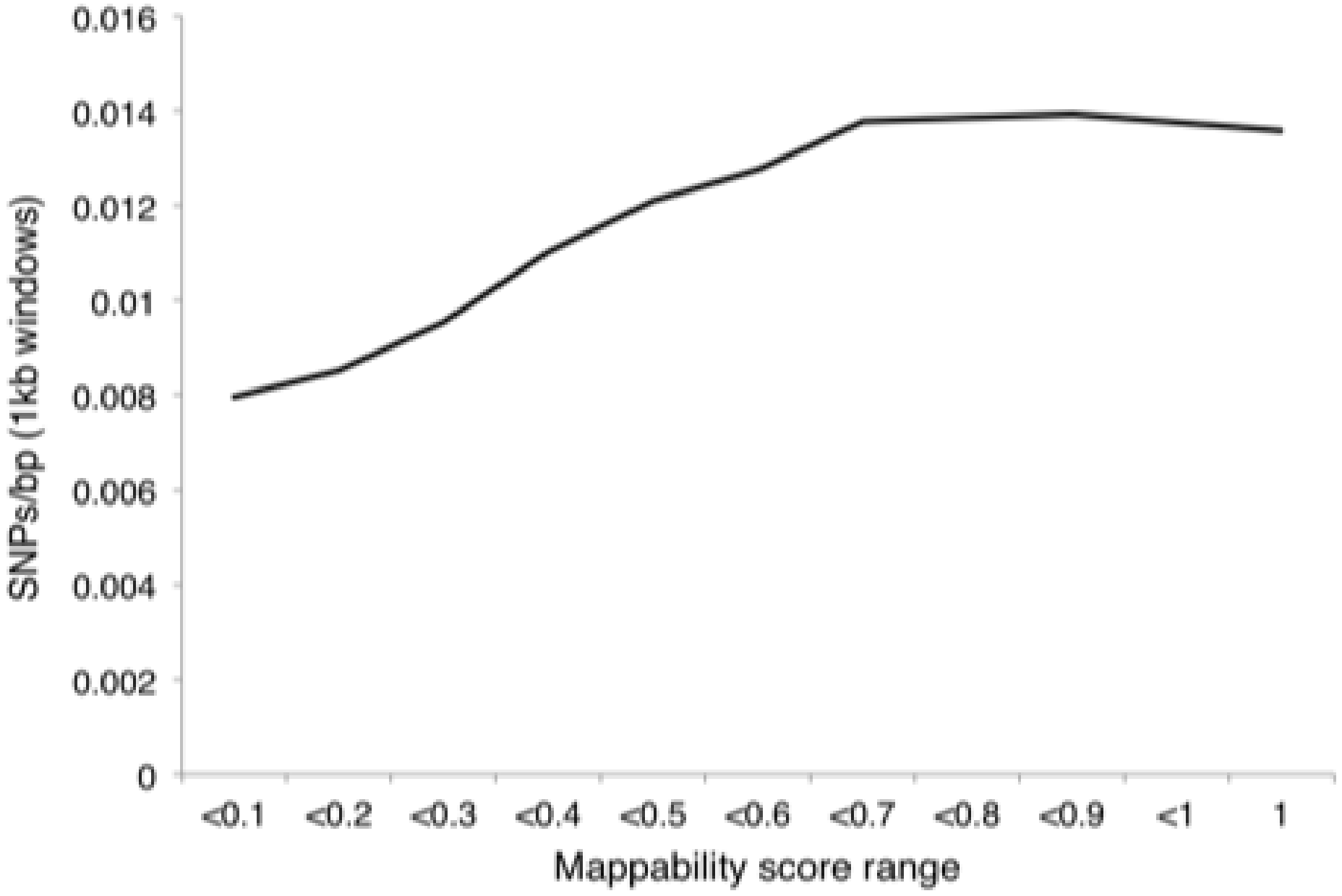

**supplementary fig. S1. SNP density and unique read mappability.** This figure shows the relationship between the average number of SNPs per base pair and the average mappability score within 1 kilobase windows.

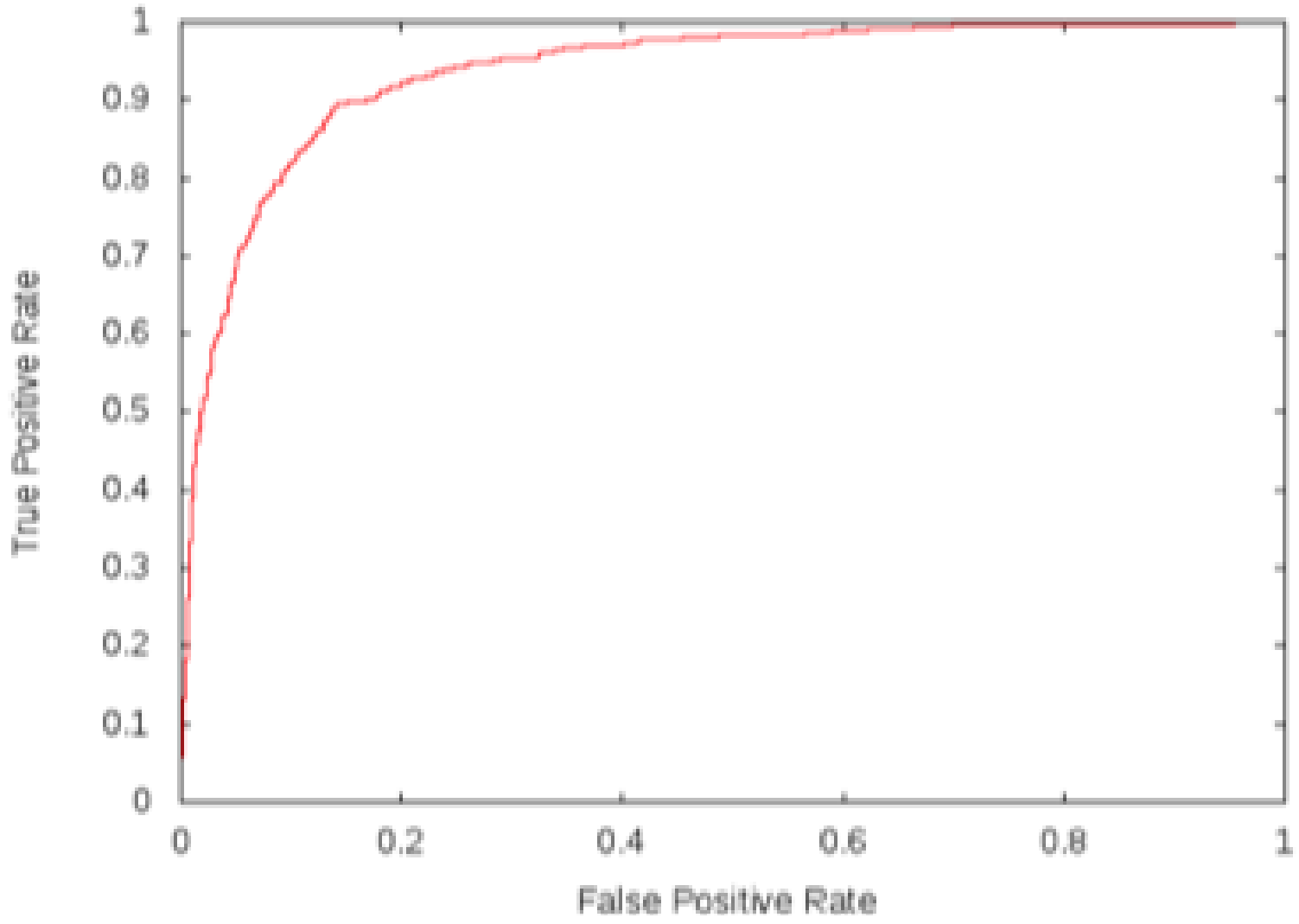

**supplementary fig. S2. An ROC curve is shown for the SVM classifier, estimated from cross validation.** The area under the curve is 0.94.

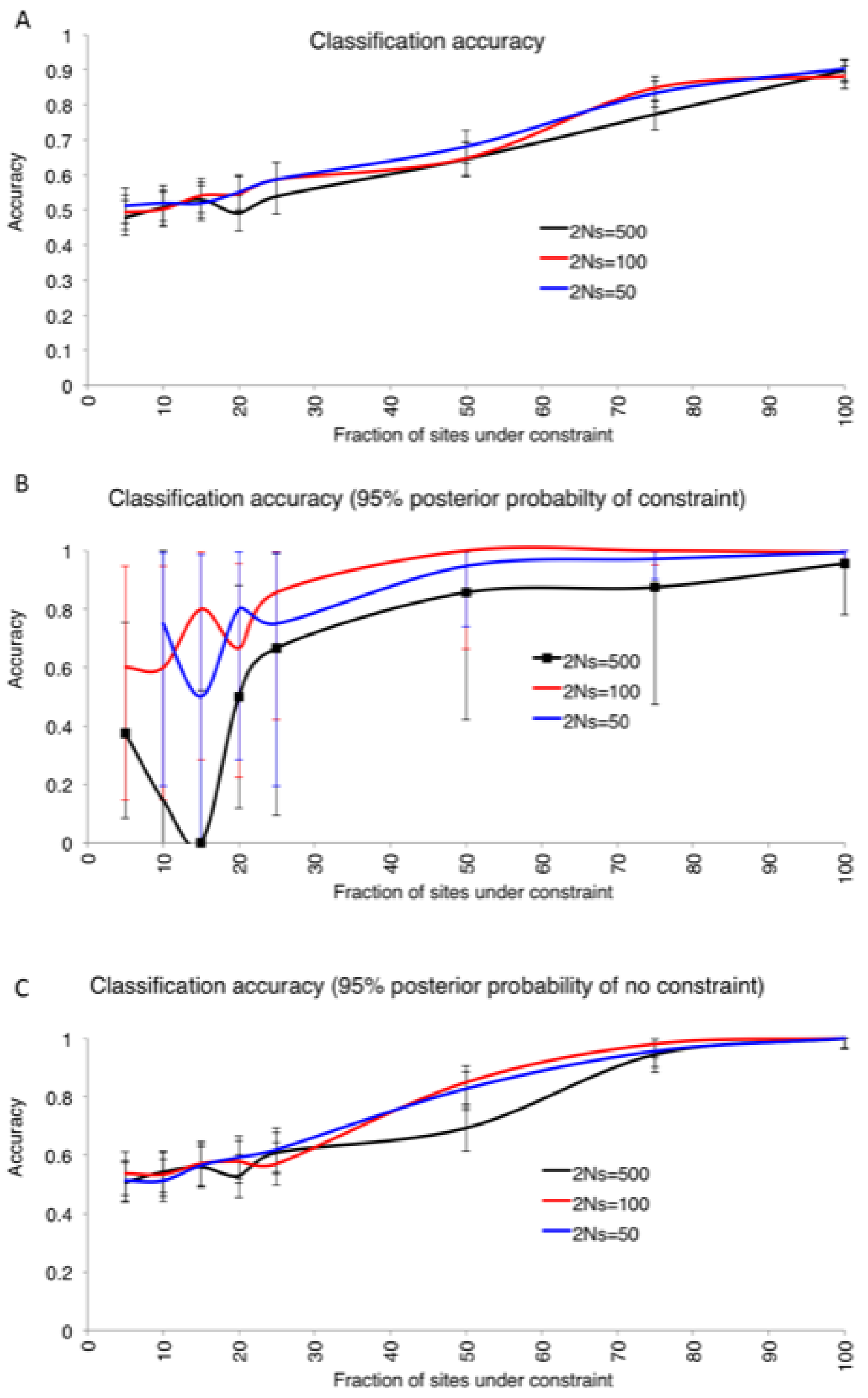

**supplementary fig. S3. Classification accuracy on simulated data.** (A) Fraction of simulated 10 kb regions correctly classified as containing or lacking selective constraint, assessed at several different fractions of sites under selection (5%, 10%, 15%, 20%, 25%, 50%, 75%, and 100%). (B) Classification accuracy on simulated 10 kb regions for which the SVM classifier's posterior probability of selective constraint was >95% (the same criterion used to define popCons elements). (C) Classification accuracy on simulated regions for which the SVM classifier's posterior probability of no constraint was >95% (the same criterion used to define popUncons elements). All of these results are from the test sets with variable mutation and recombination rates (supplementary table S6).

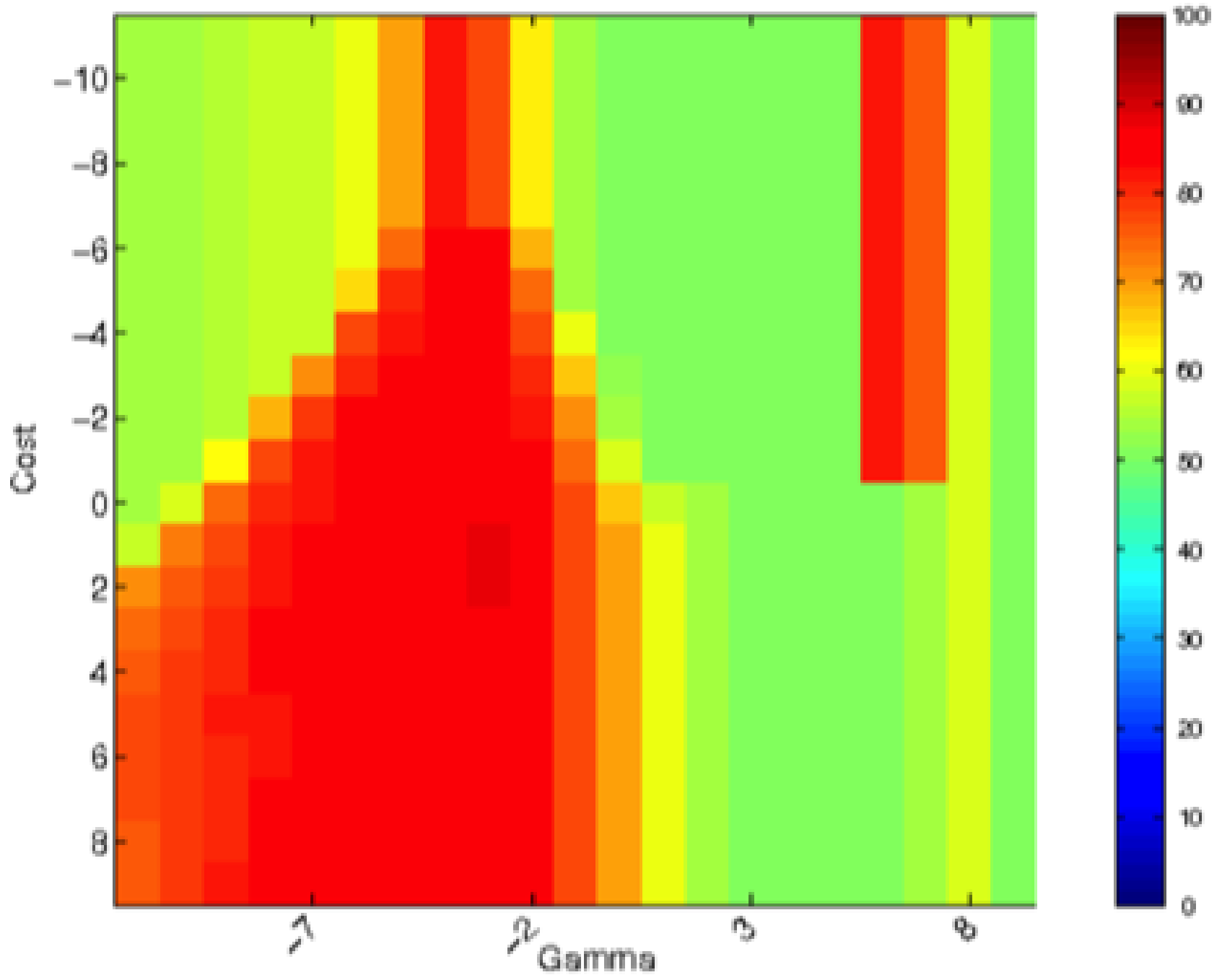

**supplementary fig. S4. Heatmap showing cross-validation accuracy (%) of each SVM hyperparameter combinations.** The optimal combination that we used to train the SVM was $C=2^1$, $\gamma=2^{-3}$.

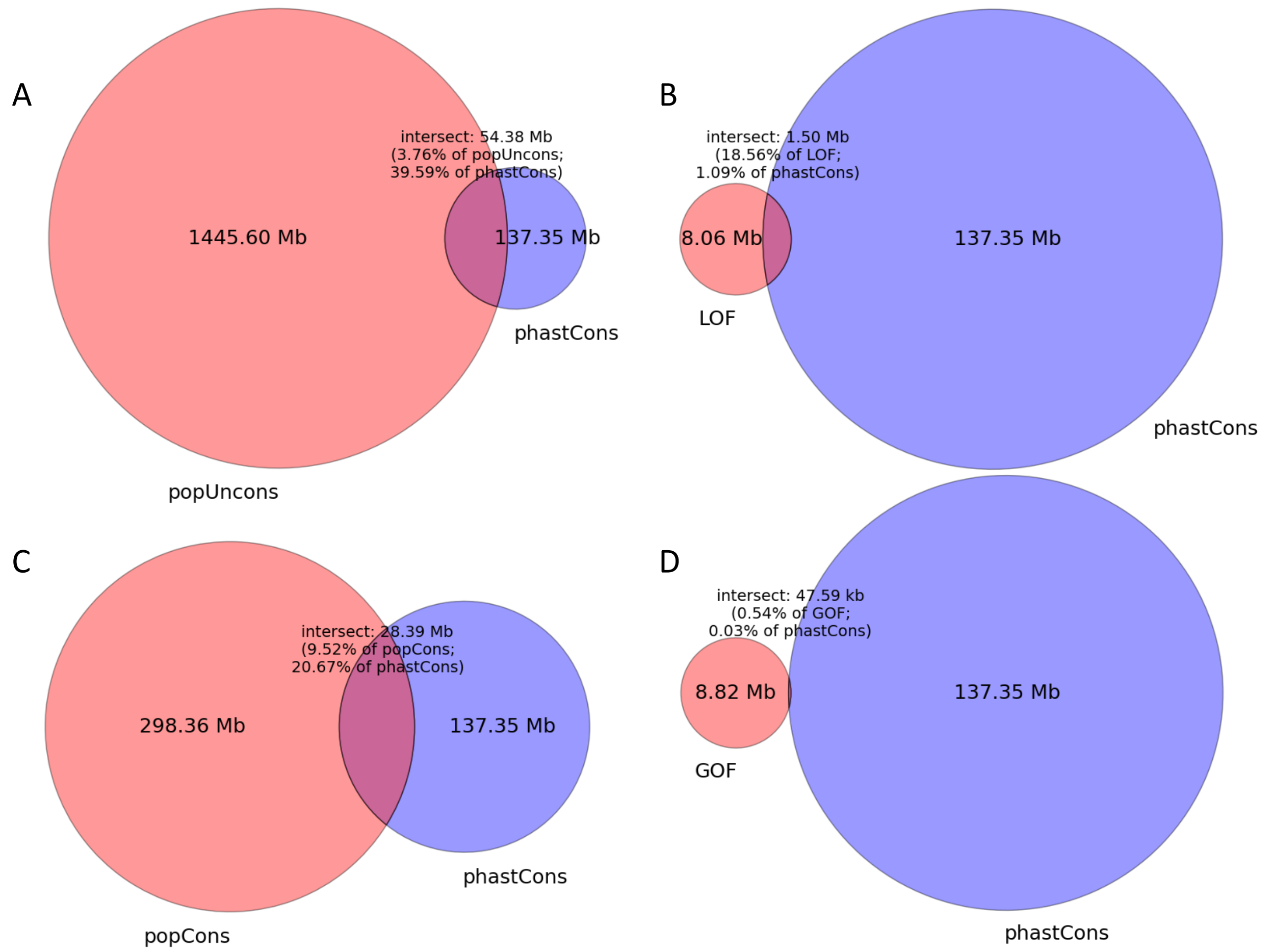

**supplementary fig. S5. Overlap between predicted elements and vertebrate phastCons elements.** (A) Venn diagram showing the number of base pairs within popUncons elements, the number of base pairs within vertebrate phastCons elements, and the number of base pairs lying within both types of elements. (B) Overlap between loss of function (LOF) candidates and phastCons. (C) Overlap between popCons and phastCons. (D) Overlap between gain of function (GOF) candidates and phastCons.

**Table S1: Individuals from the 1000 Genomes Project included in this study**

| Individual | Population Id |
|---|---|
| NA19625 | ASW |
| NA19700 | ASW |
| NA19701 | ASW |
| NA19703 | ASW |
| NA19704 | ASW |
| NA19707 | ASW |
| NA19711 | ASW |
| NA19712 | ASW |
| NA19713 | ASW |
| NA19818 | ASW |
| NA19819 | ASW |
| NA19834 | ASW |
| NA19835 | ASW |
| NA19900 | ASW |
| NA19901 | ASW |
| NA19904 | ASW |
| NA19908 | ASW |
| NA19909 | ASW |
| NA19914 | ASW |
| NA19916 | ASW |
| NA19917 | ASW |
| NA19920 | ASW |
| NA19921 | ASW |
| NA19922 | ASW |
| NA19923 | ASW |
| NA19982 | ASW |
| NA19984 | ASW |
| NA20126 | ASW |
| NA20127 | ASW |
| NA20276 | ASW |
| NA20278 | ASW |
| NA20281 | ASW |
| NA20282 | ASW |
| NA20287 | ASW |
| NA20291 | ASW |
| NA20294 | ASW |
| NA20296 | ASW |
| NA20298 | ASW |
| NA20299 | ASW |
| NA20314 | ASW |
| NA20317 | ASW |
| NA20322 | ASW |
| NA20332 | ASW |
| NA20336 | ASW |
| NA20339 | ASW |
| NA20340 | ASW |

**Population code legend:**

| | |
|---|---|
| ASW | Americans of African Ancestry in SW USA |
| CEU | Utah Residents (CEPH) with Northern and Western European ancestry |
| CHB | Han Chinese in Bejing, China |
| CHS | Southern Han Chinese |
| CLM | Colombians from Medellin, Colombia |
| FIN | Finnish in Finland |
| GBR | British in England and Scotland |
| IBS | Iberian population in Spain |
| JPT | Japanese in Tokyo, Japan |
| LWK | Luhya in Webuye, Kenya |
| MXL | Mexican Ancestry from Los Angeles USA |
| PUR | Puerto Ricans from Puerto Rico |
| TSI | Toscani in Italia |
| YRI | Yoruba in Ibadan, Nigera |
| NA20341 | ASW |
| NA20342 | ASW |
| NA20344 | ASW |
| NA20346 | ASW |
| NA20348 | ASW |
| NA20351 | ASW |
| NA20356 | ASW |
| NA20357 | ASW |
| NA20363 | ASW |
| NA20412 | ASW |
| NA06984 | CEU |
| NA06986 | CEU |
| NA06989 | CEU |
| NA06994 | CEU |
| NA07000 | CEU |
| NA07037 | CEU |
| NA07048 | CEU |
| NA07051 | CEU |
| NA07056 | CEU |
| NA07347 | CEU |
| NA07357 | CEU |
| NA10847 | CEU |
| NA10851 | CEU |
| NA11829 | CEU |
| NA11830 | CEU |
| NA11831 | CEU |
| NA11843 | CEU |
| NA11892 | CEU |
| NA11893 | CEU |
| NA11894 | CEU |
| NA11919 | CEU |
| NA11920 | CEU |
| NA11930 | CEU |
| NA11931 | CEU |
| NA11932 | CEU |
| NA11933 | CEU |
| NA11992 | CEU |
| NA11993 | CEU |
| NA11994 | CEU |
| NA11995 | CEU |
| NA12003 | CEU |
| NA12004 | CEU |
| NA12006 | CEU |
| NA12043 | CEU |
| NA12044 | CEU |
| NA12045 | CEU |
| NA12046 | CEU |
| NA12058 | CEU |
| NA12144 | CEU |
| NA12154 | CEU |
| NA12155 | CEU |
| NA12249 | CEU |
| NA12272 | CEU |
| NA12273 | CEU |
| NA12275 | CEU |
| NA12282 | CEU |
| NA12283 | CEU |
| NA12286 | CEU |
| NA12287 | CEU |
| NA12340 | CEU |
| NA12341 | CEU |
| NA12342 | CEU |
| NA12347 | CEU |
| NA12348 | CEU |
| NA12383 | CEU |
| NA12399 | CEU |
| NA12400 | CEU |
| NA12413 | CEU |
| NA12489 | CEU |
| NA12546 | CEU |
| NA12716 | CEU |
| NA12717 | CEU |
| NA12718 | CEU |
| NA12748 | CEU |
| NA12749 | CEU |
| NA12750 | CEU |
| NA12751 | CEU |
| NA12761 | CEU |
| NA12763 | CEU |
| NA12775 | CEU |
| NA12777 | CEU |
| NA12778 | CEU |
| NA12812 | CEU |
| NA12814 | CEU |
| NA12815 | CEU |
| NA12827 | CEU |
| NA12829 | CEU |
| NA12830 | CEU |
| NA12842 | CEU |
| NA12843 | CEU |
| NA12872 | CEU |
| NA12873 | CEU |
| NA12874 | CEU |
| NA12889 | CEU |
| NA12890 | CEU |
| NA18525 | CHB |
| NA18526 | CHB |
| NA18527 | CHB |
| NA18528 | CHB |
| NA18530 | CHB |
| NA18532 | CHB |
| NA18534 | CHB |
| NA18535 | CHB |
| NA18536 | CHB |
| NA18537 | CHB |
| NA18538 | CHB |
| NA18539 | CHB |
| NA18541 | CHB |
| NA18542 | CHB |
| NA18543 | CHB |
| NA18544 | CHB |
| NA18545 | CHB |
| NA18546 | CHB |
| NA18547 | CHB |
| NA18548 | CHB |
| NA18549 | CHB |
| NA18550 | CHB |
| NA18552 | CHB |
| NA18553 | CHB |
| NA18555 | CHB |
| NA18557 | CHB |
| NA18558 | CHB |
| NA18559 | CHB |
| NA18560 | CHB |
| NA18561 | CHB |
| NA18562 | CHB |
| NA18563 | CHB |
| NA18564 | CHB |
| NA18565 | CHB |
| NA18566 | CHB |
| NA18567 | CHB |
| NA18570 | CHB |
| NA18571 | CHB |
| NA18572 | CHB |
| NA18573 | CHB |
| NA18574 | CHB |
| NA18576 | CHB |
| NA18577 | CHB |
| NA18579 | CHB |
| NA18582 | CHB |
| NA18592 | CHB |
| NA18593 | CHB |
| NA18595 | CHB |
| NA18596 | CHB |
| NA18597 | CHB |
| NA18599 | CHB |
| NA18602 | CHB |
| NA18603 | CHB |
| NA18605 | CHB |
| NA18606 | CHB |
| NA18608 | CHB |
| NA18609 | CHB |
| NA18610 | CHB |
| NA18611 | CHB |
| NA18612 | CHB |
| NA18613 | CHB |
| NA18614 | CHB |
| NA18615 | CHB |
| NA18616 | CHB |
| NA18617 | CHB |
| NA18618 | CHB |
| NA18619 | CHB |
| NA18620 | CHB |
| NA18621 | CHB |
| NA18622 | CHB |
| NA18623 | CHB |
| NA18624 | CHB |
| NA18626 | CHB |
| NA18627 | CHB |
| NA18628 | CHB |
| NA18630 | CHB |
| NA18631 | CHB |
| NA18632 | CHB |
| NA18633 | CHB |
| NA18634 | CHB |
| NA18635 | CHB |
| NA18636 | CHB |
| NA18637 | CHB |
| NA18638 | CHB |
| NA18639 | CHB |
| NA18640 | CHB |
| NA18641 | CHB |
| NA18642 | CHB |
| NA18643 | CHB |
| NA18645 | CHB |
| NA18647 | CHB |
| NA18740 | CHB |
| NA18745 | CHB |
| NA18747 | CHB |
| NA18748 | CHB |
| NA18749 | CHB |
| NA18757 | CHB |
| HG00403 | CHS |
| HG00404 | CHS |
| HG00406 | CHS |
| HG00407 | CHS |
| HG00418 | CHS |
| HG00419 | CHS |
| HG00421 | CHS |
| HG00422 | CHS |
| HG00428 | CHS |
| HG00436 | CHS |
| HG00437 | CHS |
| HG00442 | CHS |
| HG00443 | CHS |
| HG00445 | CHS |
| HG00446 | CHS |
| HG00448 | CHS |
| HG00449 | CHS |
| HG00451 | CHS |
| HG00452 | CHS |
| HG00457 | CHS |
| HG00458 | CHS |
| HG00463 | CHS |
| HG00464 | CHS |
| HG00472 | CHS |
| HG00473 | CHS |
| HG00475 | CHS |
| HG00476 | CHS |
| HG00478 | CHS |
| HG00479 | CHS |
| HG00500 | CHS |
| HG00512 | CHS |
| HG00513 | CHS |
| HG00525 | CHS |
| HG00530 | CHS |
| HG00531 | CHS |
| HG00533 | CHS |
| HG00534 | CHS |
| HG00536 | CHS |
| HG00537 | CHS |
| HG00542 | CHS |
| HG00543 | CHS |
| HG00556 | CHS |
| HG00557 | CHS |
| HG00559 | CHS |
| HG00560 | CHS |
| HG00565 | CHS |
| HG00566 | CHS |
| HG00577 | CHS |
| HG00580 | CHS |
| HG00583 | CHS |
| HG00589 | CHS |
| HG00590 | CHS |
| HG00592 | CHS |
| HG00593 | CHS |
| HG00595 | CHS |
| HG00596 | CHS |
| HG00607 | CHS |
| HG00608 | CHS |
| HG00610 | CHS |
| HG00611 | CHS |
| HG00613 | CHS |
| HG00614 | CHS |
| HG00619 | CHS |
| HG00620 | CHS |
| HG00625 | CHS |
| HG00626 | CHS |
| HG00628 | CHS |
| HG00629 | CHS |
| HG00634 | CHS |
| HG00635 | CHS |
| HG00650 | CHS |
| HG00651 | CHS |
| HG00653 | CHS |
| HG00654 | CHS |
| HG00656 | CHS |
| HG00657 | CHS |
| HG00662 | CHS |
| HG00663 | CHS |
| HG00671 | CHS |
| HG00672 | CHS |
| HG00683 | CHS |
| HG00684 | CHS |
| HG00689 | CHS |
| HG00690 | CHS |
| HG00692 | CHS |
| HG00693 | CHS |
| HG00698 | CHS |
| HG00699 | CHS |
| HG00701 | CHS |
| HG00704 | CHS |
| HG00705 | CHS |
| HG00707 | CHS |
| HG00708 | CHS |
| HG01112 | CLM |
| HG01113 | CLM |
| HG01124 | CLM |
| HG01125 | CLM |
| HG01133 | CLM |
| HG01134 | CLM |
| HG01136 | CLM |
| HG01137 | CLM |
| HG01140 | CLM |
| HG01148 | CLM |
| HG01149 | CLM |
| HG01250 | CLM |
| HG01251 | CLM |
| HG01257 | CLM |
| HG01259 | CLM |
| HG01271 | CLM |
| HG01272 | CLM |
| HG01274 | CLM |
| HG01275 | CLM |
| HG01277 | CLM |
| HG01278 | CLM |
| HG01342 | CLM |
| HG01344 | CLM |
| HG01345 | CLM |
| HG01350 | CLM |
| HG01351 | CLM |
| HG01353 | CLM |
| HG01354 | CLM |
| HG01356 | CLM |
| HG01357 | CLM |
| HG01359 | CLM |
| HG01360 | CLM |
| HG01365 | CLM |
| HG01366 | CLM |
| HG01374 | CLM |
| HG01375 | CLM |
| HG01377 | CLM |
| HG01378 | CLM |
| HG01383 | CLM |
| HG01384 | CLM |
| HG01389 | CLM |
| HG01390 | CLM |
| HG01437 | CLM |
| HG01440 | CLM |
| HG01441 | CLM |
| HG01455 | CLM |
| HG01456 | CLM |
| HG01461 | CLM |
| HG01462 | CLM |
| HG01465 | CLM |
| HG01488 | CLM |
| HG01489 | CLM |
| HG01491 | CLM |
| HG01492 | CLM |
| HG01494 | CLM |
| HG01495 | CLM |
| HG01497 | CLM |
| HG01498 | CLM |
| HG01550 | CLM |
| HG01551 | CLM |
| HG00171 | FIN |
| HG00173 | FIN |
| HG00174 | FIN |
| HG00176 | FIN |
| HG00177 | FIN |
| HG00178 | FIN |
| HG00179 | FIN |
| HG00180 | FIN |
| HG00182 | FIN |
| HG00183 | FIN |
| HG00185 | FIN |
| HG00186 | FIN |
| HG00187 | FIN |
| HG00188 | FIN |
| HG00189 | FIN |
| HG00190 | FIN |
| HG00266 | FIN |
| HG00267 | FIN |
| HG00268 | FIN |
| HG00269 | FIN |
| HG00270 | FIN |
| HG00271 | FIN |
| HG00272 | FIN |
| HG00273 | FIN |
| HG00274 | FIN |
| HG00275 | FIN |
| HG00276 | FIN |
| HG00277 | FIN |
| HG00278 | FIN |
| HG00280 | FIN |
| HG00281 | FIN |
| HG00282 | FIN |
| HG00284 | FIN |
| HG00285 | FIN |
| HG00306 | FIN |
| HG00309 | FIN |
| HG00310 | FIN |
| HG00311 | FIN |
| HG00312 | FIN |
| HG00313 | FIN |
| HG00315 | FIN |
| HG00318 | FIN |
| HG00319 | FIN |
| HG00320 | FIN |
| HG00321 | FIN |
| HG00323 | FIN |
| HG00324 | FIN |
| HG00325 | FIN |
| HG00326 | FIN |
| HG00327 | FIN |
| HG00328 | FIN |
| HG00329 | FIN |
| HG00330 | FIN |
| HG00331 | FIN |
| HG00332 | FIN |
| HG00334 | FIN |
| HG00335 | FIN |
| HG00336 | FIN |
| HG00337 | FIN |
| HG00338 | FIN |
| HG00339 | FIN |
| HG00341 | FIN |
| HG00342 | FIN |
| HG00343 | FIN |
| HG00344 | FIN |
| HG00345 | FIN |
| HG00346 | FIN |
| HG00349 | FIN |
| HG00350 | FIN |
| HG00351 | FIN |
| HG00353 | FIN |
| HG00355 | FIN |
| HG00356 | FIN |
| HG00357 | FIN |
| HG00358 | FIN |
| HG00359 | FIN |
| HG00360 | FIN |
| HG00361 | FIN |
| HG00362 | FIN |
| HG00364 | FIN |
| HG00366 | FIN |
| HG00367 | FIN |
| HG00369 | FIN |
| HG00372 | FIN |
| HG00373 | FIN |
| HG00375 | FIN |
| HG00376 | FIN |
| HG00377 | FIN |
| HG00378 | FIN |
| HG00381 | FIN |
| HG00382 | FIN |
| HG00383 | FIN |
| HG00384 | FIN |
| HG00096 | GBR |
| HG00097 | GBR |
| HG00099 | GBR |
| HG00100 | GBR |
| HG00101 | GBR |
| HG00102 | GBR |
| HG00103 | GBR |
| HG00104 | GBR |
| HG00106 | GBR |
| HG00108 | GBR |
| HG00109 | GBR |
| HG00110 | GBR |
| HG00111 | GBR |
| HG00112 | GBR |
| HG00113 | GBR |
| HG00114 | GBR |
| HG00116 | GBR |
| HG00117 | GBR |
| HG00118 | GBR |
| HG00119 | GBR |
| HG00120 | GBR |
| HG00121 | GBR |
| HG00122 | GBR |
| HG00123 | GBR |
| HG00124 | GBR |
| HG00125 | GBR |
| HG00126 | GBR |
| HG00127 | GBR |
| HG00128 | GBR |
| HG00129 | GBR |
| HG00130 | GBR |
| HG00131 | GBR |
| HG00133 | GBR |
| HG00134 | GBR |
| HG00135 | GBR |
| HG00136 | GBR |
| HG00137 | GBR |
| HG00138 | GBR |
| HG00139 | GBR |
| HG00140 | GBR |
| HG00141 | GBR |
| HG00142 | GBR |
| HG00143 | GBR |
| HG00146 | GBR |
| HG00148 | GBR |
| HG00149 | GBR |
| HG00150 | GBR |
| HG00151 | GBR |
| HG00152 | GBR |
| HG00154 | GBR |
| HG00155 | GBR |
| HG00156 | GBR |
| HG00158 | GBR |
| HG00159 | GBR |
| HG00160 | GBR |
| HG00231 | GBR |
| HG00232 | GBR |
| HG00233 | GBR |
| HG00234 | GBR |
| HG00235 | GBR |
| HG00236 | GBR |
| HG00237 | GBR |
| HG00238 | GBR |
| HG00239 | GBR |
| HG00240 | GBR |
| HG00242 | GBR |
| HG00243 | GBR |
| HG00244 | GBR |
| HG00245 | GBR |
| HG00246 | GBR |
| HG00247 | GBR |
| HG00249 | GBR |
| HG00250 | GBR |
| HG00251 | GBR |
| HG00252 | GBR |
| HG00253 | GBR |
| HG00254 | GBR |
| HG00255 | GBR |
| HG00256 | GBR |
| HG00257 | GBR |
| HG00258 | GBR |
| HG00259 | GBR |
| HG00260 | GBR |
| HG00261 | GBR |
| HG00262 | GBR |
| HG00263 | GBR |
| HG00264 | GBR |
| HG00265 | GBR |
| HG01334 | GBR |
| HG01515 | IBS |
| HG01516 | IBS |
| HG01518 | IBS |
| HG01519 | IBS |
| HG01521 | IBS |
| HG01522 | IBS |
| HG01617 | IBS |
| HG01618 | IBS |
| HG01619 | IBS |
| HG01620 | IBS |
| HG01623 | IBS |
| HG01624 | IBS |
| HG01625 | IBS |
| HG01626 | IBS |
| NA18939 | JPT |
| NA18940 | JPT |
| NA18941 | JPT |
| NA18942 | JPT |
| NA18943 | JPT |
| NA18944 | JPT |
| NA18945 | JPT |
| NA18946 | JPT |
| NA18947 | JPT |
| NA18948 | JPT |
| NA18949 | JPT |
| NA18950 | JPT |
| NA18951 | JPT |
| NA18952 | JPT |
| NA18953 | JPT |
| NA18954 | JPT |
| NA18956 | JPT |
| NA18957 | JPT |
| NA18959 | JPT |
| NA18960 | JPT |
| NA18961 | JPT |
| NA18962 | JPT |
| NA18963 | JPT |
| NA18964 | JPT |
| NA18965 | JPT |
| NA18966 | JPT |
| NA18968 | JPT |
| NA18971 | JPT |
| NA18973 | JPT |
| NA18974 | JPT |
| NA18975 | JPT |
| NA18976 | JPT |
| NA18977 | JPT |
| NA18978 | JPT |
| NA18980 | JPT |
| NA18981 | JPT |
| NA18982 | JPT |
| NA18983 | JPT |
| NA18984 | JPT |
| NA18985 | JPT |
| NA18986 | JPT |
| NA18987 | JPT |
| NA18988 | JPT |
| NA18989 | JPT |
| NA18990 | JPT |
| NA18992 | JPT |
| NA18994 | JPT |
| NA18995 | JPT |
| NA18998 | JPT |
| NA18999 | JPT |
| NA19000 | JPT |
| NA19002 | JPT |
| NA19003 | JPT |
| NA19004 | JPT |
| NA19005 | JPT |
| NA19007 | JPT |
| NA19009 | JPT |
| NA19010 | JPT |
| NA19012 | JPT |
| NA19054 | JPT |
| NA19055 | JPT |
| NA19056 | JPT |
| NA19057 | JPT |
| NA19058 | JPT |
| NA19059 | JPT |
| NA19060 | JPT |
| NA19062 | JPT |
| NA19063 | JPT |
| NA19064 | JPT |
| NA19065 | JPT |
| NA19066 | JPT |
| NA19067 | JPT |
| NA19068 | JPT |
| NA19070 | JPT |
| NA19072 | JPT |
| NA19074 | JPT |
| NA19075 | JPT |
| NA19076 | JPT |
| NA19077 | JPT |
| NA19078 | JPT |
| NA19079 | JPT |
| NA19080 | JPT |
| NA19081 | JPT |
| NA19082 | JPT |
| NA19083 | JPT |
| NA19084 | JPT |
| NA19085 | JPT |
| NA19087 | JPT |
| NA19088 | JPT |
| NA19020 | LWK |
| NA19028 | LWK |
| NA19035 | LWK |
| NA19036 | LWK |
| NA19038 | LWK |
| NA19041 | LWK |
| NA19044 | LWK |
| NA19046 | LWK |
| NA19307 | LWK |
| NA19308 | LWK |
| NA19309 | LWK |
| NA19310 | LWK |
| NA19311 | LWK |
| NA19312 | LWK |
| NA19315 | LWK |
| NA19316 | LWK |
| NA19317 | LWK |
| NA19318 | LWK |
| NA19319 | LWK |
| NA19321 | LWK |
| NA19324 | LWK |
| NA19327 | LWK |
| NA19328 | LWK |
| NA19331 | LWK |
| NA19332 | LWK |
| NA19338 | LWK |
| NA19346 | LWK |
| NA19350 | LWK |
| NA19351 | LWK |
| NA19352 | LWK |
| NA19355 | LWK |
| NA19359 | LWK |
| NA19360 | LWK |
| NA19371 | LWK |
| NA19372 | LWK |
| NA19374 | LWK |
| NA19375 | LWK |
| NA19376 | LWK |
| NA19377 | LWK |
| NA19379 | LWK |
| NA19380 | LWK |
| NA19381 | LWK |
| NA19383 | LWK |
| NA19384 | LWK |
| NA19385 | LWK |
| NA19390 | LWK |
| NA19391 | LWK |
| NA19393 | LWK |
| NA19394 | LWK |
| NA19395 | LWK |
| NA19397 | LWK |
| NA19398 | LWK |
| NA19399 | LWK |
| NA19401 | LWK |
| NA19403 | LWK |
| NA19404 | LWK |
| NA19428 | LWK |
| NA19429 | LWK |
| NA19430 | LWK |
| NA19431 | LWK |
| NA19435 | LWK |
| NA19436 | LWK |
| NA19437 | LWK |
| NA19438 | LWK |
| NA19439 | LWK |
| NA19440 | LWK |
| NA19444 | LWK |
| NA19445 | LWK |
| NA19446 | LWK |
| NA19448 | LWK |
| NA19449 | LWK |
| NA19451 | LWK |
| NA19452 | LWK |
| NA19455 | LWK |
| NA19456 | LWK |
| NA19457 | LWK |
| NA19461 | LWK |
| NA19462 | LWK |
| NA19463 | LWK |
| NA19466 | LWK |
| NA19467 | LWK |
| NA19468 | LWK |
| NA19469 | LWK |
| NA19471 | LWK |
| NA19472 | LWK |
| NA19473 | LWK |
| NA19474 | LWK |
| NA19648 | MXL |
| NA19651 | MXL |
| NA19652 | MXL |
| NA19654 | MXL |
| NA19655 | MXL |
| NA19657 | MXL |
| NA19661 | MXL |
| NA19663 | MXL |
| NA19672 | MXL |
| NA19676 | MXL |
| NA19678 | MXL |
| NA19679 | MXL |
| NA19681 | MXL |
| NA19682 | MXL |
| NA19684 | MXL |
| NA19716 | MXL |
| NA19717 | MXL |
| NA19719 | MXL |
| NA19720 | MXL |
| NA19722 | MXL |
| NA19723 | MXL |
| NA19725 | MXL |
| NA19728 | MXL |
| NA19729 | MXL |
| NA19731 | MXL |
| NA19732 | MXL |
| NA19734 | MXL |
| NA19735 | MXL |
| NA19737 | MXL |
| NA19738 | MXL |
| NA19740 | MXL |
| NA19741 | MXL |
| NA19746 | MXL |
| NA19747 | MXL |
| NA19749 | MXL |
| NA19750 | MXL |
| NA19752 | MXL |
| NA19755 | MXL |
| NA19756 | MXL |
| NA19758 | MXL |
| NA19759 | MXL |
| NA19761 | MXL |
| NA19762 | MXL |
| NA19764 | MXL |
| NA19770 | MXL |
| NA19771 | MXL |
| NA19773 | MXL |
| NA19774 | MXL |
| NA19776 | MXL |
| NA19777 | MXL |
| NA19779 | MXL |
| NA19780 | MXL |
| NA19782 | MXL |
| NA19783 | MXL |
| NA19785 | MXL |
| NA19786 | MXL |
| NA19788 | MXL |
| NA19789 | MXL |
| NA19794 | MXL |
| NA19795 | MXL |
| HG00553 | PUR |
| HG00554 | PUR |
| HG00637 | PUR |
| HG00638 | PUR |
| HG00640 | PUR |
| HG00641 | PUR |
| HG00731 | PUR |
| HG00732 | PUR |
| HG00734 | PUR |
| HG00736 | PUR |
| HG00737 | PUR |
| HG00740 | PUR |
| HG01047 | PUR |
| HG01048 | PUR |
| HG01051 | PUR |
| HG01052 | PUR |
| HG01055 | PUR |
| HG01060 | PUR |
| HG01061 | PUR |
| HG01066 | PUR |
| HG01067 | PUR |
| HG01069 | PUR |
| HG01070 | PUR |
| HG01072 | PUR |
| HG01073 | PUR |
| HG01075 | PUR |
| HG01079 | PUR |
| HG01080 | PUR |
| HG01082 | PUR |
| HG01083 | PUR |
| HG01085 | PUR |
| HG01095 | PUR |
| HG01097 | PUR |
| HG01098 | PUR |
| HG01101 | PUR |
| HG01102 | PUR |
| HG01104 | PUR |
| HG01105 | PUR |
| HG01107 | PUR |
| HG01108 | PUR |
| HG01167 | PUR |
| HG01168 | PUR |
| HG01170 | PUR |
| HG01171 | PUR |
| HG01173 | PUR |
| HG01174 | PUR |
| HG01176 | PUR |
| HG01183 | PUR |
| HG01187 | PUR |
| HG01188 | PUR |
| HG01190 | PUR |
| HG01191 | PUR |
| HG01197 | PUR |
| HG01198 | PUR |
| HG01204 | PUR |
| NA20502 | TSI |
| NA20503 | TSI |
| NA20504 | TSI |
| NA20505 | TSI |
| NA20506 | TSI |
| NA20507 | TSI |
| NA20508 | TSI |
| NA20509 | TSI |
| NA20510 | TSI |
| NA20512 | TSI |
| NA20513 | TSI |
| NA20515 | TSI |
| NA20516 | TSI |
| NA20517 | TSI |
| NA20518 | TSI |
| NA20519 | TSI |
| NA20520 | TSI |
| NA20521 | TSI |
| NA20522 | TSI |
| NA20524 | TSI |
| NA20525 | TSI |
| NA20527 | TSI |
| NA20528 | TSI |
| NA20529 | TSI |
| NA20530 | TSI |
| NA20531 | TSI |
| NA20532 | TSI |
| NA20533 | TSI |
| NA20534 | TSI |
| NA20535 | TSI |
| NA20536 | TSI |
| NA20537 | TSI |
| NA20538 | TSI |
| NA20539 | TSI |
| NA20540 | TSI |
| NA20541 | TSI |
| NA20542 | TSI |
| NA20543 | TSI |
| NA20544 | TSI |
| NA20581 | TSI |
| NA20582 | TSI |
| NA20585 | TSI |
| NA20586 | TSI |
| NA20588 | TSI |
| NA20589 | TSI |
| NA20752 | TSI |
| NA20753 | TSI |
| NA20754 | TSI |
| NA20755 | TSI |
| NA20756 | TSI |
| NA20757 | TSI |
| NA20758 | TSI |
| NA20759 | TSI |
| NA20760 | TSI |
| NA20761 | TSI |
| NA20765 | TSI |
| NA20766 | TSI |
| NA20768 | TSI |
| NA20769 | TSI |
| NA20770 | TSI |
| NA20771 | TSI |
| NA20772 | TSI |
| NA20773 | TSI |
| NA20774 | TSI |
| NA20775 | TSI |
| NA20778 | TSI |
| NA20783 | TSI |
| NA20785 | TSI |
| NA20786 | TSI |
| NA20787 | TSI |
| NA20790 | TSI |
| NA20792 | TSI |
| NA20795 | TSI |
| NA20796 | TSI |
| NA20797 | TSI |
| NA20798 | TSI |
| NA20799 | TSI |
| NA20800 | TSI |
| NA20801 | TSI |
| NA20802 | TSI |
| NA20803 | TSI |
| NA20804 | TSI |
| NA20805 | TSI |
| NA20806 | TSI |
| NA20807 | TSI |
| NA20808 | TSI |
| NA20809 | TSI |
| NA20810 | TSI |
| NA20811 | TSI |
| NA20812 | TSI |
| NA20813 | TSI |
| NA20814 | TSI |
| NA20815 | TSI |
| NA20816 | TSI |
| NA20818 | TSI |
| NA20819 | TSI |
| NA20826 | TSI |
| NA20828 | TSI |
| NA18486 | YRI |
| NA18487 | YRI |
| NA18489 | YRI |
| NA18498 | YRI |
| NA18499 | YRI |
| NA18501 | YRI |
| NA18502 | YRI |
| NA18504 | YRI |
| NA18505 | YRI |
| NA18507 | YRI |
| NA18508 | YRI |
| NA18510 | YRI |
| NA18511 | YRI |
| NA18516 | YRI |
| NA18517 | YRI |
| NA18519 | YRI |
| NA18520 | YRI |
| NA18522 | YRI |
| NA18523 | YRI |
| NA18853 | YRI |
| NA18856 | YRI |
| NA18858 | YRI |
| NA18861 | YRI |
| NA18867 | YRI |
| NA18868 | YRI |
| NA18870 | YRI |
| NA18871 | YRI |
| NA18873 | YRI |
| NA18874 | YRI |
| NA18907 | YRI |
| NA18908 | YRI |
| NA18909 | YRI |
| NA18910 | YRI |
| NA18912 | YRI |
| NA18916 | YRI |
| NA18917 | YRI |
| NA18923 | YRI |
| NA18924 | YRI |
| NA18933 | YRI |
| NA18934 | YRI |
| NA19093 | YRI |
| NA19095 | YRI |
| NA19096 | YRI |
| NA19098 | YRI |
| NA19099 | YRI |
| NA19102 | YRI |
| NA19107 | YRI |
| NA19108 | YRI |
| NA19113 | YRI |
| NA19114 | YRI |
| NA19116 | YRI |
| NA19117 | YRI |
| NA19118 | YRI |
| NA19119 | YRI |
| NA19121 | YRI |
| NA19129 | YRI |
| NA19130 | YRI |
| NA19131 | YRI |
| NA19137 | YRI |
| NA19138 | YRI |
| NA19146 | YRI |
| NA19147 | YRI |
| NA19149 | YRI |
| NA19150 | YRI |
| NA19152 | YRI |
| NA19160 | YRI |
| NA19171 | YRI |
| NA19172 | YRI |
| NA19175 | YRI |
| NA19185 | YRI |
| NA19189 | YRI |
| NA19190 | YRI |
| NA19197 | YRI |
| NA19198 | YRI |
| NA19200 | YRI |
| NA19204 | YRI |
| NA19207 | YRI |
| NA19209 | YRI |
| NA19213 | YRI |
| NA19222 | YRI |
| NA19223 | YRI |
| NA19225 | YRI |
| NA19235 | YRI |
| NA19236 | YRI |
| NA19247 | YRI |
| NA19248 | YRI |
| NA19256 | YRI |
| NA19257 | YRI |

**Table S2: Results of SVM grid searches for various sets of training data**

| X or Autosomes? | Window size | PhastCons percentage cutoff | Number of bins | Include fixations or treat as monomorphic | Cost | Gamma | Accuracy |
|---|---|---|---|---|---|---|---|
| A | 5 kb | 33% | 10 | Fixed_derived | 2 | 32 | 77.45% |
| A | 5 kb | 33% | 10 | Monomorphic | 2 | 32 | 75.15% |
| A | 5 kb | 33% | 25 | Fixed_derived | 0.5 | 4 | 79.85% |
| A | 5 kb | 33% | 25 | Monomorphic | 0.5 | 4 | 77.10% |
| A | 5 kb | 33% | 33 | Fixed_derived | 2 | 1 | 80.40% |
| A | 5 kb | 33% | 33 | Monomorphic | 1 | 4 | 77.50% |
| A | 5 kb | 33% | 100 | Fixed_derived | 1 | 1 | 81.75% |
| A | 5 kb | 33% | 100 | Monomorphic | 8 | 1 | 80.30% |
| A | 5 kb | 33% | 250 | Fixed_derived | 16 | 0.125 | 83.05% |
| A | 5 kb | 33% | 250 | Monomorphic | 2 | 0.5 | 81.90% |
| A | 5 kb | 33% | 500 | Fixed_derived | 2 | 0.25 | 82.95% |
| A | 5 kb | 33% | 500 | Monomorphic | 1 | 0.25 | 82.25% |
| A | 5 kb | 33% | 1000 | Fixed_derived | 8 | 0.0625 | 83% |
| A | 5 kb | 33% | 2128 | Fixed_derived | 1 | 0.0625 | 82.10% |
| A | 5 kb | 33% | 2128 | Monomorphic | 4 | 0.0625 | 81.70% |
| A | 10 kb | 25% | 10 | Fixed_derived | 0.13 | 8 | 81.17% |
| A | 10 kb | 25% | 10 | Monomorphic | 0.06 | 8 | 77.46% |
| A | 10 kb | 25% | 25 | Fixed_derived | 0.5 | 4 | 82.66% |
| A | 10 kb | 25% | 25 | Monomorphic | 0.25 | 8 | 79.55% |
| A | 10 kb | 25% | 33 | Fixed_derived | 1 | 4 | 83.40% |
| A | 10 kb | 25% | 33 | Monomorphic | 0.13 | 4 | 79.96% |
| A | 10 kb | 25% | 100 | Fixed_derived | 2 | 2 | 84.35% |
| A | 10 kb | 25% | 100 | Monomorphic | 2 | 2 | 83.33% |
| A | 10 kb | 25% | 250 | Fixed_derived | 2 | 0.5 | 86.91% |
| A | 10 kb | 25% | 250 | Monomorphic | 4 | 0.5 | 86.03% |
| A | 10 kb | 25% | 500 | Fixed_derived | 2 | 0.25 | 86.17% |
| A | 10 kb | 25% | 500 | Monomorphic | 4 | 0.25 | 85.90% |
| A | 10 kb | 25% | 1000 | Fixed_derived | 4 | 0.125 | 88.26% |
| A | 10 kb | 25% | 1000 | Monomorphic | 2 | 0.125 | 87.79% |
| A | 10 kb | 33% | 10 | Fixed_derived | 2 | 8 | 82.11% |
| A | 10 kb | 33% | 10 | Monomorphic | 0.25 | 4 | 78.05% |
| A | 10 kb | 33% | 25 | Fixed_derived | 16 | 0.25 | 83.54% |
| A | 10 kb | 33% | 25 | Monomorphic | 0.03 | 2 | 79.27% |
| A | 10 kb | 33% | 33 | Fixed_derived | 1 | 2 | 81.91% |
| A | 10 kb | 33% | 33 | Monomorphic | 0 | 1 | 79.88% |
| A | 10 kb | 33% | 100 | Fixed_derived | 2 | 0.25 | 84.96% |
| A | 10 kb | 33% | 100 | Monomorphic | 8 | 1 | 83.13% |
| A | 10 kb | 33% | 250 | Fixed_derived | 4 | 0.25 | 85.57% |
| A | 10 kb | 33% | 250 | Monomorphic | 1 | 0.0625 | 84.55% |
| A | 10 kb | 33% | 500 | Fixed_derived | 4 | 0.03125 | 85.98% |
| A | 10 kb | 33% | 500 | Monomorphic | 8 | 0.015625 | 85.37% |
| A | 10 kb | 33% | 1000 | Fixed_derived | 2 | 0.03125 | 86.38% |
| A | 10 kb | 33% | 1000 | Monomorphic | 2 | 0.0625 | 86.59% |
| A | 10 kb | 33% | 2128 | Fixed_derived | 16 | 0.003906 | 83.33% |
| A | 10 kb | 33% | 2128 | Monomorphic | 2 | 0.03125 | 82.72% |
| A | 20 kb | 25% | 10 | Fixed_derived | 0.06 | 1 | 84.77% |
| A | 20 kb | 25% | 10 | Monomorphic | 1 | 8 | 82.76% |
| A | 20 kb | 25% | 25 | Fixed_derived | 1 | 0.25 | 87.07% |
| A | 20 kb | 25% | 25 | Monomorphic | 1 | 2 | 84.77% |
| A | 20 kb | 25% | 33 | Fixed_derived | 1 | 1 | 87.93% |
| A | 20 kb | 25% | 33 | Monomorphic | 2 | 2 | 86.21% |
| A | 20 kb | 25% | 100 | Fixed_derived | 2 | 0.25 | 89.37% |
| A | 20 kb | 25% | 100 | Monomorphic | 2 | 0.25 | 87.07% |
| A | 20 kb | 25% | 250 | Fixed_derived | 8 | 0.125 | 88.51% |
| A | 20 kb | 25% | 250 | Monomorphic | 4 | 0.25 | 87.64% |
| A | 20 kb | 25% | 500 | Fixed_derived | 0.06 | 0.0625 | 89.37% |
| A | 20 kb | 25% | 500 | Monomorphic | 1 | 0.0625 | 89.66% |
| A | 20 kb | 25% | 1000 | Fixed_derived | 0.5 | 0.0625 | 87.93% |
| A | 20 kb | 25% | 1000 | Monomorphic | 8 | 0.015625 | 88.22% |
| A | 20 kb | 25% | 2128 | Fixed_derived | 4 | 0.015625 | 88.22% |
| A | 20 kb | 25% | 2128 | Monomorphic | 16 | 0.003906 | 88.22% |
| X | 10 kb | 5% | 500 | Fixed_derived | 2 | 0.25 | 71.65% |
| X | 10 kb | 5% | 500 | Monomorphic | 2 | 0.25 | 71.65% |
| X | 10 kb | 5% | 1658 | Fixed_derived | 4 | 0.125 | 72.45% |
| X | 10 kb | 10% | 500 | Fixed_derived | 4 | 0.003906 | 69.44% |
| X | 10 kb | 10% | 500 | Monomorphic | 8 | 0.003906 | 67.76% |
| X | 10 kb | 10% | 1658 | Fixed_derived | 2 | 0.0625 | 72.37% |
| X | 10 kb | 15% | 500 | Fixed_derived | 4 | 0.125 | 72.45% |
| X | 10 kb | 15% | 500 | Monomorphic | 2 | 0.125 | 72.70% |
| X | 10 kb | 15% | 1658 | Fixed_derived | 8 | 0.003906 | 67.76% |
| X | 10 kb | 20% | 500 | Fixed_derived | 2 | 0.015625 | 68.52% |
| X | 10 kb | 20% | 500 | Monomorphic | 2 | 0.015625 | 68.52% |
| X | 10 kb | 20% | 1658 | Fixed_derived | 4 | 0.003906 | 69.44% |
| X | 10 kb | 75% | 500 | Fixed_derived | 2 | 0.125 | 74.82% |
| X | 10 kb | 75% | 500 | Monomorphic | 2 | 0.0625 | 72.82% |
| X | 10 kb | 75% | 2128 | Fixed_derived | 2 | 0.0625 | 72.82% |

**Table S3: Tests for enrichment/depletion of various genomic features in popCons and popUncons elements (all data found on UCSC Table Browser)**

| Genomic feature | popCons enrichment | popCons P-value (one-sided test for enrichment) | popUncons enrichment | popUncons P-value (one-sided test for depletion) |
|---|---|---|---|---|
| CNVs in Coriell's inherited disorder and chromosomal aberration cell lines | 0.938969967 | 1 | 1.019364017 | 1 |
| COSMIC mutations (Catalogue Of Somatic Mutations In Cancer)[1,2] | 1.747335664 | <0.001 | 0.802299356 | <0.001 |
| Enhancers/promoters present in humans but not mice[3] | 1.461226038 | <0.001 | 0.67360754 | <0.001 |
| Enhancers/promoters present in humans but not mice[3] | 1.994627786 | 0.095 | 1.715806099 | 1 |
| Genetic Association Database (GAD) disease-associated genes[4] | 1.222935588 | <0.001 | 0.771507586 | <0.001 |
| Gencode exons[5] | 2.087529643 | <0.001 | 0.57111197 | <0.001 |
| GWAS SNPs[6] | 0.839870752 | 1 | 0.968436967 | 0.003 |
| Human QTLs[7] | 1.027885184 | <0.001 | 0.995664621 | <0.001 |
| lincRNAs[8,9] | 0.681400899 | 1 | 1.134831695 | 1 |
| miRNAs, snoRNAs, and scaRNAs[10,11,12,13,14] | 1.902130625 | <0.001 | 0.509152073 | <0.001 |
| OMIM Genes[15,16] | 1.44542909 | <0.001 | 0.748919359 | <0.001 |
| OMIM SNPs[15,16] | 1.976802733 | 0.005 | 0.721701212 | <0.001 |
| ORegAnno regulatory elements[17,18] | 1.345238119 | <0.001 | 0.724761364 | <0.001 |
| Transcription factor binding sites from ENCODE ChIP-Seq[20] | 1.211871882 | <0.001 | 0.831242902 | <0.001 |

References

1. Forbes SA, et al. The Catalogue of Somatic Mutations in Cancer (COSMIC). Curr Protoc Hum Genet. 2008 Apr 1;57:10.11.1-10.11.26.
2. Forbes SA, et al. COSMIC: mining complete cancer genomes in the Catalogue of Somatic Mutations in Cancer. Nucleic Acids Res. 2011 Jan;39(Database issue):D945-50. Epub 2010 Oct 15.
3. Cotney J, et al. The evolution of lineage-specific regulatory activities in the human embryonic limb. Cell 2013 Jul; 154(1):185-196.
4. Becker KG, et al. The Genetic Association Database. Nature Genetics 2004 May; 36(5):431-432.
5. Harrow J et al. GENCODE: the reference human genome annotation for The ENCODE Project. Genome Reserach 2012 Sep; 22(9):1760-1774.
6. Hindorff et al. Potential etiologic and functional implications of genome-wide association loci for human diseases and traits. PNAS. 2009 Jun 9;106(23):9362-7.
7. Rapp, JP. Genetic Analysis of Inherited Hypertension in the Rat. Physiol. Rev. 2000 Jan;90(1):135-172.
8. Cabili MN et al. Integrative annotation of human large intergenic noncoding RNAs reveals global properties and specific subclasses. Genes and Development. 2011 Sep 15;25:1915-1927.
9. Trapnell C et al. Transcript assembly and quantification by RNA-Seq reveals unannotated transcripts and isoform switching during cell differentiation. Nature Biotechnology. 2010 May 2;28:511-515.
10. Griffiths-Jones S. The microRNA Registry. Nucleic Acids Res. 2004 Jan 1;32(Database issue):D109-11.
11. Griffiths-Jones S et al. miRBase: microRNA sequences, targets and gene nomenclature. Nucleic Acids Res. 2006 Jan 1;334(Database issue):D14-4.
12. Griffiths-Jones S et al. miRBase: tools for microRNA genomics. Nucleic Acids Res. 2008 Jan;36(Database issue):D154-8.
13. Lestrade L et al. snoRNA-LBME-db, a comprehensive database of human H/ACA and C/D box snoRNAs. Nucleic Acids Res. 2006 Jan 1;34(Database issue):D158- 62.
14. Weber MJ. New human and mouse microRNA genes found by homology search. Febs J. 2005 Jan;272(1):59-73.
15. Amberger J et al. Nucleic Acids Res. 2009 Jan;37(Database issue):D793-6. Epub 2008 Oct 8.
16. Hamosh A et al. Online Mendelian Inheritance in Man (OMIM), a knowledgebase of human genes and genetic disorders. Nucleic Acids Res. 2005 Jan 1;33(Database issue):D514-7.
17. Griffith OL et al. ORegAnno: an open-access community-driven resource for regulatory annotation. Nucleic Acids Res. 2008 Jan;36(Database issue):D107-13.
18. Montgomery SB et al. ORegAnno: an open access database and curation system for literature-derived promoters, transcription factor binding sites and regulatory variation. Bioinformatics. 2006 Mar 1;22(5):637-40.
19. Siepel A et al. Evolutionarily conserved elements in vertebrate, insect, worm, and yeast genomes. Genome Res. 2005 Aug;15(8):1034-50.
20. ENCODE Project Consortium. An integrated encyclopedia of DNA elements in the human genome. Nature. 2012 Sep 6;489(7414):57-74.

**Table S4: Tests for enrichment/depletion of various genomic features in popCons GOF candidates popUncons LOF candidates (all data except ref. 3 found on UCSC Table Browser)**

| Genomic feature | popCons GOF fold-enrichment | GOF P-value (one-sided test for enrichment) | popUncons LOF fold-enrichment | LOF P-value (one-sided test for depletion) |
|---|---|---|---|---|
| CNVs in Coriell's inherited disorder and chromosomal aberration cell lines | 0 | 1 | 0.252999551 | 0.185 |
| COSMIC mutations (Catalogue Of Somatic Mutations In Cancer)[1,2] | 0.549174305 | 0.87 | 0.345739086 | 0 |
| Enhancers/promoters present in humans but not mice[3] | 2.041275835 | 0.05 | 0.227876927 | 0 |
| Enhancers/promoters present in mice but not humans[3] | 0 | 1 | 0 | 0.367 |
| Genetic Association Database (GAD) disease-associated genes[4] | 0 | 1 | 0.39410206 | 0.001 |
| Gencode exons[5] | 0.693098357 | 0.982 | 0.371065201 | 0 |
| GWAS SNPs[6] | 0.371043746 | 1 | 0.710011157 | 0.229 |
| Human QTLs[7] | NA* | NA* | NA* | NA* |
| lincRNAs[8,9] | 0.796253398 | 0.864 | 0.440963667 | 0 |
| miRNAs, snoRNAs, and scaRNAs[10,11,12,13,14] | 7.603695396 | 0.003 | 0.214847904 | 0 |
| OMIM Genes[15,16] | 0.607402324 | 0.84 | 0.402109854 | 0 |
| OMIM SNPs[15,16] | 0 | 1 | 0.280668271 | 0 |
| ORegAnno regulatory elements[17,18] | 0.506722542 | 0.984 | 0.230686062 | 0 |
| Transcription factor binding sites from ENCODE ChIP-Seq[20] | 0.941034787 | 0.812 | 0.370711249 | 0 |

**References**

1. Forbes SA, et al. The Catalogue of Somatic Mutations in Cancer (COSMIC). Curr Protoc Hum Genet. 2008 Apr 1;57:10.11.1-10.11.26.
2. Forbes SA, et al. COSMIC: mining complete cancer genomes in the Catalogue of Somatic Mutations in Cancer. Nucleic Acids Res. 2011 Jan;39(Database issue):D945-50. Epub 2010 Oct 15.
3. Cotney J, et al. The evolution of lineage-specific regulatory activities in the human embryonic limb. Cell 2013 Jul; 154(1):185-196.
4. Becker KG, et al. The Genetic Association Database. Nature Genetics 2004 May; 36(5):431-432.
5. Harrow J et al. GENCODE: the reference human genome annotation for The ENCODE Project. Genome Reserach 2012 Sep; 22(9):1760-1774.
6. Hindorff et al. Potential etiologic and functional implications of genome-wide association loci for human diseases and traits. PNAS. 2009 Jun 9;106(23):9362-7.
7. Rapp, JP. Genetic Analysis of Inherited Hypertension in the Rat. Physiol. Rev. 2000 Jan;90(1):135-172.
8. Cabili MN et al. Integrative annotation of human large intergenic noncoding RNAs reveals global properties and specific subclasses. Genes and Development. 2011 Sep 15;25:1915-1927.
9. Trapnell C et al. Transcript assembly and quantification by RNA-Seq reveals unannotated transcripts and isoform switching during cell differentiation. Nature Biotechnology. 2010 May 2;28:511-515.
10. Griffiths-Jones S. The microRNA Registry. Nucleic Acids Res. 2004 Jan 1;32(Database issue):D109-11.
11. Griffiths-Jones S et al. miRBase: microRNA sequences, targets and gene nomenclature. Nucleic Acids Res. 2006 Jan 1;334(Database issue):D14-4.
12. Griffiths-Jones S et al. miRBase: tools for microRNA genomics. Nucleic Acids Res. 2008 Jan;36(Database issue):D154-8.
13. Lestrade L et al. snoRNA-LBME-db, a comprehensive database of human H/ACA and C/D box snoRNAs. Nucleic Acids Res. 2006 Jan 1;34(Database issue):D158- 62.
14. Weber MJ. New human and mouse microRNA genes found by homology search. Febs J. 2005 Jan;272(1):59-73.
15. Amberger J et al. Nucleic Acids Res. 2009 Jan;37(Database issue):D793-6. Epub 2008 Oct 8.
16. Hamosh A et al. Online Mendelian Inheritance in Man (OMIM), a knowledgebase of human genes and genetic disorders. Nucleic Acids Res. 2005 Jan 1;33(Database issue):D514-7.
17. Griffith OL et al. ORegAnno: an open-access community-driven resource for regulatory annotation. Nucleic Acids Res. 2008 Jan;36(Database issue):D107-13.
18. Montgomery SB et al. ORegAnno: an open access database and curation system for literature-derived promoters, transcription factor binding sites and regulatory variation. Bioinformatics. 2006 Mar 1;22(5):637-40.
19. Siepel A et al. Evolutionarily conserved elements in vertebrate, insect, worm, and yeast genomes. Genome Res. 2005 Aug;15(8):1034-50.
20. ENCODE Project Consortium. An integrated encyclopedia of DNA elements in the human genome. Nature. 2012 Sep 6;489(7414):57-74.

*No human QTL regions were made up of <1% phastCons bases for testing GOF enrichment or >15% phastCons bases for testing LOF enrichment.

**Table S5: Accuracy assessments of SVMs trained on variableious simulated data sets**

| Fraction of selected sites | Strength of selection (2*Ns*) | Mutation rate* | Recombination rate** | Cross validation accuracy | Test set accuracy | Accuracy on unconstrained data | Accuracy on unconstrained data after >95% cutoff | Accuracy on constrained data | Accuracy on constrained data after >95% cutoff | Fraction of positively selected simulations classified as constrained |
|---|---|---|---|---|---|---|---|---|---|---|
| 5% | 50 | medium | medium | 0.568333 | 0.5425 | 87/157 (0.55414 | NA | 130/243 (0.53497 | NA | 0.23 |
| 5% | 50 | medium | variable | 0.515 | 0.5 | NA | NA | 201/402 (0.5) | NA | 1 |
| 5% | 50 | variable | medium | 0.568333 | 0.544776119 | 97/176 (0.55113 | NA | 122/226 (0.53982 | NA | 0.5975 |
| 5% | 50 | variable | variable | 0.534314 | 0.526570048 | 87/163 (0.53374 | NA | 131/251 (0.52191 | NA | 0.565 |
| 10% | 50 | medium | medium | 0.595 | 0.5725 | 118/207 (0.5700 | NA | 111/193 (0.57512 | NA | 0.22 |
| 10% | 50 | medium | variable | 0.585 | 0.542288557 | 114/211 (0.5402 | NA | 104/191 (0.54450 | NA | 0.155 |
| 10% | 50 | variable | medium | 0.538333 | 0.52238806 | 98/187 (0.52406 | NA | 112/215 (0.52093 | NA | 0.46 |
| 10% | 50 | variable | variable | 0.552288 | 0.52173913 | 95/181 (0.52486 | NA | 121/233 (0.51931 | NA | 0.5175 |
| 15% | 50 | medium | medium | 0.586667 | 0.58 | 135/238 (0.5672 | NA | 97/162 (0.598765 | NA | 0.1 |
| 15% | 50 | medium | variable | 0.621667 | 0.60199005 | 138/235 (0.5872 | NA | 104/167 (0.62275 | NA | 0.105 |
| 15% | 50 | variable | medium | 0.573333 | 0.572139303 | 100/171 (0.5847 | NA | 130/231 (0.56277 | NA | 0.48 |
| 15% | 50 | variable | variable | 0.555556 | 0.541062802 | 108/199 (0.5427 | NA | 116/215 (0.53953 | NA | 0.455 |
| 20% | 50 | medium | medium | 0.653333 | 0.6225 | 133/217 (0.6129 | NA | 116/183 (0.63387 | NA | 0.1725 |
| 20% | 50 | medium | variable | 0.66 | 0.634328358 | 131/208 (0.6298 | NA | 124/194 (0.63917 | NA | 0.0975 |
| 20% | 50 | variable | medium | 0.546667 | 0.549751244 | 115/210 (0.5476 | NA | 106/192 (0.55208 | NA | 0.2325 |
| 20% | 50 | variable | variable | 0.573529 | 0.52173913 | 102/195 (0.5230 | NA | 114/219 (0.52054 | NA | 0.4375 |
| 25% | 50 | medium | medium | 0.688333 | 0.67 | 132/196 (0.6734 | NA | 136/204 (0.66666 | NA | 0.055 |
| 25% | 50 | medium | variable | 0.7 | 0.68159204 | 141/209 (0.6746 | 4/4 (1.0) | 133/193 (0.68911 | NA | 0.0525 |
| 25% | 50 | variable | medium | 0.615 | 0.544776119 | 93/168 (0.55357 | NA | 126/234 (0.53846 | NA | 0.435 |
| 25% | 50 | variable | variable | 0.580065 | 0.579710145 | 108/183 (0.5901 | NA | 132/231 (0.57142 | NA | 0.385 |
| 50% | 50 | medium | medium | 0.893333 | 0.88 | 175/198 (0.8838 | 98/98 (1.0) | 177/202 (0.87623 | 74/74 (1.0) | 0.03 |
| 50% | 50 | medium | variable | 0.855 | 0.890547264 | 178/199 (0.8944 | 71/71 (1.0) | 180/203 (0.88669 | 70/71 (0.985915 | 0.035 |
| 50% | 50 | variable | medium | 0.696667 | 0.7039801 | 134/186 (0.7204 | NA | 149/216 (0.68981 | NA | 0.23 |
| 50% | 50 | variable | variable | 0.712418 | 0.70531401 | 138/191 (0.7225 | 11/11 (1.0) | 154/223 (0.69058 | 2/2 (1.0) | 0.225 |
| 75% | 50 | medium | medium | 0.975 | 0.9625 | 193/201 (0.9601 | 168/168 (1.0) | 192/199 (0.96482 | 161/161 (1.0) | 0.0025 |
| 75% | 50 | medium | variable | 0.971667 | 0.990049751 | 199/201 (0.9900 | 178/178 (1.0) | 199/201 (0.99004 | 171/171 (1.0) | 0.0025 |
| 75% | 50 | variable | medium | 0.86 | 0.855721393 | 170/197 (0.8629 | 90/92 (0.978260869 | 174/205 (0.84878 | 66/69 (0.956521 | 0.0525 |
| 75% | 50 | variable | variable | 0.848039 | 0.835748792 | 174/209 (0.8325 | 97/99 (0.979797979 | 172/205 (0.83902 | 80/81 (0.987654 | 0.0675 |
| 100% | 50 | medium | medium | 0.993333 | 1 | 200/200 (1.0) | 199/199 (1.0) | 200/200 (1.0) | 196/196 (1.0) | 0 |
| 100% | 50 | medium | variable | 0.991667 | 1 | 201/201 (1.0) | 199/199 (1.0) | 201/201 (1.0) | 195/195 (1.0) | 0 |
| 100% | 50 | variable | medium | 0.99 | 0.990049751 | 198/199 (0.9949 | 192/192 (1.0) | 200/203 (0.98522 | 196/197 (0.9949 | 0 |
| 100% | 50 | variable | variable | 0.99183 | 0.990338164 | 205/207 (0.9903 | 198/198 (1.0) | 205/207 (0.99033 | 192/192 (1.0) | 0 |
| 5% | 100 | medium | medium | 0.576667 | 0.5125 | 90/175 (0.51428 | NA | 115/225 (0.51111 | NA | 0.2475 |
| 5% | 100 | medium | variable | 0.553333 | 0.502487562 | 84/167 (0.50299 | NA | 118/235 (0.50212 | NA | 0.55 |
| 5% | 100 | variable | medium | 0.535 | 0.502487562 | 114/227 (0.5022 | NA | 88/175 (0.502857 | NA | 0.4175 |
| 5% | 100 | variable | variable | 0.527778 | 0.485507246 | 96/198 (0.48484 | NA | 105/216 (0.48611 | NA | 0.76 |
| 10% | 100 | medium | medium | 0.571667 | 0.5275 | 110/209 (0.5263 | NA | 101/191 (0.52879 | NA | 0.16 |
| 10% | 100 | medium | variable | 0.588333 | 0.532338308 | 25/37 (0.675675 | NA | 189/365 (0.51780 | NA | 0.56 |
| 10% | 100 | variable | medium | 0.538333 | 0.547263682 | 39/59 (0.661016 | NA | 181/343 (0.52769 | NA | 0.9875 |
| 10% | 100 | variable | variable | 0.562092 | 0.487922705 | 109/223 (0.4887 | NA | 93/191 (0.486910 | NA | 0.4375 |
| 15% | 100 | medium | medium | 0.623333 | 0.595 | 115/192 (0.5989 | NA | 123/208 (0.59134 | NA | 0.085 |
| 15% | 100 | medium | variable | 0.653333 | 0.616915423 | 116/185 (0.6270 | NA | 132/217 (0.60829 | NA | 0.0825 |
| 15% | 100 | variable | medium | 0.556667 | 0.542288557 | 82/147 (0.55782 | NA | 136/255 (0.53333 | NA | 0.505 |
| 15% | 100 | variable | variable | 0.550654 | 0.543478261 | 98/178 (0.55056 | NA | 127/236 (0.53813 | NA | 0.35 |
| 20% | 100 | medium | medium | 0.676667 | 0.6575 | 125/187 (0.6684 | NA | 138/213 (0.64788 | NA | 0.075 |
| 20% | 100 | medium | variable | 0.681667 | 0.694029851 | 151/224 (0.6741 | NA | 128/178 (0.71910 | NA | 0.085 |
| 20% | 100 | variable | medium | 0.6 | 0.544776119 | 98/178 (0.55056 | NA | 121/224 (0.54017 | NA | 0.38 |
| 20% | 100 | variable | variable | 0.552288 | 0.572463768 | 103/176 (0.5852 | NA | 134/238 (0.56302 | NA | 0.4175 |
| 25% | 100 | medium | medium | 0.738333 | 0.73 | 138/184 (0.75) | 7/7 (1.0) | 154/216 (0.71296 | 5/5 (1.0) | 0.06 |
| 25% | 100 | medium | variable | 0.755 | 0.753731343 | 149/196 (0.7602 | NA | 154/206 (0.74757 | 5/5 (1.0) | 0.06 |
| 25% | 100 | variable | medium | 0.596667 | 0.572139303 | 102/175 (0.5828 | NA | 128/227 (0.56387 | NA | 0.3825 |
| 25% | 100 | variable | variable | 0.589869 | 0.586956522 | 100/164 (0.6097 | NA | 143/250 (0.572) | NA | 0.4025 |
| 50% | 100 | medium | medium | 0.92 | 0.9175 | 184/201 (0.9154 | 125/126 (0.99206349 | 183/199 (0.91959 | 133/133 (1.0) | 0.0075 |
| 50% | 100 | medium | variable | 0.906667 | 0.915422886 | 188/209 (0.8995 | 110/111 (0.99099099 | 180/193 (0.93264 | 101/102 (0.9901 | 0.01 |
| 50% | 100 | variable | medium | 0.696667 | 0.671641791 | 135/201 (0.6716 | NA | 135/201 (0.67164 | NA | 0.185 |
| 50% | 100 | variable | variable | 0.715686 | 0.712560386 | 144/200 (0.72) | 6/7 (0.857142857143 | 151/214 (0.70560 | NA | 0.18 |
| 75% | 100 | medium | medium | 0.983333 | 0.99 | 198/200 (0.99) | 184/184 (1.0) | 198/200 (0.99) | 188/188 (1.0) | 0 |
| 75% | 100 | medium | variable | 0.985 | 1 | 201/201 (1.0) | 197/197 (1.0) | 201/201 (1.0) | 183/183 (1.0) | 0 |
| 75% | 100 | variable | medium | 0.841667 | 0.805970149 | 155/187 (0.8288 | 86/88 (0.977272727 | 169/215 (0.78604 | 54/57 (0.947368 | 0.0175 |
| 75% | 100 | variable | variable | 0.857843 | 0.847826087 | 160/176 (0.90909 | 104/106 (0.9811320 | 191/238 (0.80252 | 74/74 (1.0) | 0.0475 |
| 100% | 100 | medium | medium | 0.995 | 0.995 | 200/202 (0.99009 | 200/200 (1.0) | 198/198 (1.0) | 195/195 (1.0) | 0 |
| 100% | 100 | medium | variable | 0.998333 | 0.997512438 | 201/202 (0.99504 | 201/201 (1.0) | 200/200 (1.0) | 194/194 (1.0) | 0 |
| 100% | 100 | variable | medium | 1 | 1 | 201/201 (1.0) | 199/199 (1.0) | 201/201 (1.0) | 199/199 (1.0) | 0 |
| 100% | 100 | variable | variable | 0.998366 | 0.997584541 | 207/208 (0.99519 | 205/205 (1.0) | 206/206 (1.0) | 197/197 (1.0) | 0 |
| 5% | 500 | medium | medium | 0.54 | 0.495 | 83/168 (0.49404 | NA | 115/232 (0.49568 | NA | 0.2375 |
| 5% | 500 | medium | variable | 0.543333 | 0.472636816 | 59/129 (0.45736 | NA | 131/273 (0.47985 | NA | 0.815 |
| 5% | 500 | variable | medium | 0.546667 | 0.475124378 | 106/222 (0.4774 | NA | 85/180 (0.47222 | NA | 0.38 |
| 5% | 500 | variable | variable | 0.504902 | 0.473429952 | 56/123 (0.45528 | NA | 140/291 (0.48109 | NA | 0.5925 |
| 10% | 500 | medium | medium | 0.578333 | 0.6125 | 121/197 (0.6142 | NA | 124/203 (0.61083 | NA | 0.0675 |
| 10% | 500 | medium | variable | 0.611667 | 0.592039801 | 118/199 (0.5929 | NA | 120/203 (0.59113 | NA | 0.115 |
| 10% | 500 | variable | medium | 0.568333 | 0.517412935 | 81/155 (0.52258 | NA | 127/247 (0.51417 | NA | 0.3625 |
| 10% | 500 | variable | variable | 0.544118 | 0.548309179 | 62/104 (0.59615 | NA | 165/310 (0.53225 | NA | 0.665 |
| 15% | 500 | medium | medium | 0.67 | 0.655 | 127/192 (0.6614 | NA | 135/208 (0.64903 | NA | 0.0525 |
| 15% | 500 | medium | variable | 0.65 | 0.646766169 | 127/195 (0.6512 | NA | 133/207 (0.64251 | NA | 0.0525 |
| 15% | 500 | variable | medium | 0.57 | 0.552238806 | 131/241 (0.5435 | NA | 91/161 (0.565217 | NA | 0.0925 |
| 15% | 500 | variable | variable | 0.553922 | 0.553140097 | 119/216 (0.5509 | NA | 110/198 (0.55555 | NA | 0.33 |
| 20% | 500 | medium | medium | 0.726667 | 0.7075 | 140/197 (0.7106 | NA | 143/203 (0.70443 | 5/5 (1.0) | 0.045 |
| 20% | 500 | medium | variable | 0.746667 | 0.773631841 | 155/200 (0.775) | 16/17 (0.941176470 | 156/202 (0.77227 | 7/7 (1.0) | 0.0225 |
| 20% | 500 | variable | medium | 0.573333 | 0.534825871 | 97/180 (0.53888 | NA | 118/222 (0.53153 | NA | 0.3725 |
| 20% | 500 | variable | variable | 0.560458 | 0.524154589 | 145/280 (0.5178 | NA | 72/134 (0.537313 | NA | 0.0875 |
| 25% | 500 | medium | medium | 0.805 | 0.7725 | 155/201 (0.7711 | 48/51 (0.941176470 | 154/199 (0.77386 | 36/37 (0.972972 | 0.0125 |
| 25% | 500 | medium | variable | 0.781667 | 0.788557214 | 160/204 (0.7843 | 36/37 (0.972972972 | 157/198 (0.79292 | 26/26 (1.0) | 0.0175 |
| 25% | 500 | variable | medium | 0.606667 | 0.542288557 | 70/123 (0.56910 | NA | 148/279 (0.53046 | NA | 0.6675 |
| 25% | 500 | variable | variable | 0.593137 | 0.553140097 | 105/188 (0.5585 | NA | 124/226 (0.54867 | NA | 0.2875 |
| 50% | 500 | medium | medium | 0.958333 | 0.965 | 192/198 (0.9696 | 161/161 (1.0) | 194/202 (0.96039 | 159/160 (0.9937 | 0 |
| 50% | 500 | medium | variable | 0.96 | 0.955223881 | 195/207 (0.9420 | 172/173 (0.9942196 | 189/195 (0.96923 | 152/152 (1.0) | 0 |
| 50% | 500 | variable | medium | 0.686667 | 0.664179104 | 123/180 (0.6833 | 4/4 (1.0) | 144/222 (0.64864 | NA | 0.025 |
| 50% | 500 | variable | variable | 0.686275 | 0.623188406 | 126/201 (0.6268 | NA | 132/213 (0.61971 | NA | 0.2325 |
| 75% | 500 | medium | medium | 0.996667 | 1 | 200/200 (1.0) | 196/196 (1.0) | 200/200 (1.0) | 185/185 (1.0) | 0 |
| 75% | 500 | medium | variable | 0.991667 | 0.997512438 | 201/202 (0.99504 | 197/197 (1.0) | 200/200 (1.0) | 187/187 (1.0) | 0 |
| 75% | 500 | variable | medium | 0.815 | 0.778606965 | 149/186 (0.8010 | 71/71 (1.0) | 164/216 (0.75925 | 49/49 (1.0) | 0 |
| 75% | 500 | variable | variable | 0.815359 | 0.833333333 | 155/172 (0.9011 | 92/92 (1.0) | 190/242 (0.78512 | NA | 0 |
| 100% | 500 | medium | medium | 1 | 1 | 200/200 (1.0) | 200/200 (1.0) | 200/200 (1.0) | 200/200 (1.0) | 0 |
| 100% | 500 | medium | variable | 1 | 1 | 201/201 (1.0) | 201/201 (1.0) | 201/201 (1.0) | 201/201 (1.0) | 0 |
| 100% | 500 | variable | medium | 1 | 1 | 201/201 (1.0) | 200/200 (1.0) | 201/201 (1.0) | 201/201 (1.0) | 0 |
| 100% | 500 | variable | variable | 1 | 1 | 207/207 (1.0) | 207/207 (1.0) | 207/207 (1.0) | 207/207 (1.0) | 0 |

*medium=$1.2\times10^{-8}$; variable=uniformly drawn from [$6.0\times10^{-9}$, $1.2\times10^{-8}$, $2.4\times10^{-8}$]

**medium=$1.0\times10^{-8}$; variable=uniformly drawn from [$1\times10^{-9}$, $1\times10^{-8}$, $1\times10^{-7}$]

**Table S6: Accuracy assessments of a single SVM tested against various simulated data sets.**
**The SVM was trained from the set where constrained examples experience selection at 75% sites, with 2Ns=100, and variable mutation and recombination rates.**

| Fraction of selected sites | Strength of selection (2*Ns*) | Mutation rate* | Recombin-ation rate** | Test set accuracy | Accuracy on unconstrained data | Accuracy on unconstrained data after >95% cutoff | Accuracy on constrained data | Accuracy on constrained data after >95% cutoff |
|---|---|---|---|---|---|---|---|---|
| 5 | 50 | high | medium | 0.5025 | 200/399 (0.501 | 189/375 (0.504 | NA | NA |
| 5 | 50 | low | medium | 0.4925 | 94/191 (0.4921 | 11/26 (0.42307 | 103/209 (0.49 | 5/10 (0.5) |
| 5 | 50 | medium | high | 0.5 | 188/376 (0.5) | 99/199 (0.4974 | 12/24 (0.5) | 2/2 (1.0) |
| 5 | 50 | medium | low | 0.5075 | 186/369 (0.504 | 115/198 (0.580 | 17/31 (0.5483 | NA |
| 5 | 50 | medium | medium | 0.5075 | 185/367 (0.504 | 111/216 (0.513 | 18/33 (0.5454 | 3/3 (1.0) |
| 5 | 50 | medium | variable | 0.532338 | 193/373 (0.517 | 115/208 (0.552 | 21/29 (0.7241 | 2/2 (1.0) |
| 5 | 50 | variable | medium | 0.5 | 156/312 (0.5) | 103/196 (0.525 | 45/90 (0.5) | NA |
| 5 | 50 | variable | variable | 0.512077 | 164/323 (0.507 | 107/209 (0.511 | 48/91 (0.5274 | NA |
| 10 | 50 | high | medium | 0.505 | 200/398 (0.502 | 189/365 (0.517 | 2/2 (1.0) | NA |
| 10 | 50 | low | medium | 0.505 | 94/186 (0.5053 | 9/17 (0.529411 | 108/214 (0.50 | 3/8 (0.375) |
| 10 | 50 | medium | high | 0.5125 | 188/371 (0.506 | 104/194 (0.536 | 17/29 (0.5862 | NA |
| 10 | 50 | medium | low | 0.525 | 187/364 (0.513 | 118/190 (0.621 | 23/36 (0.6388 | NA |
| 10 | 50 | medium | medium | 0.495 | 186/374 (0.497 | 118/219 (0.538 | 12/26 (0.4615 | NA |
| 10 | 50 | medium | variable | 0.524876 | 186/362 (0.513 | 105/183 (0.573 | 25/40 (0.625 | NA |
| 10 | 50 | variable | medium | 0.534826 | 168/322 (0.521 | 102/191 (0.534 | 47/80 (0.5875 | NA |
| 10 | 50 | variable | variable | 0.519324 | 169/330 (0.512 | 100/195 (0.512 | 46/84 (0.5476 | 3/4 (0.75) |
| 15 | 50 | high | medium | 0.5075 | 200/397 (0.503 | 190/367 (0.517 | 3/3 (1.0) | NA |
| 15 | 50 | low | medium | 0.53 | 93/174 (0.5344 | 9/14 (0.642857 | 119/226 (0.52 | 7/12 (0.5833) |
| 15 | 50 | medium | high | 0.5275 | 187/363 (0.515 | 104/180 (0.577 | 24/37 (0.6486 | NA |
| 15 | 50 | medium | low | 0.54 | 188/360 (0.522 | 119/193 (0.616 | 28/40 (0.7) | NA |
| 15 | 50 | medium | medium | 0.5175 | 186/365 (0.509 | 110/189 (0.582 | 21/35 (0.6) | 2/2 (1.0) |
| 15 | 50 | medium | variable | 0.549751 | 192/364 (0.527 | 112/188 (0.595 | 29/38 (0.7631 | NA |
| 15 | 50 | variable | medium | 0.529851 | 168/324 (0.518 | 100/180 (0.555 | 45/78 (0.5769 | 2/4 (0.5) |
| 15 | 50 | variable | variable | 0.519324 | 172/336 (0.511 | 104/184 (0.565 | 43/78 (0.5512 | 1/2 (0.5) |
| 20 | 50 | high | medium | 0.5125 | 200/395 (0.506 | 189/349 (0.541 | 5/5 (1.0) | NA |
| 20 | 50 | low | medium | 0.5725 | 95/161 (0.5900 | 9/13 (0.692307 | 134/239 (0.56 | 11/16 (0.6875) |
| 20 | 50 | medium | high | 0.55 | 188/356 (0.528 | 103/161 (0.639 | 32/44 (0.7272 | NA |
| 20 | 50 | medium | low | 0.54 | 186/356 (0.522 | 109/161 (0.677 | 30/44 (0.6818 | NA |
| 20 | 50 | medium | medium | 0.53 | 186/360 (0.516 | 115/183 (0.628 | 26/40 (0.65) | NA |
| 20 | 50 | medium | variable | 0.529851 | 189/366 (0.516 | 109/171 (0.637 | 24/36 (0.6666 | NA |
| 20 | 50 | variable | medium | 0.522388 | 158/307 (0.514 | 95/168 (0.5654 | 52/95 (0.5473 | 2/3 (0.6667) |
| 20 | 50 | variable | variable | 0.550725 | 172/323 (0.532 | 112/189 (0.592 | 56/91 (0.6153 | 4/5 (0.8) |
| 25 | 50 | high | medium | 0.5125 | 200/395 (0.506 | 189/324 (0.583 | 5/5 (1.0) | NA |
| 25 | 50 | low | medium | 0.57 | 94/160 (0.5875 | 9/12 (0.75) | 134/240 (0.55 | 5/9 (0.55556) |
| 25 | 50 | medium | high | 0.5825 | 187/341 (0.548 | 106/161 (0.658 | 46/59 (0.7796 | 4/4 (1.0) |
| 25 | 50 | medium | low | 0.575 | 185/340 (0.544 | 119/177 (0.672 | 45/60 (0.75) | 2/2 (1.0) |
| 25 | 50 | medium | medium | 0.5675 | 187/347 (0.538 | 109/175 (0.622 | 40/53 (0.7547 | 2/2 (1.0) |
| 25 | 50 | medium | variable | 0.58209 | 187/341 (0.548 | 96/145 (0.6620 | 47/61 (0.7704 | 2/2 (1.0) |
| 25 | 50 | variable | medium | 0.574627 | 165/300 (0.55) | 95/165 (0.5757 | 66/102 (0.647 | 7/9 (0.777777777778) |
| 25 | 50 | variable | variable | 0.586957 | 178/320 (0.556 | 104/168 (0.619 | 65/94 (0.6914 | 3/4 (0.75) |
| 50 | 50 | high | medium | 0.62 | 200/352 (0.568 | 189/255 (0.741 | 48/48 (1.0) | 4/4 (1.0) |
| 50 | 50 | low | medium | 0.6825 | 94/115 (0.8173 | 8/8 (1.0) | 179/285 (0.62 | 20/25 (0.8) |
| 50 | 50 | medium | high | 0.7525 | 188/275 (0.683 | 104/113 (0.920 | 113/125 (0.90 | 22/22 (1.0) |
| 50 | 50 | medium | low | 0.7225 | 187/285 (0.656 | 112/119 (0.941 | 102/115 (0.88 | 12/12 (1.0) |
| 50 | 50 | medium | medium | 0.7225 | 185/281 (0.658 | 109/126 (0.865 | 104/119 (0.87 | 14/14 (1.0) |
| 50 | 50 | medium | variable | 0.726368 | 190/289 (0.657 | 110/121 (0.909 | 102/113 (0.90 | 14/14 (1.0) |
| 50 | 50 | variable | medium | 0.686567 | 161/247 (0.651 | 105/137 (0.766 | 115/155 (0.74 | 16/18 (0.888888888889) |
| 50 | 50 | variable | variable | 0.681159 | 169/263 (0.642 | 115/139 (0.827 | 113/151 (0.74 | 18/19 (0.947368421053) |
| 75 | 50 | high | medium | 0.865 | 200/254 (0.787 | 189/191 (0.989 | 146/146 (1.0) | 43/43 (1.0) |
| 75 | 50 | low | medium | 0.715 | 94/102 (0.9215 | 9/9 (1.0) | 192/298 (0.64 | 65/70 (0.928571428571) |
| 75 | 50 | medium | high | 0.9175 | 188/209 (0.899 | 100/102 (0.980 | 179/191 (0.93 | 74/74 (1.0) |
| 75 | 50 | medium | low | 0.915 | 186/206 (0.902 | 116/116 (1.0) | 180/194 (0.92 | 67/67 (1.0) |
| 75 | 50 | medium | medium | 0.895 | 187/216 (0.865 | 110/111 (0.990 | 171/184 (0.92 | 69/69 (1.0) |
| 75 | 50 | medium | variable | 0.893035 | 183/208 (0.879 | 112/114 (0.982 | 176/194 (0.90 | 62/62 (1.0) |
| 75 | 50 | variable | medium | 0.838308 | 165/194 (0.850 | 96/96 (1.0) | 172/208 (0.82 | 76/78 (0.974358974359) |
| 75 | 50 | variable | variable | 0.833333 | 171/204 (0.838 | 109/114 (0.956 | 174/210 (0.82 | 69/71 (0.971830985915) |
| 100 | 50 | high | medium | 1 | 200/200 (1.0) | 188/188 (1.0) | 200/200 (1.0) | 198/198 (1.0) |
| 100 | 50 | low | medium | 0.7375 | 95/95 (1.0) | 9/9 (1.0) | 200/305 (0.65 | 158/163 (0.969325153374) |
| 100 | 50 | medium | high | 0.97 | 188/188 (1.0) | 100/100 (1.0) | 200/212 (0.94 | 187/187 (1.0) |
| 100 | 50 | medium | low | 0.9675 | 187/187 (1.0) | 114/114 (1.0) | 200/213 (0.93 | 182/182 (1.0) |
| 100 | 50 | medium | medium | 0.9675 | 187/187 (1.0) | 109/109 (1.0) | 200/213 (0.93 | 181/181 (1.0) |
| 100 | 50 | medium | variable | 0.967662 | 188/188 (1.0) | 101/101 (1.0) | 201/214 (0.93 | 182/182 (1.0 |
| 100 | 50 | variable | medium | 0.895522 | 159/159 (1.0) | 102/102 (1.0) | 201/243 (0.82 | 178/178 (1.0 |
| 100 | 50 | variable | variable | 0.903382 | 167/167 (1.0) | 107/107 (1.0) | 207/247 (0.83 | 181/182 (0.994505494505) |
| 5 | 100 | high | medium | 0.5025 | 200/399 (0.501 | 190/376 (0.505 | NA | NA |
| 5 | 100 | low | medium | 0.5225 | 95/181 (0.5248 | 9/15 (0.6) | 114/219 (0.52 | 4/9 (0.444444444444) |
| 5 | 100 | medium | high | 0.5075 | 187/371 (0.504 | 100/194 (0.515 | 16/29 (0.5517 | NA |
| 5 | 100 | medium | low | 0.5 | 189/378 (0.5) | 112/199 (0.562 | 11/22 (0.5) | NA |
| 5 | 100 | medium | medium | 0.5125 | 184/363 (0.506 | 107/209 (0.511 | 21/37 (0.5675 | NA |
| 5 | 100 | medium | variable | 0.514925 | 188/370 (0.508 | 95/178 (0.5337 | 19/32 (0.5937 | NA |
| 5 | 100 | variable | medium | 0.49005 | 153/310 (0.493 | 91/185 (0.4918 | 44/92 (0.4782 | 3/8 (0.375) |
| 5 | 100 | variable | variable | 0.492754 | 156/315 (0.495 | 100/186 (0.537 | 48/99 (0.4848 | 3/5 (0.6) |
| 10 | 100 | high | medium | 0.505 | 200/398 (0.502 | 191/371 (0.514 | 2/2 (1.0) | NA |
| 10 | 100 | low | medium | 0.485 | 95/196 (0.4846 | 10/25 (0.4) | 99/204 (0.485 | 5/10 (0.5) |
| 10 | 100 | medium | high | 0.4975 | 188/377 (0.498 | 103/185 (0.556 | 11/23 (0.4782 | NA |
| 10 | 100 | medium | low | 0.525 | 187/364 (0.513 | 114/202 (0.564 | 23/36 (0.6388 | 3/3 (1.0) |
| 10 | 100 | medium | medium | 0.5225 | 185/361 (0.512 | 111/196 (0.566 | 24/39 (0.6153 | NA |
| 10 | 100 | medium | variable | 0.524876 | 187/364 (0.513 | 113/193 (0.585 | 24/38 (0.6315 | 3/3 (1.0) |
| 10 | 100 | variable | medium | 0.534826 | 161/308 (0.522 | 109/197 (0.553 | 54/94 (0.5744 | 4/6 (0.666666666667) |
| 10 | 100 | variable | variable | 0.502415 | 164/327 (0.501 | 95/178 (0.5337 | 44/87 (0.5057 | 3/5 (0.6) |
| 15 | 100 | high | medium | 0.5025 | 200/399 (0.501 | 190/363 (0.523 | NA | NA |
| 15 | 100 | low | medium | 0.5425 | 94/171 (0.5497 | 9/18 (0.5) | 123/229 (0.53 | 10/15 (0.666666666667) |
| 15 | 100 | medium | high | 0.5375 | 188/361 (0.520 | 100/177 (0.564 | 27/39 (0.6923 | NA |
| 15 | 100 | medium | low | 0.5425 | 187/357 (0.523 | 114/176 (0.647 | 30/43 (0.6976 | 5/5 (1.0) |
| 15 | 100 | medium | medium | 0.53 | 186/360 (0.516 | 114/195 (0.584 | 26/40 (0.65) | 4/4 (1.0) |
| 15 | 100 | medium | variable | 0.527363 | 188/365 (0.515 | 110/195 (0.564 | 24/37 (0.6486 | 2/3 (0.666666666667) |
| 15 | 100 | variable | medium | 0.512438 | 157/309 (0.508 | 102/185 (0.551 | 49/93 (0.5268 | 4/4 (1.0) |
| 15 | 100 | variable | variable | 0.541063 | 168/319 (0.526 | 100/175 (0.571 | 56/95 (0.5894 | 4/5 (0.8) |
| 20 | 100 | high | medium | 0.5075 | 200/397 (0.503 | 190/355 (0.535 | 3/3 (1.0) | NA |
| 20 | 100 | low | medium | 0.55 | 94/168 (0.5595 | 9/14 (0.642857 | 126/232 (0.54 | 10/15 (0.666666666667) |
| 20 | 100 | medium | high | 0.535 | 187/360 (0.519 | 103/169 (0.609 | 27/40 (0.675) | NA |
| 20 | 100 | medium | low | 0.555 | 186/350 (0.531 | 111/163 (0.680 | 36/50 (0.72) | 3/3 (1.0) |
| 20 | 100 | medium | medium | 0.5725 | 185/341 (0.542 | 107/165 (0.648 | 44/59 (0.7457 | 3/3 (1.0) |
| 20 | 100 | medium | variable | 0.567164 | 191/355 (0.538 | 124/185 (0.670 | 37/47 (0.7872 | 2/2 (1.0) |
| 20 | 100 | variable | medium | 0.539801 | 158/300 (0.526 | 97/174 (0.5574 | 59/102 (0.578 | 4/6 (0.666666666667) |
| 20 | 100 | variable | variable | 0.545894 | 166/313 (0.530 | 110/190 (0.578 | 60/101 (0.594 | 4/6 (0.666666666667) |
| 25 | 100 | high | medium | 0.515 | 200/394 (0.507 | 188/345 (0.544 | 6/6 (1.0) | NA |
| 25 | 100 | low | medium | 0.59 | 95/154 (0.6168 | 9/12 (0.75) | 141/246 (0.57 | 13/18 (0.722222222222) |
| 25 | 100 | medium | high | 0.585 | 188/342 (0.549 | 103/151 (0.682 | 46/58 (0.7931 | 5/5 (1.0) |
| 25 | 100 | medium | low | 0.5525 | 188/355 (0.529 | 115/159 (0.723 | 33/45 (0.7333 | 3/3 (1.0) |
| 25 | 100 | medium | medium | 0.58 | 185/338 (0.547 | 111/145 (0.765 | 47/62 (0.7580 | 3/3 (1.0) |
| 25 | 100 | medium | variable | 0.542289 | 191/365 (0.523 | 100/153 (0.653 | 27/37 (0.7297 | 2/2 (1.0) |
| 25 | 100 | variable | medium | 0.562189 | 165/305 (0.540 | 101/165 (0.612 | 61/97 (0.6288 | 6/11 (0.545454545455) |
| 25 | 100 | variable | variable | 0.586957 | 175/314 (0.557 | 109/191 (0.570 | 68/100 (0.68 | 6/7 (0.857142857143) |
| 50 | 100 | high | medium | 0.595 | 200/362 (0.552 | 190/258 (0.736 | 38/38 (1.0) | 2/2 (1.0) |
| 50 | 100 | low | medium | 0.66 | 95/126 (0.7539 | 9/9 (1.0) | 169/274 (0.61 | 22/27 (0.814814814815) |
| 50 | 100 | medium | high | 0.76 | 188/272 (0.691 | 103/117 (0.880 | 116/128 (0.90 | 21/21 (1.0) |
| 50 | 100 | medium | low | 0.7125 | 191/297 (0.643 | 112/126 (0.888 | 94/103 (0.912 | 10/10 (1.0) |
| 50 | 100 | medium | medium | 0.755 | 187/272 (0.687 | 109/116 (0.939 | 115/128 (0.89 | 13/13 (1.0) |
| 50 | 100 | medium | variable | 0.748756 | 187/274 (0.682 | 100/111 (0.900 | 114/128 (0.89 | 14/14 (1.0) |
| 50 | 100 | variable | medium | 0.669154 | 159/250 (0.636 | 99/122 (0.8114 | 110/152 (0.72 | 22/22 (1.0) |
| 50 | 100 | variable | variable | 0.647343 | 167/273 (0.611 | 107/126 (0.849 | 101/141 (0.71 | 9/9 (1.0) |
| 75 | 100 | high | medium | 0.8825 | 200/247 (0.809 | 189/192 (0.984 | 153/153 (1.0) | 48/48 (1.0) |
| 75 | 100 | low | medium | 0.73 | 95/98 (0.96938 | 9/9 (1.0) | 197/302 (0.65 | 72/77 (0.935064935065) |
| 75 | 100 | medium | high | 0.9225 | 188/207 (0.908 | 100/101 (0.990 | 181/193 (0.93 | 79/79 (1.0) |
| 75 | 100 | medium | low | 0.925 | 188/206 (0.912 | 111/111 (1.0) | 182/194 (0.93 | 54/54 (1.0) |
| 75 | 100 | medium | medium | 0.905 | 187/212 (0.882 | 108/110 (0.981 | 175/188 (0.93 | 82/82 (1.0) |
| 75 | 100 | medium | variable | 0.90796 | 187/210 (0.890 | 89/89 (1.0) | 178/192 (0.92 | 75/76 (0.986842105263) |
| 75 | 100 | variable | medium | 0.848259 | 162/184 (0.880 | 96/99 (0.96969 | 179/218 (0.82 | 63/64 (0.984375) |
| 75 | 100 | variable | variable | 0.847826 | 160/176 (0.909 | 104/106 (0.981 | 191/238 (0.80 | 74/74 (1.0) |
| 100 | 100 | high | medium | 1 | 200/200 (1.0) | 188/188 (1.0) | 200/200 (1.0) | 200/200 (1.0) |
| 100 | 100 | low | medium | 0.7375 | 95/95 (1.0) | 9/9 (1.0) | 200/305 (0.65 | 171/176 (0.971590909091) |
| 100 | 100 | medium | high | 0.97 | 188/188 (1.0) | 100/100 (1.0) | 200/212 (0.94 | 197/197 (1.0) |
| 100 | 100 | medium | low | 0.9675 | 187/187 (1.0) | 114/114 (1.0) | 200/213 (0.93 | 195/195 (1.0) |
| 100 | 100 | medium | medium | 0.9675 | 187/187 (1.0) | 109/109 (1.0) | 200/213 (0.93 | 196/196 (1.0) |
| 100 | 100 | medium | variable | 0.970149 | 189/189 (1.0) | 111/111 (1.0) | 201/213 (0.94 | 196/196 (1.0 |
| 100 | 100 | variable | medium | 0.900498 | 161/161 (1.0) | 101/101 (1.0) | 201/241 (0.83 | 183/183 (1.0 |
| 100 | 100 | variable | variable | 0.881643 | 158/158 (1.0) | 99/99 (1.0) | 207/256 (0.80 | 190/191 (0.994764397906) |
| 5 | 500 | high | medium | 0.5 | 200/400 (0.5) | 189/376 (0.502 | NA | NA |
| 5 | 500 | low | medium | 0.49 | 96/196 (0.4897 | 9/22 (0.409090 | 100/204 (0.49 | 4/9 (0.444444444444) |
| 5 | 500 | medium | high | 0.4975 | 188/377 (0.498 | 104/205 (0.507 | 11/23 (0.4782 | NA |
| 5 | 500 | medium | low | 0.5225 | 188/367 (0.512 | 115/202 (0.569 | 21/33 (0.6363 | NA |
| 5 | 500 | medium | medium | 0.4925 | 183/369 (0.495 | 112/224 (0.5) | 14/31 (0.4516 | NA |
| 5 | 500 | medium | variable | 0.497512 | 187/375 (0.498 | 110/213 (0.516 | 13/27 (0.4814 | NA |
| 5 | 500 | variable | medium | 0.504975 | 162/322 (0.503 | 103/194 (0.530 | 41/80 (0.5125 | 1/4 (0.25) |
| 5 | 500 | variable | variable | 0.478261 | 166/341 (0.486 | 110/217 (0.506 | 32/73 (0.4383 | 3/8 (0.375) |
| 10 | 500 | high | medium | 0.5 | 200/400 (0.5) | 189/367 (0.514 | NA | NA |
| 10 | 500 | low | medium | 0.5225 | 94/179 (0.5251 | 11/18 (0.61111 | 115/221 (0.52 | 7/12 (0.583333333333) |
| 10 | 500 | medium | high | 0.535 | 187/360 (0.519 | 105/193 (0.544 | 27/40 (0.675) | NA |
| 10 | 500 | medium | low | 0.5025 | 185/369 (0.501 | 112/202 (0.554 | 16/31 (0.5161 | NA |
| 10 | 500 | medium | medium | 0.52 | 184/360 (0.511 | 114/203 (0.561 | 24/40 (0.6) | NA |
| 10 | 500 | medium | variable | 0.517413 | 188/369 (0.509 | 102/192 (0.531 | 20/33 (0.6060 | 2/2 (1.0) |
| 10 | 500 | variable | medium | 0.502488 | 156/311 (0.501 | 95/191 (0.4973 | 46/91 (0.5054 | 0/4 (0.0) |
| 10 | 500 | variable | variable | 0.507246 | 171/339 (0.504 | 108/199 (0.542 | 39/75 (0.52) | NA |
| 15 | 500 | high | medium | 0.5125 | 199/393 (0.506 | 189/367 (0.514 | 6/7 (0.857142 | NA |
| 15 | 500 | low | medium | 0.5325 | 95/177 (0.5367 | 8/15 (0.533333 | 118/223 (0.52 | 6/11 (0.545454545455) |
| 15 | 500 | medium | high | 0.5075 | 187/371 (0.504 | 107/182 (0.587 | 16/29 (0.5517 | NA |
| 15 | 500 | medium | low | 0.5325 | 185/357 (0.518 | 112/187 (0.598 | 28/43 (0.6511 | 2/2 (1.0) |
| 15 | 500 | medium | medium | 0.525 | 186/362 (0.513 | 105/189 (0.555 | 24/38 (0.6315 | NA |
| 15 | 500 | medium | variable | 0.507463 | 186/369 (0.504 | 105/184 (0.570 | 18/33 (0.5454 | NA |
| 15 | 500 | variable | medium | 0.564677 | 174/322 (0.540 | 105/195 (0.538 | 53/80 (0.6625 | 3/3 (1.0) |
| 15 | 500 | variable | variable | 0.528986 | 169/326 (0.518 | 117/209 (0.559 | 50/88 (0.5681 | 0/5 (0.0) |
| 20 | 500 | high | medium | 0.505 | 200/398 (0.502 | 189/361 (0.523 | 2/2 (1.0) | NA |
| 20 | 500 | low | medium | 0.5525 | 94/167 (0.5628 | 9/11 (0.818181 | 127/233 (0.54 | 8/13 (0.615384615385) |
| 20 | 500 | medium | high | 0.5525 | 188/355 (0.529 | 101/171 (0.590 | 33/45 (0.7333 | 2/2 (1.0) |
| 20 | 500 | medium | low | 0.5325 | 187/361 (0.518 | 113/183 (0.617 | 26/39 (0.6666 | NA |
| 20 | 500 | medium | medium | 0.5325 | 186/359 (0.518 | 112/185 (0.605 | 27/41 (0.6585 | NA |
| 20 | 500 | medium | variable | 0.549751 | 186/352 (0.528 | 109/172 (0.633 | 35/50 (0.7) | NA |
| 20 | 500 | variable | medium | 0.534826 | 161/308 (0.522 | 97/180 (0.5388 | 54/94 (0.5744 | NA |
| 20 | 500 | variable | variable | 0.490338 | 161/326 (0.493 | 101/191 (0.528 | 42/88 (0.4772 | 3/6 (0.5) |
| 25 | 500 | high | medium | 0.5125 | 200/395 (0.506 | 189/349 (0.541 | 5/5 (1.0) | NA |
| 25 | 500 | low | medium | 0.5675 | 95/163 (0.5828 | 9/12 (0.75) | 132/237 (0.55 | 10/15 (0.666666666667) |
| 25 | 500 | medium | high | 0.5475 | 188/357 (0.526 | 99/155 (0.6387 | 31/43 (0.7209 | NA |
| 25 | 500 | medium | low | 0.55 | 188/356 (0.528 | 116/163 (0.711 | 32/44 (0.7272 | NA |
| 25 | 500 | medium | medium | 0.54 | 185/354 (0.522 | 114/175 (0.651 | 31/46 (0.6739 | 1/2 (0.5) |
| 25 | 500 | medium | variable | 0.569652 | 192/356 (0.539 | 110/166 (0.662 | 37/46 (0.8043 | NA |
| 25 | 500 | variable | medium | 0.529851 | 156/300 (0.52) | 94/173 (0.5433 | 57/102 (0.558 | 4/7 (0.571428571429) |
| 25 | 500 | variable | variable | 0.538647 | 165/314 (0.525 | 115/189 (0.608 | 58/100 (0.58 | 2/3 (0.666666666667) |
| 50 | 500 | high | medium | 0.5475 | 200/381 (0.524 | 188/298 (0.630 | 19/19 (1.0) | NA |
| 50 | 500 | low | medium | 0.635 | 95/136 (0.6985 | 10/13 (0.76923 | 159/264 (0.60 | 13/18 (0.722222222222) |
| 50 | 500 | medium | high | 0.695 | 188/298 (0.630 | 104/119 (0.873 | 90/102 (0.882 | 4/4 (1.0) |
| 50 | 500 | medium | low | 0.67 | 186/304 (0.611 | 116/134 (0.865 | 82/96 (0.8541 | 3/3 (1.0) |
| 50 | 500 | medium | medium | 0.64 | 187/318 (0.588 | 110/132 (0.833 | 69/82 (0.8414 | 8/8 (1.0) |
| 50 | 500 | medium | variable | 0.659204 | 187/310 (0.603 | 110/129 (0.852 | 78/92 (0.8478 | 3/3 (1.0) |
| 50 | 500 | variable | medium | 0.624378 | 156/262 (0.595 | 102/137 (0.744 | 95/140 (0.678 | 8/9 (0.888888888889) |
| 50 | 500 | variable | variable | 0.644928 | 174/288 (0.604 | 106/153 (0.692 | 93/126 (0.738 | 6/7 (0.857142857143) |
| 75 | 500 | high | medium | 0.7275 | 200/309 (0.647 | 189/209 (0.904 | 91/91 (1.0) | 2/2 (1.0) |
| 75 | 500 | low | medium | 0.7025 | 96/111 (0.8648 | 9/9 (1.0) | 185/289 (0.64 | 23/28 (0.821428571429) |
| 75 | 500 | medium | high | 0.8575 | 188/233 (0.806 | 101/104 (0.971 | 155/167 (0.92 | 11/11 (1.0) |
| 75 | 500 | medium | low | 0.8575 | 186/229 (0.812 | 112/113 (0.991 | 157/171 (0.91 | 12/12 (1.0) |
| 75 | 500 | medium | medium | 0.8575 | 186/229 (0.812 | 109/111 (0.981 | 157/171 (0.91 | 11/11 (1.0) |
| 75 | 500 | medium | variable | 0.885572 | 188/221 (0.850 | 97/97 (1.0) | 168/181 (0.92 | 11/11 (1.0) |
| 75 | 500 | variable | medium | 0.748756 | 157/214 (0.733 | 96/102 (0.9411 | 144/188 (0.76 | 12/14 (0.857142857143) |
| 75 | 500 | variable | variable | 0.772947 | 174/235 (0.740 | 102/108 (0.944 | 146/179 (0.81 | 7/8 (0.875) |
| 100 | 500 | high | medium | 1 | 200/200 (1.0) | 188/188 (1.0) | 200/200 (1.0) | 52/52 (1.0) |
| 100 | 500 | low | medium | 0.7375 | 95/95 (1.0) | 9/9 (1.0) | 200/305 (0.65 | 34/39 (0.871794871795) |
| 100 | 500 | medium | high | 0.97 | 188/188 (1.0) | 100/100 (1.0) | 200/212 (0.94 | 70/70 (1.0) |
| 100 | 500 | medium | low | 0.9675 | 187/187 (1.0) | 114/114 (1.0) | 200/213 (0.93 | 12/12 (1.0) |
| 100 | 500 | medium | medium | 0.9675 | 187/187 (1.0) | 109/109 (1.0) | 200/213 (0.93 | 63/63 (1.0) |
| 100 | 500 | medium | variable | 0.9801 | 193/193 (1.0) | 102/102 (1.0) | 201/209 (0.96 | 69/69 (1.0) |
| 100 | 500 | variable | medium | 0.900498 | 161/161 (1.0) | 103/103 (1.0) | 201/241 (0.83 | 74/75 (0.986666666667) |
| 100 | 500 | variable | variable | 0.898551 | 165/165 (1.0) | 103/103 (1.0) | 207/249 (0.83 | 22/23 (0.95652173913) |

*low=$6.0\times10^{-9}$; medium=$1.2\times10^{-8}$; high=$2.4\times10^{-8}$; variable=uniformly drawn from [$6.0\times10^{-9}$, $1.2\times10^{-8}$, $2.4\times10^{-8}$]

**low=$1\times10^{-9}$; medium=$1.0\times10^{-8}$; high=$1\times10^{-7}$; variable=uniformly drawn from [$1\times10^{-9}$, $1\times10^{-8}$, $1\times10^{-7}$]

**Table S7: Mouse anatomical structure-developmental stage pairs (from the Mouse Genome Informatics database) whose expressed genes are enriched for nearby LOF candidate regions (calculated using GREAT)**

| Term Name | Enrichment p-value | Enrichment FDR q value | Fold-enrichment | P-value from permutation test |
|---|---|---|---|---|
| TS15_telencephalon; roof plate | 2.74E-04 | 2.93E-02 | 6.7712 | 0.000 |
| TS19_myelencephalon | 5.34E-04 | 4.54E-02 | 2.9739 | 0.000 |
| TS21_pelvic urethra dorsal mesenchyme | 8.01E-05 | 1.17E-02 | 11.2854 | 0.000 |
| TS21_pelvic urethra ventral mesenchyme | 3.17E-04 | 3.22E-02 | 8.464 | 0.000 |
| TS23_muscle layer of dorsal pelvic urethra of male | 8.01E-05 | 1.17E-02 | 11.2854 | 0.000 |
| TS23_muscle layer of pelvic urethra of female | 6.98E-06 | 1.49E-03 | 5.45 | 0.000 |
| TS23_muscle layer of ventral pelvic urethra of male | 1.21E-04 | 1.60E-02 | 10.3641 | 0.000 |
| TS24_medulla oblongata | 1.52E-06 | 4.37E-04 | 6.3843 | 0.000 |
| TS24_medulla oblongata; lateral wall; basal plate; medullary raphe | 5.90E-05 | 9.11E-03 | 18.467 | 0.000 |
| TS26_heart; ventricle | 2.78E-04 | 2.94E-02 | 5.5983 | 0.000 |
| TS28_inner medulla loop of Henle thin descending limb | 3.73E-05 | 5.99E-03 | 9.6732 | 0.000 |
| TS28_outer medulla loop of Henle thin descending limb | 5.31E-05 | 8.35E-03 | 9.0957 | 0.000 |
| TS16_lens pit | 5.73E-04 | 4.77E-02 | 3.5764 | 0.001 |
| TS23_muscle layer of pelvic urethra of male | 1.62E-04 | 2.07E-02 | 4.197 | 0.001 |
| TS23_remnant of Rathke's pouch | 5.82E-04 | 4.80E-02 | 17.9239 | 0.001 |
| TS28_fourth ventricle choroid plexus | 4.74E-04 | 4.20E-02 | 10.9804 | 0.001 |
| TS16_future forebrain | 4.21E-04 | 3.90E-02 | 2.2018 | 0.002 |
| TS16_optic cup | 4.53E-04 | 4.11E-02 | 3.4167 | 0.003 |
| TS16_diencephalon | 2.26E-04 | 2.69E-02 | 2.9469 | 0.004 |
| TS20_optic stalk | 5.65E-04 | 4.76E-02 | 5.9166 | 0.007 |
| TS18_brain | 1.43E-05 | 2.71E-03 | 2.2025 | 0.008 |
| TS28_inner renal medulla loop of henle | 4.20E-04 | 3.93E-02 | 3.4483 | 0.008 |
| TS18_telencephalon | 3.37E-04 | 3.39E-02 | 2.8371 | 0.011 |
| TS18_forebrain | 2.72E-04 | 2.94E-02 | 2.2724 | 0.012 |
| TS18_hindbrain | 1.65E-05 | 2.92E-03 | 2.7451 | 0.013 |
| TS23_mesenchymal layer of pelvic urethra of female | 2.49E-04 | 2.88E-02 | 2.5292 | 0.024 |
| TS23_pelvic urethra of female | 2.26E-04 | 2.73E-02 | 2.255 | 0.045 |
| TS23_urinary bladder fundus urothelium | 4.39E-04 | 4.02E-02 | 2.3511 | 0.047 |
| TS23_urinary bladder trigone urothelium | 1.99E-04 | 2.52E-02 | 2.5017 | 0.072 |
| TS21_peripheral nervous system | 1.10E-06 | 3.29E-04 | 2.0068 | 0.258 |
| TS21_peripheral nervous system; spinal; ganglion | 8.32E-08 | 4.33E-05 | 2.1817 | 0.269 |
| TS21_dorsal root ganglion | 1.51E-07 | 6.62E-05 | 2.1602 | 0.272 |
| TS21_peripheral nervous system; spinal | 1.74E-07 | 6.89E-05 | 2.1316 | 0.274 |
| TS21_hypothalamus | 2.41E-08 | 2.87E-05 | 2.1519 | 0.287 |
| TS21_thalamus | 1.16E-08 | 1.94E-05 | 2.0836 | 0.294 |
| TS21_midbrain | 2.94E-08 | 2.72E-05 | 2.0338 | 0.298 |
| TS21_telencephalon; olfactory lobe | 4.56E-08 | 3.17E-05 | 2.2028 | 0.338 |
| TS21_retina | 2.42E-08 | 2.52E-05 | 2.0582 | 0.556 |